\documentclass[aps,prd,twocolumn,showpacs,superscriptaddress,groupedaddress]{revtex4}  
\usepackage{graphicx}  
\usepackage{dcolumn}   
\usepackage{bm}        
\usepackage{amssymb}   
\usepackage{amsmath}
\usepackage{appendix}
\usepackage{footnote}
\usepackage{setspace}
\usepackage{enumerate}
\usepackage{color}
\usepackage{multirow} 

\def \ifb    {\ensuremath{ \rm fb^{-1}                          }}

\def \etmisssc {\mbox{\ensuremath{E\kern-0.6em\slash_T^{\rm\kern+0.1em Sc}}}}
\def \etmiss {\mbox{\ensuremath{E\kern-0.6em\slash_T}}}

\newcommand{\Eslash}{\mbox{$E \kern-0.6em\slash$                }}

\def \hww {\ensuremath{ H\to W^+W^- }}

\def\met{{\mbox{$E\kern-0.57em\raise0.19ex\hbox{/}_{T}$}}}
\def\DZ{D0 }

\def\ifb{fb$^{-1}$}
\def\pp{$p\bar{p}$}

\def\lmet{$WH\rightarrow \ell\kern-0.45em\raise0.19ex\hbox{/} \nu b\bar{b}$}

\def\hww{$H\rightarrow W^+ W^-$}

\def\hbb{$H\rightarrow b\bar{b}$}

\def\tevE{$\sqrt{s}=1.96$~TeV}

\newcommand{\MCFM}       {{\sc mcfm}}
\newcommand{\PYTHIA}     {{\sc pythia}}
\newcommand{\ALPGEN}       {{\sc alpgen}}
\newcommand{\HERWIG}     {{\sc herwig}}

\newcommand{\HDECAY}       {{\sc hdecay}}

\newcommand{\rar}       {\rightarrow}

\def\hww{$H\rightarrow W^+ W^-$}
\def\hbb{$H\rightarrow b\bar{b}$}

\def \met  {\mbox{$\not\!\!E_T$}}

\def \nL0  {N$_{\mathrm{L0}}$}

\begin{document}


\newcommand{\SMLexplow}{{\color{black}90}} 
\newcommand{\SMLexphigh}{{\color{black}120}} 
\newcommand{\SMLobslow}{{\color{black}90}} 
\newcommand{\SMLobshigh}{{\color{black}109}} 

\newcommand{\SMHexplow}{{\color{black}140}} 
\newcommand{\SMHexphigh}{{\color{black}184}} 
\newcommand{\SMHobslow}{{\color{black}149}} 
\newcommand{\SMHobshigh}{{\color{black}182}} 

\newcommand{\FPobs}{{\color{black}116}} 
\newcommand{\FPexp}{{\color{black}135}} 
\newcommand{\FPobslow}{{\color{black}100}} 
\newcommand{\FPobshigh}{{\color{black}116}} 
\newcommand{\FPexplow}{{\color{black}100}} 
\newcommand{\FPexphigh}{{\color{black}135}} 

\newcommand{\SMFLobslow}{{\color{black}121}} 
\newcommand{\SMFLobshigh}{{\color{black}225}} 

\newcommand{\SMFLexplow}{{\color{black}118}} 
\newcommand{\SMFLexphigh}{{\color{black}270}} 

\newcommand{\SMFHobslow}{{\color{black}121}} 
\newcommand{\SMFHobshigh}{{\color{black}232}} 

\newcommand{\SMFHexplow}{{\color{black}118}} 
\newcommand{\SMFHexphigh}{{\color{black}290}} 

\widetext
\hspace{5.2in} \mbox{FERMILAB-PUB-13-081-E}

\title{Higgs Boson Studies at the Tevatron\\}

\affiliation{LAFEX, Centro Brasileiro de Pesquisas F\'{i}sicas, Rio de Janeiro, Brazil}
\affiliation{Universidade do Estado do Rio de Janeiro, Rio de Janeiro, Brazil}
\affiliation{Universidade Federal do ABC, Santo Andr\'e, Brazil}
\affiliation{Institute of Particle Physics: McGill University, Montr\'{e}al, Qu\'{e}bec, Canada H3A~2T8; Simon Fraser University, Burnaby, British Columbia, Canada V5A~1S6; University of Toronto, Toronto, Ontario, Canada M5S~1A7; and TRIUMF, Vancouver, British Columbia, V6T~2A3, Canada}
\affiliation{University of Science and Technology of China, Hefei, People's Republic of China}
\affiliation{Institute of Physics, Academia Sinica, Taipei, Taiwan 11529, Republic of China}
\affiliation{Universidad de los Andes, Bogot\'a, Colombia}
\affiliation{Charles University, Faculty of Mathematics and Physics, Center for Particle Physics, Prague, Czech Republic}
\affiliation{Czech Technical University in Prague, Prague, Czech Republic}
\affiliation{Center for Particle Physics, Institute of Physics, Academy of Sciences of the Czech Republic, Prague, Czech Republic}
\affiliation{Universidad San Francisco de Quito, Quito, Ecuador}
\affiliation{Division of High Energy Physics, Department of Physics, University of Helsinki and Helsinki Institute of Physics, FIN-00014, Helsinki, Finland}
\affiliation{LPC, Universit\'e Blaise Pascal, CNRS/IN2P3, Clermont, France}
\affiliation{LPSC, Universit\'e Joseph Fourier Grenoble 1, CNRS/IN2P3, Institut National Polytechnique de Grenoble, Grenoble, France}
\affiliation{CPPM, Aix-Marseille Universit\'e, CNRS/IN2P3, Marseille, France}
\affiliation{LAL, Universit\'e Paris-Sud, CNRS/IN2P3, Orsay, France}
\affiliation{LPNHE, Universit\'es Paris VI and VII, CNRS/IN2P3, Paris, France}
\affiliation{CEA, Irfu, SPP, Saclay, France}
\affiliation{IPHC, Universit\'e de Strasbourg, CNRS/IN2P3, Strasbourg, France}
\affiliation{IPNL, Universit\'e Lyon 1, CNRS/IN2P3, Villeurbanne, France and Universit\'e de Lyon, Lyon, France}
\affiliation{III. Physikalisches Institut A, RWTH Aachen University, Aachen, Germany}
\affiliation{Physikalisches Institut, Universit\"at Freiburg, Freiburg, Germany}
\affiliation{II. Physikalisches Institut, Georg-August-Universit\"at G\"ottingen, G\"ottingen, Germany}
\affiliation{Institut f\"{u}r Experimentelle Kernphysik, Karlsruhe Institute of Technology, D-76131 Karlsruhe, Germany}
\affiliation{Institut f\"ur Physik, Universit\"at Mainz, Mainz, Germany}
\affiliation{Ludwig-Maximilians-Universit\"at M\"unchen, M\"unchen, Germany}
\affiliation{Fachbereich Physik, Bergische Universit\"at Wuppertal, Wuppertal, Germany}
\affiliation{University of Athens, 157 71 Athens, Greece}
\affiliation{Panjab University, Chandigarh, India}
\affiliation{Delhi University, Delhi, India}
\affiliation{Tata Institute of Fundamental Research, Mumbai, India}
\affiliation{University College Dublin, Dublin, Ireland}
\affiliation{Istituto Nazionale di Fisica Nucleare Bologna, $^{\sharp{a}}$University of Bologna, I-40127 Bologna, Italy}
\affiliation{Laboratori Nazionali di Frascati, Istituto Nazionale di Fisica Nucleare, I-00044 Frascati, Italy}
\affiliation{Istituto Nazionale di Fisica Nucleare, Sezione di Padova-Trento, $^{\sharp{b}}$University of Padova, I-35131 Padova, Italy}
\affiliation{Istituto Nazionale di Fisica Nucleare Pisa, $^{\sharp{c}}$University of Pisa, $^{\sharp{d}}$University of Siena and $^{\sharp{e}}$Scuola Normale Superiore, I-56127 Pisa, Italy, $^{\sharp{h}}$INFN Pavia and University of Pavia, I-27100 Pavia, Italy}
\affiliation{Istituto Nazionale di Fisica Nucleare, Sezione di Roma 1, $^{\sharp{f}}$Sapienza Universit\`{a} di Roma, I-00185 Roma, Italy}
\affiliation{Istituto Nazionale di Fisica Nucleare Trieste/Udine; $^{\sharp{i}}$University of Trieste, I-34127 Trieste, Italy; $^{\sharp{g}}$University of Udine, I-33100 Udine, Italy}
\affiliation{Okayama University, Okayama 700-8530, Japan}
\affiliation{Osaka City University, Osaka 588, Japan}
\affiliation{Waseda University, Tokyo 169, Japan}
\affiliation{University of Tsukuba, Tsukuba, Ibaraki 305, Japan}
\affiliation{Center for High Energy Physics: Kyungpook National University, Daegu 702-701, Korea; Seoul National University, Seoul 151-742, Korea; Sungkyunkwan University, Suwon 440-746, Korea; Korea Institute of Science and Technology Information, Daejeon 305-806, Korea; Chonnam National University, Gwangju 500-757, Korea; Chonbuk National University, Jeonju 561-756, Korea; Ewha Womans University, Seoul, 120-750, Korea}
\affiliation{Korea Detector Laboratory, Korea University, Seoul, Korea}
\affiliation{CINVESTAV, Mexico City, Mexico}
\affiliation{Nikhef, Science Park, Amsterdam, the Netherlands}
\affiliation{Radboud University Nijmegen, Nijmegen, the Netherlands}
\affiliation{Joint Institute for Nuclear Research, Dubna, Russia}
\affiliation{Institution for Theoretical and Experimental Physics, ITEP, Moscow 117259, Russia}
\affiliation{Moscow State University, Moscow, Russia}
\affiliation{Institute for High Energy Physics, Protvino, Russia}
\affiliation{Petersburg Nuclear Physics Institute, St. Petersburg, Russia}
\affiliation{Comenius University, 842 48 Bratislava, Slovakia; Institute of Experimental Physics, 040 01 Kosice, Slovakia}
\affiliation{Institut de Fisica d'Altes Energies, ICREA, Universitat Autonoma de Barcelona, E-08193, Bellaterra (Barcelona), Spain}
\affiliation{Instituci\'{o} Catalana de Recerca i Estudis Avan\c{c}ats (ICREA) and Institut de F\'{i}sica d'Altes Energies (IFAE), Barcelona, Spain}
\affiliation{Centro de Investigaciones Energeticas Medioambientales y Tecnologicas, E-28040 Madrid, Spain}
\affiliation{Instituto de Fisica de Cantabria, CSIC-University of Cantabria, 39005 Santander, Spain}
\affiliation{Uppsala University, Uppsala, Sweden}
\affiliation{University of Geneva, CH-1211 Geneva 4, Switzerland}
\affiliation{Glasgow University, Glasgow G12 8QQ, United Kingdom}
\affiliation{Lancaster University, Lancaster LA1 4YB, United Kingdom}
\affiliation{University of Liverpool, Liverpool L69 7ZE, United Kingdom}
\affiliation{Imperial College London, London SW7 2AZ, United Kingdom}
\affiliation{University College London, London WC1E 6BT, United Kingdom}
\affiliation{The University of Manchester, Manchester M13 9PL, United Kingdom}
\affiliation{University of Oxford, Oxford OX1 3RH, United Kingdom}
\affiliation{University of Arizona, Tucson, Arizona 85721, USA}
\affiliation{Ernest Orlando Lawrence Berkeley National Laboratory, Berkeley, California 94720, USA}
\affiliation{University of California, Davis, Davis, California 95616, USA}
\affiliation{University of California, Los Angeles, Los Angeles, California 90024, USA}
\affiliation{University of California Riverside, Riverside, California 92521, USA}
\affiliation{Yale University, New Haven, Connecticut 06520, USA}
\affiliation{University of Florida, Gainesville, Florida 32611, USA}
\affiliation{Florida State University, Tallahassee, Florida 32306, USA}
\affiliation{Argonne National Laboratory, Argonne, Illinois 60439, USA}
\affiliation{Fermi National Accelerator Laboratory, Batavia, Illinois 60510, USA}
\affiliation{Enrico Fermi Institute, University of Chicago, Chicago, Illinois 60637, USA}
\affiliation{University of Illinois at Chicago, Chicago, Illinois 60607, USA}
\affiliation{Northern Illinois University, DeKalb, Illinois 60115, USA}
\affiliation{Northwestern University, Evanston, Illinois 60208, USA}
\affiliation{University of Illinois, Urbana, Illinois 61801, USA}
\affiliation{Indiana University, Bloomington, Indiana 47405, USA}
\affiliation{Purdue University Calumet, Hammond, Indiana 46323, USA}
\affiliation{University of Notre Dame, Notre Dame, Indiana 46556, USA}
\affiliation{Purdue University, West Lafayette, Indiana 47907, USA}
\affiliation{Iowa State University, Ames, Iowa 50011, USA}
\affiliation{University of Kansas, Lawrence, Kansas 66045, USA}
\affiliation{Louisiana Tech University, Ruston, Louisiana 71272, USA}
\affiliation{The Johns Hopkins University, Baltimore, Maryland 21218, USA}
\affiliation{Northeastern University, Boston, Massachusetts 02115, USA}
\affiliation{Harvard University, Cambridge, Massachusetts 02138, USA}
\affiliation{Massachusetts Institute of Technology, Cambridge, Massachusetts 02139, USA}
\affiliation{Tufts University, Medford, Massachusetts 02155, USA}
\affiliation{University of Michigan, Ann Arbor, Michigan 48109, USA}
\affiliation{Wayne State University, Detroit, Michigan 48201, USA}
\affiliation{Michigan State University, East Lansing, Michigan 48824, USA}
\affiliation{University of Mississippi, University, Mississippi 38677, USA}
\affiliation{University of Nebraska, Lincoln, Nebraska 68588, USA}
\affiliation{Princeton University, Princeton, New Jersey 08544, USA}
\affiliation{University of New Mexico, Albuquerque, New Mexico 87131, USA}
\affiliation{State University of New York, Buffalo, New York 14260, USA}
\affiliation{The Rockefeller University, New York, New York 10065, USA}
\affiliation{University of Rochester, Rochester, New York 14627, USA}
\affiliation{State University of New York, Stony Brook, New York 11794, USA}
\affiliation{Brookhaven National Laboratory, Upton, New York 11973, USA}
\affiliation{Duke University, Durham, North Carolina 27708, USA}
\affiliation{The Ohio State University, Columbus, Ohio 43210, USA}
\affiliation{Langston University, Langston, Oklahoma 73050, USA}
\affiliation{University of Oklahoma, Norman, Oklahoma 73019, USA}
\affiliation{Oklahoma State University, Stillwater, Oklahoma 74078, USA}
\affiliation{University of Pennsylvania, Philadelphia, Pennsylvania 19104, USA}
\affiliation{Carnegie Mellon University, Pittsburgh, Pennsylvania 15213, USA}
\affiliation{University of Pittsburgh, Pittsburgh, Pennsylvania 15260, USA}
\affiliation{Brown University, Providence, Rhode Island 02912, USA}
\affiliation{University of Texas, Arlington, Texas 76019, USA}
\affiliation{Mitchell Institute for Fundamental Physics and Astronomy, Texas A\&M University, College Station, Texas 77843, USA}
\affiliation{Southern Methodist University, Dallas, Texas 75275, USA}
\affiliation{Rice University, Houston, Texas 77005, USA}
\affiliation{Baylor University, Waco, Texas 76798, USA}
\affiliation{University of Virginia, Charlottesville, Virginia 22904, USA}
\affiliation{University of Washington, Seattle, Washington 98195, USA}
\affiliation{University of Wisconsin, Madison, Wisconsin 53706, USA}
\author{T.~Aaltonen$^{\dag}$}~\affiliation{Division of High Energy Physics, Department of Physics, University of Helsinki and Helsinki Institute of Physics, FIN-00014, Helsinki, Finland}
\author{V.M.~Abazov$^{\ddag}$}~\affiliation{Joint Institute for Nuclear Research, Dubna, Russia}
\author{B.~Abbott$^{\ddag}$}~\affiliation{University of Oklahoma, Norman, Oklahoma 73019, USA}
\author{B.S.~Acharya$^{\ddag}$}~\affiliation{Tata Institute of Fundamental Research, Mumbai, India}
\author{M.~Adams$^{\ddag}$}~\affiliation{University of Illinois at Chicago, Chicago, Illinois 60607, USA}
\author{T.~Adams$^{\ddag}$}~\affiliation{Florida State University, Tallahassee, Florida 32306, USA}
\author{G.D.~Alexeev$^{\ddag}$}~\affiliation{Joint Institute for Nuclear Research, Dubna, Russia}
\author{G.~Alkhazov$^{\ddag}$}~\affiliation{Petersburg Nuclear Physics Institute, St. Petersburg, Russia}
\author{A.~Alton$^{\ddag a}$}~\affiliation{University of Michigan, Ann Arbor, Michigan 48109, USA}
\author{S.~Amerio$^{\dag}$}~\affiliation{Istituto Nazionale di Fisica Nucleare, Sezione di Padova-Trento, $^{\sharp{b}}$University of Padova, I-35131 Padova, Italy}
\author{D.~Amidei$^{\dag}$}~\affiliation{University of Michigan, Ann Arbor, Michigan 48109, USA}
\author{A.~Anastassov$^{\dag b}$}~\affiliation{Fermi National Accelerator Laboratory, Batavia, Illinois 60510, USA}
\author{A.~Annovi$^{\dag}$}~\affiliation{Laboratori Nazionali di Frascati, Istituto Nazionale di Fisica Nucleare, I-00044 Frascati, Italy}
\author{J.~Antos$^{\dag}$}~\affiliation{Comenius University, 842 48 Bratislava, Slovakia; Institute of Experimental Physics, 040 01 Kosice, Slovakia}
\author{G.~Apollinari$^{\dag}$}~\affiliation{Fermi National Accelerator Laboratory, Batavia, Illinois 60510, USA}
\author{J.A.~Appel$^{\dag}$}~\affiliation{Fermi National Accelerator Laboratory, Batavia, Illinois 60510, USA}
\author{T.~Arisawa$^{\dag}$}~\affiliation{Waseda University, Tokyo 169, Japan}
\author{A.~Artikov$^{\dag}$}~\affiliation{Joint Institute for Nuclear Research, Dubna, Russia}
\author{J.~Asaadi$^{\dag}$}~\affiliation{Mitchell Institute for Fundamental Physics and Astronomy, Texas A\&M University, College Station, Texas 77843, USA}
\author{W.~Ashmanskas$^{\dag}$}~\affiliation{Fermi National Accelerator Laboratory, Batavia, Illinois 60510, USA}
\author{A.~Askew$^{\ddag}$}~\affiliation{Florida State University, Tallahassee, Florida 32306, USA}
\author{S.~Atkins$^{\ddag}$}~\affiliation{Louisiana Tech University, Ruston, Louisiana 71272, USA}
\author{B.~Auerbach$^{\dag}$}~\affiliation{Argonne National Laboratory, Argonne, Illinois 60439, USA}
\author{K.~Augsten$^{\ddag}$}~\affiliation{Czech Technical University in Prague, Prague, Czech Republic}
\author{A.~Aurisano$^{\dag}$}~\affiliation{Mitchell Institute for Fundamental Physics and Astronomy, Texas A\&M University, College Station, Texas 77843, USA}
\author{C.~Avila$^{\ddag}$}~\affiliation{Universidad de los Andes, Bogot\'a, Colombia}
\author{F.~Azfar$^{\dag}$}~\affiliation{University of Oxford, Oxford OX1 3RH, United Kingdom}
\author{F.~Badaud$^{\ddag}$}~\affiliation{LPC, Universit\'e Blaise Pascal, CNRS/IN2P3, Clermont, France}
\author{W.~Badgett$^{\dag}$}~\affiliation{Fermi National Accelerator Laboratory, Batavia, Illinois 60510, USA}
\author{T.~Bae$^{\dag}$}~\affiliation{Center for High Energy Physics: Kyungpook National University, Daegu 702-701, Korea; Seoul National University, Seoul 151-742, Korea; Sungkyunkwan University, Suwon 440-746, Korea; Korea Institute of Science and Technology Information, Daejeon 305-806, Korea; Chonnam National University, Gwangju 500-757, Korea; Chonbuk National University, Jeonju 561-756, Korea; Ewha Womans University, Seoul, 120-750, Korea}
\author{L.~Bagby$^{\ddag}$}~\affiliation{Fermi National Accelerator Laboratory, Batavia, Illinois 60510, USA}
\author{B.~Baldin$^{\ddag}$}~\affiliation{Fermi National Accelerator Laboratory, Batavia, Illinois 60510, USA}
\author{D.V.~Bandurin$^{\ddag}$}~\affiliation{Florida State University, Tallahassee, Florida 32306, USA}
\author{S.~Banerjee$^{\ddag}$}~\affiliation{Tata Institute of Fundamental Research, Mumbai, India}
\author{A.~Barbaro-Galtieri$^{\dag}$}~\affiliation{Ernest Orlando Lawrence Berkeley National Laboratory, Berkeley, California 94720, USA}
\author{E.~Barberis$^{\ddag}$}~\affiliation{Northeastern University, Boston, Massachusetts 02115, USA}
\author{P.~Baringer$^{\ddag}$}~\affiliation{University of Kansas, Lawrence, Kansas 66045, USA}
\author{V.E.~Barnes$^{\dag}$}~\affiliation{Purdue University, West Lafayette, Indiana 47907, USA}
\author{B.A.~Barnett$^{\dag}$}~\affiliation{The Johns Hopkins University, Baltimore, Maryland 21218, USA}
\author{P.~Barria$^{\dag\sharp{d}}$}~\affiliation{Istituto Nazionale di Fisica Nucleare Pisa, $^{\sharp{c}}$University of Pisa, $^{\sharp{d}}$University of Siena and $^{\sharp{e}}$Scuola Normale Superiore, I-56127 Pisa, Italy, $^{\sharp{h}}$INFN Pavia and University of Pavia, I-27100 Pavia, Italy}
\author{J.F.~Bartlett$^{\ddag}$}~\affiliation{Fermi National Accelerator Laboratory, Batavia, Illinois 60510, USA}
\author{P.~Bartos$^{\dag}$}~\affiliation{Comenius University, 842 48 Bratislava, Slovakia; Institute of Experimental Physics, 040 01 Kosice, Slovakia}
\author{U.~Bassler$^{\ddag}$}~\affiliation{CEA, Irfu, SPP, Saclay, France}
\author{M.~Bauce$^{\dag\sharp{b}}$}~\affiliation{Istituto Nazionale di Fisica Nucleare, Sezione di Padova-Trento, $^{\sharp{b}}$University of Padova, I-35131 Padova, Italy}
\author{V.~Bazterra$^{\ddag}$}~\affiliation{University of Illinois at Chicago, Chicago, Illinois 60607, USA}
\author{A.~Bean$^{\ddag}$}~\affiliation{University of Kansas, Lawrence, Kansas 66045, USA}
\author{F.~Bedeschi$^{\dag}$}~\affiliation{Istituto Nazionale di Fisica Nucleare Pisa, $^{\sharp{c}}$University of Pisa, $^{\sharp{d}}$University of Siena and $^{\sharp{e}}$Scuola Normale Superiore, I-56127 Pisa, Italy, $^{\sharp{h}}$INFN Pavia and University of Pavia, I-27100 Pavia, Italy}
\author{M.~Begalli$^{\ddag}$}~\affiliation{Universidade do Estado do Rio de Janeiro, Rio de Janeiro, Brazil}
\author{S.~Behari$^{\dag}$}~\affiliation{Fermi National Accelerator Laboratory, Batavia, Illinois 60510, USA}
\author{L.~Bellantoni$^{\ddag}$}~\affiliation{Fermi National Accelerator Laboratory, Batavia, Illinois 60510, USA}
\author{G.~Bellettini$^{\dag\sharp{c}}$}~\affiliation{Istituto Nazionale di Fisica Nucleare Pisa, $^{\sharp{c}}$University of Pisa, $^{\sharp{d}}$University of Siena and $^{\sharp{e}}$Scuola Normale Superiore, I-56127 Pisa, Italy, $^{\sharp{h}}$INFN Pavia and University of Pavia, I-27100 Pavia, Italy}
\author{J.~Bellinger$^{\dag}$}~\affiliation{University of Wisconsin, Madison, Wisconsin 53706, USA}
\author{D.~Benjamin$^{\dag}$}~\affiliation{Duke University, Durham, North Carolina 27708, USA}
\author{A.~Beretvas$^{\dag}$}~\affiliation{Fermi National Accelerator Laboratory, Batavia, Illinois 60510, USA}
\author{S.B.~Beri$^{\ddag}$}~\affiliation{Panjab University, Chandigarh, India}
\author{G.~Bernardi$^{\ddag}$}~\affiliation{LPNHE, Universit\'es Paris VI and VII, CNRS/IN2P3, Paris, France}
\author{R.~Bernhard$^{\ddag}$}~\affiliation{Physikalisches Institut, Universit\"at Freiburg, Freiburg, Germany}
\author{I.~Bertram$^{\ddag}$}~\affiliation{Lancaster University, Lancaster LA1 4YB, United Kingdom}
\author{M.~Besan\c{c}on$^{\ddag}$}~\affiliation{CEA, Irfu, SPP, Saclay, France}
\author{R.~Beuselinck$^{\ddag}$}~\affiliation{Imperial College London, London SW7 2AZ, United Kingdom}
\author{P.C.~Bhat$^{\ddag}$}~\affiliation{Fermi National Accelerator Laboratory, Batavia, Illinois 60510, USA}
\author{S.~Bhatia$^{\ddag}$}~\affiliation{University of Mississippi, University, Mississippi 38677, USA}
\author{V.~Bhatnagar$^{\ddag}$}~\affiliation{Panjab University, Chandigarh, India}
\author{A.~Bhatti$^{\dag}$}~\affiliation{The Rockefeller University, New York, New York 10065, USA}
\author{K.R.~Bland$^{\dag}$}~\affiliation{Baylor University, Waco, Texas 76798, USA}
\author{G.~Blazey$^{\ddag}$}~\affiliation{Northern Illinois University, DeKalb, Illinois 60115, USA}
\author{S.~Blessing$^{\ddag}$}~\affiliation{Florida State University, Tallahassee, Florida 32306, USA}
\author{K.~Bloom$^{\ddag}$}~\affiliation{University of Nebraska, Lincoln, Nebraska 68588, USA}
\author{B.~Blumenfeld$^{\dag}$}~\affiliation{The Johns Hopkins University, Baltimore, Maryland 21218, USA}
\author{A.~Bocci$^{\dag}$}~\affiliation{Duke University, Durham, North Carolina 27708, USA}
\author{A.~Bodek$^{\dag}$}~\affiliation{University of Rochester, Rochester, New York 14627, USA}
\author{A.~Boehnlein$^{\ddag}$}~\affiliation{Fermi National Accelerator Laboratory, Batavia, Illinois 60510, USA}
\author{D.~Boline$^{\ddag}$}~\affiliation{State University of New York, Stony Brook, New York 11794, USA}
\author{E.E.~Boos$^{\ddag}$}~\affiliation{Moscow State University, Moscow, Russia}
\author{G.~Borissov$^{\ddag}$}~\affiliation{Lancaster University, Lancaster LA1 4YB, United Kingdom}
\author{D.~Bortoletto$^{\dag}$}~\affiliation{Purdue University, West Lafayette, Indiana 47907, USA}
\author{J.~Boudreau$^{\dag}$}~\affiliation{University of Pittsburgh, Pittsburgh, Pennsylvania 15260, USA}
\author{A.~Boveia$^{\dag}$}~\affiliation{Enrico Fermi Institute, University of Chicago, Chicago, Illinois 60637, USA}
\author{A.~Brandt$^{\ddag}$}~\affiliation{University of Texas, Arlington, Texas 76019, USA}
\author{O.~Brandt$^{\ddag}$}~\affiliation{II. Physikalisches Institut, Georg-August-Universit\"at G\"ottingen, G\"ottingen, Germany}
\author{L.~Brigliadori$^{\dag\sharp{a}}$}~\affiliation{Istituto Nazionale di Fisica Nucleare Bologna, $^{\sharp{a}}$University of Bologna, I-40127 Bologna, Italy}
\author{R.~Brock$^{\ddag}$}~\affiliation{Michigan State University, East Lansing, Michigan 48824, USA}
\author{C.~Bromberg$^{\dag}$}~\affiliation{Michigan State University, East Lansing, Michigan 48824, USA}
\author{A.~Bross$^{\ddag}$}~\affiliation{Fermi National Accelerator Laboratory, Batavia, Illinois 60510, USA}
\author{D.~Brown$^{\ddag}$}~\affiliation{LPNHE, Universit\'es Paris VI and VII, CNRS/IN2P3, Paris, France}
\author{E.~Brucken$^{\dag}$}~\affiliation{Division of High Energy Physics, Department of Physics, University of Helsinki and Helsinki Institute of Physics, FIN-00014, Helsinki, Finland}
\author{J.~Budagov$^{\dag}$}~\affiliation{Joint Institute for Nuclear Research, Dubna, Russia}
\author{X.B.~Bu$^{\ddag}$}~\affiliation{Fermi National Accelerator Laboratory, Batavia, Illinois 60510, USA}
\author{H.S.~Budd$^{\dag}$}~\affiliation{University of Rochester, Rochester, New York 14627, USA}
\author{M.~Buehler$^{\ddag}$}~\affiliation{Fermi National Accelerator Laboratory, Batavia, Illinois 60510, USA}
\author{V.~Buescher$^{\ddag}$}~\affiliation{Institut f\"ur Physik, Universit\"at Mainz, Mainz, Germany}
\author{V.~Bunichev$^{\ddag}$}~\affiliation{Moscow State University, Moscow, Russia}
\author{S.~Burdin$^{\ddag b}$}~\affiliation{Lancaster University, Lancaster LA1 4YB, United Kingdom}
\author{K.~Burkett$^{\dag}$}~\affiliation{Fermi National Accelerator Laboratory, Batavia, Illinois 60510, USA}
\author{G.~Busetto$^{\dag\sharp{b}}$}~\affiliation{Istituto Nazionale di Fisica Nucleare, Sezione di Padova-Trento, $^{\sharp{b}}$University of Padova, I-35131 Padova, Italy}
\author{P.~Bussey$^{\dag}$}~\affiliation{Glasgow University, Glasgow G12 8QQ, United Kingdom}
\author{C.P.~Buszello$^{\ddag}$}~\affiliation{Uppsala University, Uppsala, Sweden}
\author{P.~Butti$^{\dag\sharp{c}}$}~\affiliation{Istituto Nazionale di Fisica Nucleare Pisa, $^{\sharp{c}}$University of Pisa, $^{\sharp{d}}$University of Siena and $^{\sharp{e}}$Scuola Normale Superiore, I-56127 Pisa, Italy, $^{\sharp{h}}$INFN Pavia and University of Pavia, I-27100 Pavia, Italy}
\author{A.~Buzatu$^{\dag}$}~\affiliation{Glasgow University, Glasgow G12 8QQ, United Kingdom}
\author{A.~Calamba$^{\dag}$}~\affiliation{Carnegie Mellon University, Pittsburgh, Pennsylvania 15213, USA}
\author{E.~Camacho-P\'erez$^{\ddag}$}~\affiliation{CINVESTAV, Mexico City, Mexico}
\author{S.~Camarda$^{\dag}$}~\affiliation{Institut de Fisica d'Altes Energies, ICREA, Universitat Autonoma de Barcelona, E-08193, Bellaterra (Barcelona), Spain}
\author{M.~Campanelli$^{\dag}$}~\affiliation{University College London, London WC1E 6BT, United Kingdom}
\author{F.~Canelli$^{\dag oo}$}~\affiliation{Enrico Fermi Institute, University of Chicago, Chicago, Illinois 60637, USA}
\author{B.~Carls$^{\dag}$}~\affiliation{University of Illinois, Urbana, Illinois 61801, USA}
\author{D.~Carlsmith$^{\dag}$}~\affiliation{University of Wisconsin, Madison, Wisconsin 53706, USA}
\author{R.~Carosi$^{\dag}$}~\affiliation{Istituto Nazionale di Fisica Nucleare Pisa, $^{\sharp{c}}$University of Pisa, $^{\sharp{d}}$University of Siena and $^{\sharp{e}}$Scuola Normale Superiore, I-56127 Pisa, Italy, $^{\sharp{h}}$INFN Pavia and University of Pavia, I-27100 Pavia, Italy}
\author{S.~Carrillo$^{\dag c}$}~\affiliation{University of Florida, Gainesville, Florida 32611, USA}
\author{B.~Casal$^{\dag d}$}~\affiliation{Instituto de Fisica de Cantabria, CSIC-University of Cantabria, 39005 Santander, Spain}
\author{M.~Casarsa$^{\dag}$}~\affiliation{Istituto Nazionale di Fisica Nucleare Trieste/Udine; $^{\sharp{i}}$University of Trieste, I-34127 Trieste, Italy; $^{\sharp{g}}$University of Udine, I-33100 Udine, Italy}
\author{B.C.K.~Casey$^{\ddag}$}~\affiliation{Fermi National Accelerator Laboratory, Batavia, Illinois 60510, USA}
\author{H.~Castilla-Valdez$^{\ddag}$}~\affiliation{CINVESTAV, Mexico City, Mexico}
\author{A.~Castro$^{\dag\sharp{a}}$}~\affiliation{Istituto Nazionale di Fisica Nucleare Bologna, $^{\sharp{a}}$University of Bologna, I-40127 Bologna, Italy}
\author{P.~Catastini$^{\dag}$}~\affiliation{Harvard University, Cambridge, Massachusetts 02138, USA}
\author{S.~Caughron$^{\ddag}$}~\affiliation{Michigan State University, East Lansing, Michigan 48824, USA}
\author{D.~Cauz$^{\dag}$}~\affiliation{Istituto Nazionale di Fisica Nucleare Trieste/Udine; $^{\sharp{i}}$University of Trieste, I-34127 Trieste, Italy; $^{\sharp{g}}$University of Udine, I-33100 Udine, Italy}
\author{V.~Cavaliere$^{\dag}$}~\affiliation{University of Illinois, Urbana, Illinois 61801, USA}
\author{M.~Cavalli-Sforza$^{\dag}$}~\affiliation{Institut de Fisica d'Altes Energies, ICREA, Universitat Autonoma de Barcelona, E-08193, Bellaterra (Barcelona), Spain}
\author{A.~Cerri$^{\dag e}$}~\affiliation{Ernest Orlando Lawrence Berkeley National Laboratory, Berkeley, California 94720, USA}
\author{L.~Cerrito$^{\dag f}$}~\affiliation{University College London, London WC1E 6BT, United Kingdom}
\author{S.~Chakrabarti$^{\ddag}$}~\affiliation{State University of New York, Stony Brook, New York 11794, USA}
\author{D.~Chakraborty$^{\ddag}$}~\affiliation{Northern Illinois University, DeKalb, Illinois 60115, USA}
\author{K.M.~Chan$^{\ddag}$}~\affiliation{University of Notre Dame, Notre Dame, Indiana 46556, USA}
\author{A.~Chandra$^{\ddag}$}~\affiliation{Rice University, Houston, Texas 77005, USA}
\author{E.~Chapon$^{\ddag}$}~\affiliation{CEA, Irfu, SPP, Saclay, France}
\author{G.~Chen$^{\ddag}$}~\affiliation{University of Kansas, Lawrence, Kansas 66045, USA}
\author{Y.C.~Chen$^{\dag}$}~\affiliation{Institute of Physics, Academia Sinica, Taipei, Taiwan 11529, Republic of China}
\author{M.~Chertok$^{\dag}$}~\affiliation{University of California, Davis, Davis, California 95616, USA}
\author{G.~Chiarelli$^{\dag}$}~\affiliation{Istituto Nazionale di Fisica Nucleare Pisa, $^{\sharp{c}}$University of Pisa, $^{\sharp{d}}$University of Siena and $^{\sharp{e}}$Scuola Normale Superiore, I-56127 Pisa, Italy, $^{\sharp{h}}$INFN Pavia and University of Pavia, I-27100 Pavia, Italy}
\author{G.~Chlachidze$^{\dag}$}~\affiliation{Fermi National Accelerator Laboratory, Batavia, Illinois 60510, USA}
\author{K.~Cho$^{\dag}$}~\affiliation{Center for High Energy Physics: Kyungpook National University, Daegu 702-701, Korea; Seoul National University, Seoul 151-742, Korea; Sungkyunkwan University, Suwon 440-746, Korea; Korea Institute of Science and Technology Information, Daejeon 305-806, Korea; Chonnam National University, Gwangju 500-757, Korea; Chonbuk National University, Jeonju 561-756, Korea; Ewha Womans University, Seoul, 120-750, Korea}
\author{S.W.~Cho$^{\ddag}$}~\affiliation{Korea Detector Laboratory, Korea University, Seoul, Korea}
\author{S.~Choi$^{\ddag}$}~\affiliation{Korea Detector Laboratory, Korea University, Seoul, Korea}
\author{D.~Chokheli$^{\dag}$}~\affiliation{Joint Institute for Nuclear Research, Dubna, Russia}
\author{B.~Choudhary$^{\ddag}$}~\affiliation{Delhi University, Delhi, India}
\author{S.~Cihangir$^{\ddag}$}~\affiliation{Fermi National Accelerator Laboratory, Batavia, Illinois 60510, USA}
\author{M.A.~Ciocci$^{\dag\sharp{d}}$}~\affiliation{Istituto Nazionale di Fisica Nucleare Pisa, $^{\sharp{c}}$University of Pisa, $^{\sharp{d}}$University of Siena and $^{\sharp{e}}$Scuola Normale Superiore, I-56127 Pisa, Italy, $^{\sharp{h}}$INFN Pavia and University of Pavia, I-27100 Pavia, Italy}
\author{D.~Claes$^{\ddag}$}~\affiliation{University of Nebraska, Lincoln, Nebraska 68588, USA}
\author{A.~Clark$^{\dag}$}~\affiliation{University of Geneva, CH-1211 Geneva 4, Switzerland}
\author{C.~Clarke$^{\dag}$}~\affiliation{Wayne State University, Detroit, Michigan 48201, USA}
\author{J.~Clutter$^{\ddag}$}~\affiliation{University of Kansas, Lawrence, Kansas 66045, USA}
\author{M.E.~Convery$^{\dag}$}~\affiliation{Fermi National Accelerator Laboratory, Batavia, Illinois 60510, USA}
\author{J.~Conway$^{\dag}$}~\affiliation{University of California, Davis, Davis, California 95616, USA}
\author{M.~Cooke$^{\ddag}$}~\affiliation{Fermi National Accelerator Laboratory, Batavia, Illinois 60510, USA}
\author{W.E.~Cooper$^{\ddag}$}~\affiliation{Fermi National Accelerator Laboratory, Batavia, Illinois 60510, USA}
\author{M.~Corbo$^{\dag}$}~\affiliation{Fermi National Accelerator Laboratory, Batavia, Illinois 60510, USA}
\author{M.~Corcoran$^{\ddag}$}~\affiliation{Rice University, Houston, Texas 77005, USA}
\author{M.~Cordelli$^{\dag}$}~\affiliation{Laboratori Nazionali di Frascati, Istituto Nazionale di Fisica Nucleare, I-00044 Frascati, Italy}
\author{F.~Couderc$^{\ddag}$}~\affiliation{CEA, Irfu, SPP, Saclay, France}
\author{M.-C.~Cousinou$^{\ddag}$}~\affiliation{CPPM, Aix-Marseille Universit\'e, CNRS/IN2P3, Marseille, France}
\author{C.A.~Cox$^{\dag}$}~\affiliation{University of California, Davis, Davis, California 95616, USA}
\author{D.J.~Cox$^{\dag}$}~\affiliation{University of California, Davis, Davis, California 95616, USA}
\author{M.~Cremonesi$^{\dag}$}~\affiliation{Istituto Nazionale di Fisica Nucleare Pisa, $^{\sharp{c}}$University of Pisa, $^{\sharp{d}}$University of Siena and $^{\sharp{e}}$Scuola Normale Superiore, I-56127 Pisa, Italy, $^{\sharp{h}}$INFN Pavia and University of Pavia, I-27100 Pavia, Italy}
\author{D.~Cruz$^{\dag}$}~\affiliation{Mitchell Institute for Fundamental Physics and Astronomy, Texas A\&M University, College Station, Texas 77843, USA}
\author{J.~Cuevas$^{\dag a}$}~\affiliation{Instituto de Fisica de Cantabria, CSIC-University of Cantabria, 39005 Santander, Spain}
\author{R.~Culbertson$^{\dag}$}~\affiliation{Fermi National Accelerator Laboratory, Batavia, Illinois 60510, USA}
\author{D.~Cutts$^{\ddag}$}~\affiliation{Brown University, Providence, Rhode Island 02912, USA}
\author{N.~d'Ascenzo$^{\dag g}$}~\affiliation{Fermi National Accelerator Laboratory, Batavia, Illinois 60510, USA}
\author{A.~Das$^{\ddag}$}~\affiliation{University of Arizona, Tucson, Arizona 85721, USA}
\author{M.~Datta$^{\dag qq}$}~\affiliation{Fermi National Accelerator Laboratory, Batavia, Illinois 60510, USA}
\author{G.~Davies$^{\ddag}$}~\affiliation{Imperial College London, London SW7 2AZ, United Kingdom}
\author{P.~De~Barbaro$^{\dag}$}~\affiliation{University of Rochester, Rochester, New York 14627, USA}
\author{S.J.~de~Jong$^{\ddag}$}~\affiliation{Nikhef, Science Park, Amsterdam, the Netherlands}~\affiliation{Radboud University Nijmegen, Nijmegen, the Netherlands}
\author{E.~De~La~Cruz-Burelo$^{\ddag}$}~\affiliation{CINVESTAV, Mexico City, Mexico}
\author{F.~D\'eliot$^{\ddag}$}~\affiliation{CEA, Irfu, SPP, Saclay, France}
\author{R.~Demina$^{\ddag}$}~\affiliation{University of Rochester, Rochester, New York 14627, USA}
\author{L.~Demortier$^{\dag}$}~\affiliation{The Rockefeller University, New York, New York 10065, USA}
\author{M.~Deninno$^{\dag}$}~\affiliation{Istituto Nazionale di Fisica Nucleare Bologna, $^{\sharp{a}}$University of Bologna, I-40127 Bologna, Italy}
\author{D.~Denisov$^{\ddag}$}~\affiliation{Fermi National Accelerator Laboratory, Batavia, Illinois 60510, USA}
\author{S.P.~Denisov$^{\ddag}$}~\affiliation{Institute for High Energy Physics, Protvino, Russia}
\author{M.~d'Errico$^{\dag\sharp{b}}$}~\affiliation{Istituto Nazionale di Fisica Nucleare, Sezione di Padova-Trento, $^{\sharp{b}}$University of Padova, I-35131 Padova, Italy}
\author{S.~Desai$^{\ddag}$}~\affiliation{Fermi National Accelerator Laboratory, Batavia, Illinois 60510, USA}
\author{C.~Deterre$^{\ddag d}$}~\affiliation{II. Physikalisches Institut, Georg-August-Universit\"at G\"ottingen, G\"ottingen, Germany}
\author{K.~DeVaughan$^{\ddag}$}~\affiliation{University of Nebraska, Lincoln, Nebraska 68588, USA}
\author{F.~Devoto$^{\dag}$}~\affiliation{Division of High Energy Physics, Department of Physics, University of Helsinki and Helsinki Institute of Physics, FIN-00014, Helsinki, Finland}
\author{A.~Di~Canto$^{\dag\sharp{c}}$}~\affiliation{Istituto Nazionale di Fisica Nucleare Pisa, $^{\sharp{c}}$University of Pisa, $^{\sharp{d}}$University of Siena and $^{\sharp{e}}$Scuola Normale Superiore, I-56127 Pisa, Italy, $^{\sharp{h}}$INFN Pavia and University of Pavia, I-27100 Pavia, Italy}
\author{B.~Di~Ruzza$^{\dag rr}$}~\affiliation{Fermi National Accelerator Laboratory, Batavia, Illinois 60510, USA}
\author{H.T.~Diehl$^{\ddag}$}~\affiliation{Fermi National Accelerator Laboratory, Batavia, Illinois 60510, USA}
\author{M.~Diesburg$^{\ddag}$}~\affiliation{Fermi National Accelerator Laboratory, Batavia, Illinois 60510, USA}
\author{P.F.~Ding$^{\ddag}$}~\affiliation{The University of Manchester, Manchester M13 9PL, United Kingdom}
\author{J.R.~Dittmann$^{\dag}$}~\affiliation{Baylor University, Waco, Texas 76798, USA}
\author{A.~Dominguez$^{\ddag}$}~\affiliation{University of Nebraska, Lincoln, Nebraska 68588, USA}
\author{S.~Donati$^{\dag\sharp{c}}$}~\affiliation{Istituto Nazionale di Fisica Nucleare Pisa, $^{\sharp{c}}$University of Pisa, $^{\sharp{d}}$University of Siena and $^{\sharp{e}}$Scuola Normale Superiore, I-56127 Pisa, Italy, $^{\sharp{h}}$INFN Pavia and University of Pavia, I-27100 Pavia, Italy}
\author{M.~D'Onofrio$^{\dag}$}~\affiliation{University of Liverpool, Liverpool L69 7ZE, United Kingdom}
\author{M.~Dorigo$^{\dag \sharp{i}}$}~\affiliation{Istituto Nazionale di Fisica Nucleare Trieste/Udine; $^{\sharp{i}}$University of Trieste, I-34127 Trieste, Italy; $^{\sharp{g}}$University of Udine, I-33100 Udine, Italy}
\author{A.~Driutti$^{\dag}$}~\affiliation{Istituto Nazionale di Fisica Nucleare Trieste/Udine; $^{\sharp{i}}$University of Trieste, I-34127 Trieste, Italy; $^{\sharp{g}}$University of Udine, I-33100 Udine, Italy}
\author{A.~Dubey$^{\ddag}$}~\affiliation{Delhi University, Delhi, India}
\author{L.V.~Dudko$^{\ddag}$}~\affiliation{Moscow State University, Moscow, Russia}
\author{A.~Duperrin$^{\ddag}$}~\affiliation{CPPM, Aix-Marseille Universit\'e, CNRS/IN2P3, Marseille, France}
\author{S.~Dutt$^{\ddag}$}~\affiliation{Panjab University, Chandigarh, India}
\author{A.~Dyshkant$^{\ddag}$}~\affiliation{Northern Illinois University, DeKalb, Illinois 60115, USA}
\author{M.~Eads$^{\ddag}$}~\affiliation{Northern Illinois University, DeKalb, Illinois 60115, USA}
\author{K.~Ebina$^{\dag}$}~\affiliation{Waseda University, Tokyo 169, Japan}
\author{R.~Edgar$^{\dag}$}~\affiliation{University of Michigan, Ann Arbor, Michigan 48109, USA}
\author{D.~Edmunds$^{\ddag}$}~\affiliation{Michigan State University, East Lansing, Michigan 48824, USA}
\author{A.~Elagin$^{\dag}$}~\affiliation{Mitchell Institute for Fundamental Physics and Astronomy, Texas A\&M University, College Station, Texas 77843, USA}
\author{J.~Ellison$^{\ddag}$}~\affiliation{University of California Riverside, Riverside, California 92521, USA}
\author{V.D.~Elvira$^{\ddag}$}~\affiliation{Fermi National Accelerator Laboratory, Batavia, Illinois 60510, USA}
\author{Y.~Enari$^{\ddag}$}~\affiliation{LPNHE, Universit\'es Paris VI and VII, CNRS/IN2P3, Paris, France}
\author{R.~Erbacher$^{\dag}$}~\affiliation{University of California, Davis, Davis, California 95616, USA}
\author{S.~Errede$^{\dag}$}~\affiliation{University of Illinois, Urbana, Illinois 61801, USA}
\author{B.~Esham$^{\dag}$}~\affiliation{University of Illinois, Urbana, Illinois 61801, USA}
\author{R.~Eusebi$^{\dag}$}~\affiliation{Mitchell Institute for Fundamental Physics and Astronomy, Texas A\&M University, College Station, Texas 77843, USA}
\author{H.~Evans$^{\ddag}$}~\affiliation{Indiana University, Bloomington, Indiana 47405, USA}
\author{V.N.~Evdokimov$^{\ddag}$}~\affiliation{Institute for High Energy Physics, Protvino, Russia}
\author{G.~Facini$^{\ddag}$}~\affiliation{Northeastern University, Boston, Massachusetts 02115, USA}
\author{S.~Farrington$^{\dag}$}~\affiliation{University of Oxford, Oxford OX1 3RH, United Kingdom}
\author{A.~Faur\'e$^{\ddag}$}~\affiliation{CEA, Irfu, SPP, Saclay, France}
\author{L.~Feng$^{\ddag}$}~\affiliation{Northern Illinois University, DeKalb, Illinois 60115, USA}
\author{T.~Ferbel$^{\ddag}$}~\affiliation{University of Rochester, Rochester, New York 14627, USA}
\author{J.P.~Fern\'{a}ndez~Ramos$^{\dag}$}~\affiliation{Centro de Investigaciones Energeticas Medioambientales y Tecnologicas, E-28040 Madrid, Spain}
\author{F.~Fiedler$^{\ddag}$}~\affiliation{Institut f\"ur Physik, Universit\"at Mainz, Mainz, Germany}
\author{R.~Field$^{\dag}$}~\affiliation{University of Florida, Gainesville, Florida 32611, USA}
\author{F.~Filthaut$^{\ddag}$}~\affiliation{Nikhef, Science Park, Amsterdam, the Netherlands}~\affiliation{Radboud University Nijmegen, Nijmegen, the Netherlands}
\author{W.~Fisher$^{\ddag}$}~\affiliation{Michigan State University, East Lansing, Michigan 48824, USA}
\author{H.E.~Fisk$^{\ddag}$}~\affiliation{Fermi National Accelerator Laboratory, Batavia, Illinois 60510, USA}
\author{G.~Flanagan$^{\dag i}$}~\affiliation{Fermi National Accelerator Laboratory, Batavia, Illinois 60510, USA}
\author{R.~Forrest$^{\dag}$}~\affiliation{University of California, Davis, Davis, California 95616, USA}
\author{M.~Fortner$^{\ddag}$}~\affiliation{Northern Illinois University, DeKalb, Illinois 60115, USA}
\author{H.~Fox$^{\ddag}$}~\affiliation{Lancaster University, Lancaster LA1 4YB, United Kingdom}
\author{M.~Franklin$^{\dag}$}~\affiliation{Harvard University, Cambridge, Massachusetts 02138, USA}
\author{J.C.~Freeman$^{\dag}$}~\affiliation{Fermi National Accelerator Laboratory, Batavia, Illinois 60510, USA}
\author{H.~Frisch$^{\dag}$}~\affiliation{Enrico Fermi Institute, University of Chicago, Chicago, Illinois 60637, USA}
\author{S.~Fuess$^{\ddag}$}~\affiliation{Fermi National Accelerator Laboratory, Batavia, Illinois 60510, USA}
\author{Y.~Funakoshi$^{\dag}$}~\affiliation{Waseda University, Tokyo 169, Japan}
\author{A.~Garcia-Bellido$^{\ddag}$}~\affiliation{University of Rochester, Rochester, New York 14627, USA}
\author{J.A.~Garc\'{\i}a-Gonz\'alez$^{\ddag}$}~\affiliation{CINVESTAV, Mexico City, Mexico}
\author{G.A.~Garc\'ia-Guerra$^{\ddag c}$}~\affiliation{CINVESTAV, Mexico City, Mexico}
\author{A.F.~Garfinkel$^{\dag}$}~\affiliation{Purdue University, West Lafayette, Indiana 47907, USA}
\author{P.~Garosi$^{\dag\sharp{d}}$}~\affiliation{Istituto Nazionale di Fisica Nucleare Pisa, $^{\sharp{c}}$University of Pisa, $^{\sharp{d}}$University of Siena and $^{\sharp{e}}$Scuola Normale Superiore, I-56127 Pisa, Italy, $^{\sharp{h}}$INFN Pavia and University of Pavia, I-27100 Pavia, Italy}
\author{V.~Gavrilov$^{\ddag}$}~\affiliation{Institution for Theoretical and Experimental Physics, ITEP, Moscow 117259, Russia}
\author{W.~Geng$^{\ddag}$}~\affiliation{CPPM, Aix-Marseille Universit\'e, CNRS/IN2P3, Marseille, France}~\affiliation{Michigan State University, East Lansing, Michigan 48824, USA}
\author{C.E.~Gerber$^{\ddag}$}~\affiliation{University of Illinois at Chicago, Chicago, Illinois 60607, USA}
\author{H.~Gerberich$^{\dag}$}~\affiliation{University of Illinois, Urbana, Illinois 61801, USA}
\author{E.~Gerchtein$^{\dag}$}~\affiliation{Fermi National Accelerator Laboratory, Batavia, Illinois 60510, USA}
\author{S.~Giagu$^{\dag}$}~\affiliation{Istituto Nazionale di Fisica Nucleare, Sezione di Roma 1, $^{\sharp{f}}$Sapienza Universit\`{a} di Roma, I-00185 Roma, Italy}
\author{V.~Giakoumopoulou$^{\dag}$}~\affiliation{University of Athens, 157 71 Athens, Greece}
\author{K.~Gibson$^{\dag}$}~\affiliation{University of Pittsburgh, Pittsburgh, Pennsylvania 15260, USA}
\author{C.M.~Ginsburg$^{\dag}$}~\affiliation{Fermi National Accelerator Laboratory, Batavia, Illinois 60510, USA}
\author{G.~Ginther$^{\ddag}$}~\affiliation{Fermi National Accelerator Laboratory, Batavia, Illinois 60510, USA}~\affiliation{University of Rochester, Rochester, New York 14627, USA}
\author{N.~Giokaris$^{\dag}$}~\affiliation{University of Athens, 157 71 Athens, Greece}
\author{P.~Giromini$^{\dag}$}~\affiliation{Laboratori Nazionali di Frascati, Istituto Nazionale di Fisica Nucleare, I-00044 Frascati, Italy}
\author{G.~Giurgiu$^{\dag}$}~\affiliation{The Johns Hopkins University, Baltimore, Maryland 21218, USA}
\author{V.~Glagolev$^{\dag}$}~\affiliation{Joint Institute for Nuclear Research, Dubna, Russia}
\author{D.~Glenzinski$^{\dag}$}~\affiliation{Fermi National Accelerator Laboratory, Batavia, Illinois 60510, USA}
\author{M.~Gold$^{\dag}$}~\affiliation{University of New Mexico, Albuquerque, New Mexico 87131, USA}
\author{D.~Goldin$^{\dag}$}~\affiliation{Mitchell Institute for Fundamental Physics and Astronomy, Texas A\&M University, College Station, Texas 77843, USA}
\author{A.~Golossanov$^{\dag}$}~\affiliation{Fermi National Accelerator Laboratory, Batavia, Illinois 60510, USA}
\author{G.~Golovanov$^{\ddag}$}~\affiliation{Joint Institute for Nuclear Research, Dubna, Russia}
\author{G.~Gomez$^{\dag}$}~\affiliation{Instituto de Fisica de Cantabria, CSIC-University of Cantabria, 39005 Santander, Spain}
\author{G.~Gomez-Ceballos$^{\dag}$}~\affiliation{Massachusetts Institute of Technology, Cambridge, Massachusetts 02139, USA}
\author{M.~Goncharov$^{\dag}$}~\affiliation{Massachusetts Institute of Technology, Cambridge, Massachusetts 02139, USA}
\author{O.~Gonz\'{a}lez~L\'{o}pez$^{\dag}$}~\affiliation{Centro de Investigaciones Energeticas Medioambientales y Tecnologicas, E-28040 Madrid, Spain}
\author{I.~Gorelov$^{\dag}$}~\affiliation{University of New Mexico, Albuquerque, New Mexico 87131, USA}
\author{A.T.~Goshaw$^{\dag}$}~\affiliation{Duke University, Durham, North Carolina 27708, USA}
\author{K.~Goulianos$^{\dag}$}~\affiliation{The Rockefeller University, New York, New York 10065, USA}
\author{E.~Gramellini$^{\dag}$}~\affiliation{Istituto Nazionale di Fisica Nucleare Bologna, $^{\sharp{a}}$University of Bologna, I-40127 Bologna, Italy}
\author{P.D.~Grannis$^{\ddag}$}~\affiliation{State University of New York, Stony Brook, New York 11794, USA}
\author{S.~Greder$^{\ddag}$}~\affiliation{IPHC, Universit\'e de Strasbourg, CNRS/IN2P3, Strasbourg, France}
\author{H.~Greenlee$^{\ddag}$}~\affiliation{Fermi National Accelerator Laboratory, Batavia, Illinois 60510, USA}
\author{G.~Grenier$^{\ddag}$}~\affiliation{IPNL, Universit\'e Lyon 1, CNRS/IN2P3, Villeurbanne, France and Universit\'e de Lyon, Lyon, France}
\author{S.~Grinstein$^{\dag}$}~\affiliation{Institut de Fisica d'Altes Energies, ICREA, Universitat Autonoma de Barcelona, E-08193, Bellaterra (Barcelona), Spain}
\author{Ph.~Gris$^{\ddag}$}~\affiliation{LPC, Universit\'e Blaise Pascal, CNRS/IN2P3, Clermont, France}
\author{J.-F.~Grivaz$^{\ddag}$}~\affiliation{LAL, Universit\'e Paris-Sud, CNRS/IN2P3, Orsay, France}
\author{A.~Grohsjean$^{\ddag d}$}~\affiliation{CEA, Irfu, SPP, Saclay, France}
\author{C.~Grosso-Pilcher$^{\dag}$}~\affiliation{Enrico Fermi Institute, University of Chicago, Chicago, Illinois 60637, USA}
\author{R.C.~Group$^{\dag}$}~\affiliation{University of Virginia, Charlottesville, Virginia 22904, USA}~\affiliation{Fermi National Accelerator Laboratory, Batavia, Illinois 60510, USA}
\author{S.~Gr\"unendahl$^{\ddag}$}~\affiliation{Fermi National Accelerator Laboratory, Batavia, Illinois 60510, USA}
\author{M.W.~Gr{\"u}newald$^{\ddag}$}~\affiliation{University College Dublin, Dublin, Ireland}
\author{T.~Guillemin$^{\ddag}$}~\affiliation{LAL, Universit\'e Paris-Sud, CNRS/IN2P3, Orsay, France}
\author{J.~Guimaraes~da~Costa$^{\dag}$}~\affiliation{Harvard University, Cambridge, Massachusetts 02138, USA}
\author{G.~Gutierrez$^{\ddag}$}~\affiliation{Fermi National Accelerator Laboratory, Batavia, Illinois 60510, USA}
\author{P.~Gutierrez$^{\ddag}$}~\affiliation{University of Oklahoma, Norman, Oklahoma 73019, USA}
\author{S.R.~Hahn$^{\dag}$}~\affiliation{Fermi National Accelerator Laboratory, Batavia, Illinois 60510, USA}
\author{J.~Haley$^{\ddag}$}~\affiliation{Northeastern University, Boston, Massachusetts 02115, USA}
\author{J.Y.~Han$^{\dag}$}~\affiliation{University of Rochester, Rochester, New York 14627, USA}
\author{L.~Han$^{\ddag}$}~\affiliation{University of Science and Technology of China, Hefei, People's Republic of China}
\author{F.~Happacher$^{\dag}$}~\affiliation{Laboratori Nazionali di Frascati, Istituto Nazionale di Fisica Nucleare, I-00044 Frascati, Italy}
\author{K.~Hara$^{\dag}$}~\affiliation{University of Tsukuba, Tsukuba, Ibaraki 305, Japan}
\author{K.~Harder$^{\ddag}$}~\affiliation{The University of Manchester, Manchester M13 9PL, United Kingdom}
\author{M.~Hare$^{\dag}$}~\affiliation{Tufts University, Medford, Massachusetts 02155, USA}
\author{A.~Harel$^{\ddag}$}~\affiliation{University of Rochester, Rochester, New York 14627, USA}
\author{R.F.~Harr$^{\dag}$}~\affiliation{Wayne State University, Detroit, Michigan 48201, USA}
\author{T.~Harrington-Taber$^u$}~\affiliation{Fermi National Accelerator Laboratory, Batavia, Illinois 60510, USA}
\author{K.~Hatakeyama$^{\dag}$}~\affiliation{Baylor University, Waco, Texas 76798, USA}
\author{J.M.~Hauptman$^{\ddag}$}~\affiliation{Iowa State University, Ames, Iowa 50011, USA}
\author{C.~Hays$^{\dag}$}~\affiliation{University of Oxford, Oxford OX1 3RH, United Kingdom}
\author{J.~Hays$^{\ddag}$}~\affiliation{Imperial College London, London SW7 2AZ, United Kingdom}
\author{T.~Head$^{\ddag}$}~\affiliation{The University of Manchester, Manchester M13 9PL, United Kingdom}
\author{T.~Hebbeker$^{\ddag}$}~\affiliation{III. Physikalisches Institut A, RWTH Aachen University, Aachen, Germany}
\author{D.~Hedin$^{\ddag}$}~\affiliation{Northern Illinois University, DeKalb, Illinois 60115, USA}
\author{H.~Hegab$^{\ddag}$}~\affiliation{Oklahoma State University, Stillwater, Oklahoma 74078, USA}
\author{J.~Heinrich$^{\dag}$}~\affiliation{University of Pennsylvania, Philadelphia, Pennsylvania 19104, USA}
\author{A.P.~Heinson$^{\ddag}$}~\affiliation{University of California Riverside, Riverside, California 92521, USA}
\author{U.~Heintz$^{\ddag}$}~\affiliation{Brown University, Providence, Rhode Island 02912, USA}
\author{C.~Hensel$^{\ddag}$}~\affiliation{II. Physikalisches Institut, Georg-August-Universit\"at G\"ottingen, G\"ottingen, Germany}
\author{I.~Heredia-De~La~Cruz$^{\ddag}$}~\affiliation{CINVESTAV, Mexico City, Mexico}
\author{M.~Herndon$^{\dag}$}~\affiliation{University of Wisconsin, Madison, Wisconsin 53706, USA}
\author{K.~Herner$^{\ddag}$}~\affiliation{University of Michigan, Ann Arbor, Michigan 48109, USA}
\author{G.~Hesketh$^{\ddag e}$}~\affiliation{The University of Manchester, Manchester M13 9PL, United Kingdom}
\author{M.D.~Hildreth$^{\ddag}$}~\affiliation{University of Notre Dame, Notre Dame, Indiana 46556, USA}
\author{R.~Hirosky$^{\ddag}$}~\affiliation{University of Virginia, Charlottesville, Virginia 22904, USA}
\author{T.~Hoang$^{\ddag}$}~\affiliation{Florida State University, Tallahassee, Florida 32306, USA}
\author{J.D.~Hobbs$^{\ddag}$}~\affiliation{State University of New York, Stony Brook, New York 11794, USA}
\author{A.~Hocker$^{\dag}$}~\affiliation{Fermi National Accelerator Laboratory, Batavia, Illinois 60510, USA}
\author{B.~Hoeneisen$^{\ddag}$}~\affiliation{Universidad San Francisco de Quito, Quito, Ecuador}
\author{J.~Hogan$^{\ddag}$}~\affiliation{Rice University, Houston, Texas 77005, USA}
\author{M.~Hohlfeld$^{\ddag}$}~\affiliation{Institut f\"ur Physik, Universit\"at Mainz, Mainz, Germany}
\author{Z.~Hong$^{\dag}$}~\affiliation{Mitchell Institute for Fundamental Physics and Astronomy, Texas A\&M University, College Station, Texas 77843, USA}
\author{W.~Hopkins$^{\dag j}$}~\affiliation{Fermi National Accelerator Laboratory, Batavia, Illinois 60510, USA}
\author{S.~Hou$^{\dag}$}~\affiliation{Institute of Physics, Academia Sinica, Taipei, Taiwan 11529, Republic of China}
\author{I.~Howley$^{\ddag}$}~\affiliation{University of Texas, Arlington, Texas 76019, USA}
\author{Z.~Hubacek$^{\ddag}$}~\affiliation{Czech Technical University in Prague, Prague, Czech Republic}~\affiliation{CEA, Irfu, SPP, Saclay, France}
\author{R.E.~Hughes$^{\dag}$}~\affiliation{The Ohio State University, Columbus, Ohio 43210, USA}
\author{U.~Husemann$^{\dag}$}~\affiliation{Yale University, New Haven, Connecticut 06520, USA}
\author{M.~Hussein$^{\dag h}$}~\affiliation{Michigan State University, East Lansing, Michigan 48824, USA}
\author{J.~Huston$^{\dag}$}~\affiliation{Michigan State University, East Lansing, Michigan 48824, USA}
\author{V.~Hynek$^{\ddag}$}~\affiliation{Czech Technical University in Prague, Prague, Czech Republic}
\author{I.~Iashvili$^{\ddag}$}~\affiliation{State University of New York, Buffalo, New York 14260, USA}
\author{Y.~Ilchenko$^{\ddag}$}~\affiliation{Southern Methodist University, Dallas, Texas 75275, USA}
\author{R.~Illingworth$^{\ddag}$}~\affiliation{Fermi National Accelerator Laboratory, Batavia, Illinois 60510, USA}
\author{G.~Introzzi$^{\dag \sharp{h}}$}~\affiliation{Istituto Nazionale di Fisica Nucleare Pisa, $^{\sharp{c}}$University of Pisa, $^{\sharp{d}}$University of Siena and $^{\sharp{e}}$Scuola Normale Superiore, I-56127 Pisa, Italy, $^{\sharp{h}}$INFN Pavia and University of Pavia, I-27100 Pavia, Italy}
\author{M.~Iori$^{\dag\sharp{f}}$}~\affiliation{Istituto Nazionale di Fisica Nucleare, Sezione di Roma 1, $^{\sharp{f}}$Sapienza Universit\`{a} di Roma, I-00185 Roma, Italy}
\author{A.S.~Ito$^{\ddag}$}~\affiliation{Fermi National Accelerator Laboratory, Batavia, Illinois 60510, USA}
\author{A.~Ivanov$^{\dag k}$}~\affiliation{University of California, Davis, Davis, California 95616, USA}
\author{S.~Jabeen$^{\ddag}$}~\affiliation{Brown University, Providence, Rhode Island 02912, USA}
\author{M.~Jaffr\'e$^{\ddag}$}~\affiliation{LAL, Universit\'e Paris-Sud, CNRS/IN2P3, Orsay, France}
\author{E.~James$^{\dag}$}~\affiliation{Fermi National Accelerator Laboratory, Batavia, Illinois 60510, USA}
\author{D.~Jang$^{\dag}$}~\affiliation{Carnegie Mellon University, Pittsburgh, Pennsylvania 15213, USA}
\author{A.~Jayasinghe$^{\ddag}$}~\affiliation{University of Oklahoma, Norman, Oklahoma 73019, USA}
\author{B.~Jayatilaka$^{\dag}$}~\affiliation{Fermi National Accelerator Laboratory, Batavia, Illinois 60510, USA}
\author{E.J.~Jeon$^{\dag}$}~\affiliation{Center for High Energy Physics: Kyungpook National University, Daegu 702-701, Korea; Seoul National University, Seoul 151-742, Korea; Sungkyunkwan University, Suwon 440-746, Korea; Korea Institute of Science and Technology Information, Daejeon 305-806, Korea; Chonnam National University, Gwangju 500-757, Korea; Chonbuk National University, Jeonju 561-756, Korea; Ewha Womans University, Seoul, 120-750, Korea}
\author{M.S.~Jeong$^{\ddag}$}~\affiliation{Korea Detector Laboratory, Korea University, Seoul, Korea}
\author{R.~Jesik$^{\ddag}$}~\affiliation{Imperial College London, London SW7 2AZ, United Kingdom}
\author{P.~Jiang$^{\ddag}$}~\affiliation{University of Science and Technology of China, Hefei, People's Republic of China}
\author{S.~Jindariani$^{\dag}$}~\affiliation{Fermi National Accelerator Laboratory, Batavia, Illinois 60510, USA}
\author{K.~Johns$^{\ddag}$}~\affiliation{University of Arizona, Tucson, Arizona 85721, USA}
\author{E.~Johnson$^{\ddag}$}~\affiliation{Michigan State University, East Lansing, Michigan 48824, USA}
\author{M.~Johnson$^{\ddag}$}~\affiliation{Fermi National Accelerator Laboratory, Batavia, Illinois 60510, USA}
\author{A.~Jonckheere$^{\ddag}$}~\affiliation{Fermi National Accelerator Laboratory, Batavia, Illinois 60510, USA}
\author{M.~Jones$^{\dag}$}~\affiliation{Purdue University, West Lafayette, Indiana 47907, USA}
\author{P.~Jonsson$^{\ddag}$}~\affiliation{Imperial College London, London SW7 2AZ, United Kingdom}
\author{K.K.~Joo$^{\dag}$}~\affiliation{Center for High Energy Physics: Kyungpook National University, Daegu 702-701, Korea; Seoul National University, Seoul 151-742, Korea; Sungkyunkwan University, Suwon 440-746, Korea; Korea Institute of Science and Technology Information, Daejeon 305-806, Korea; Chonnam National University, Gwangju 500-757, Korea; Chonbuk National University, Jeonju 561-756, Korea; Ewha Womans University, Seoul, 120-750, Korea}
\author{J.~Joshi$^{\ddag}$}~\affiliation{University of California Riverside, Riverside, California 92521, USA}
\author{S.Y.~Jun$^{\dag}$}~\affiliation{Carnegie Mellon University, Pittsburgh, Pennsylvania 15213, USA}
\author{A.W.~Jung$^{\ddag}$}~\affiliation{Fermi National Accelerator Laboratory, Batavia, Illinois 60510, USA}
\author{T.R.~Junk$^{\dag}$}~\affiliation{Fermi National Accelerator Laboratory, Batavia, Illinois 60510, USA}
\author{A.~Juste$^{\ddag}$}~\affiliation{Instituci\'{o} Catalana de Recerca i Estudis Avan\c{c}ats (ICREA) and Institut de F\'{i}sica d'Altes Energies (IFAE), Barcelona, Spain}
\author{E.~Kajfasz$^{\ddag}$}~\affiliation{CPPM, Aix-Marseille Universit\'e, CNRS/IN2P3, Marseille, France}
\author{M.~Kambeitz$^{\dag}$}~\affiliation{Institut f\"{u}r Experimentelle Kernphysik, Karlsruhe Institute of Technology, D-76131 Karlsruhe, Germany}
\author{T.~Kamon$^{\dag}$}~\affiliation{Center for High Energy Physics: Kyungpook National University, Daegu 702-701, Korea; Seoul National University, Seoul 151-742, Korea; Sungkyunkwan University, Suwon 440-746, Korea; Korea Institute of Science and Technology Information, Daejeon 305-806, Korea; Chonnam National University, Gwangju 500-757, Korea; Chonbuk National University, Jeonju 561-756, Korea; Ewha Womans University, Seoul, 120-750, Korea}~\affiliation{Mitchell Institute for Fundamental Physics and Astronomy, Texas A\&M University, College Station, Texas 77843, USA}
\author{P.E.~Karchin$^{\dag}$}~\affiliation{Wayne State University, Detroit, Michigan 48201, USA}
\author{D.~Karmanov$^{\ddag}$}~\affiliation{Moscow State University, Moscow, Russia}
\author{A.~Kasmi$^{\dag}$}~\affiliation{Baylor University, Waco, Texas 76798, USA}
\author{Y.~Kato$^{\dag l}$}~\affiliation{Osaka City University, Osaka 588, Japan}
\author{I.~Katsanos$^{\ddag}$}~\affiliation{University of Nebraska, Lincoln, Nebraska 68588, USA}
\author{R.~Kehoe$^{\ddag}$}~\affiliation{Southern Methodist University, Dallas, Texas 75275, USA}
\author{S.~Kermiche$^{\ddag}$}~\affiliation{CPPM, Aix-Marseille Universit\'e, CNRS/IN2P3, Marseille, France}
\author{W.~Ketchum$^{\dag ss}$ }~\affiliation{Enrico Fermi Institute, University of Chicago, Chicago, Illinois 60637, USA}
\author{J.~Keung$^{\dag}$}~\affiliation{University of Pennsylvania, Philadelphia, Pennsylvania 19104, USA}
\author{N.~Khalatyan$^{\ddag}$}~\affiliation{Fermi National Accelerator Laboratory, Batavia, Illinois 60510, USA}
\author{A.~Khanov$^{\ddag}$}~\affiliation{Oklahoma State University, Stillwater, Oklahoma 74078, USA}
\author{A.~Kharchilava$^{\ddag}$}~\affiliation{State University of New York, Buffalo, New York 14260, USA}
\author{Y.N.~Kharzheev$^{\ddag}$}~\affiliation{Joint Institute for Nuclear Research, Dubna, Russia}
\author{B.~Kilminster$^{\dag oo}$}~\affiliation{Fermi National Accelerator Laboratory, Batavia, Illinois 60510, USA}
\author{D.H.~Kim$^{\dag}$}~\affiliation{Center for High Energy Physics: Kyungpook National University, Daegu 702-701, Korea; Seoul National University, Seoul 151-742, Korea; Sungkyunkwan University, Suwon 440-746, Korea; Korea Institute of Science and Technology Information, Daejeon 305-806, Korea; Chonnam National University, Gwangju 500-757, Korea; Chonbuk National University, Jeonju 561-756, Korea; Ewha Womans University, Seoul, 120-750, Korea}
\author{H.S.~Kim$^{\dag}$}~\affiliation{Center for High Energy Physics: Kyungpook National University, Daegu 702-701, Korea; Seoul National University, Seoul 151-742, Korea; Sungkyunkwan University, Suwon 440-746, Korea; Korea Institute of Science and Technology Information, Daejeon 305-806, Korea; Chonnam National University, Gwangju 500-757, Korea; Chonbuk National University, Jeonju 561-756, Korea; Ewha Womans University, Seoul, 120-750, Korea}
\author{J.E.~Kim$^{\dag}$}~\affiliation{Center for High Energy Physics: Kyungpook National University, Daegu 702-701, Korea; Seoul National University, Seoul 151-742, Korea; Sungkyunkwan University, Suwon 440-746, Korea; Korea Institute of Science and Technology Information, Daejeon 305-806, Korea; Chonnam National University, Gwangju 500-757, Korea; Chonbuk National University, Jeonju 561-756, Korea; Ewha Womans University, Seoul, 120-750, Korea}
\author{M.J.~Kim$^{\dag}$}~\affiliation{Laboratori Nazionali di Frascati, Istituto Nazionale di Fisica Nucleare, I-00044 Frascati, Italy}
\author{S.B.~Kim$^{\dag}$}~\affiliation{Center for High Energy Physics: Kyungpook National University, Daegu 702-701, Korea; Seoul National University, Seoul 151-742, Korea; Sungkyunkwan University, Suwon 440-746, Korea; Korea Institute of Science and Technology Information, Daejeon 305-806, Korea; Chonnam National University, Gwangju 500-757, Korea; Chonbuk National University, Jeonju 561-756, Korea; Ewha Womans University, Seoul, 120-750, Korea}
\author{S.H.~Kim$^{\dag}$}~\affiliation{University of Tsukuba, Tsukuba, Ibaraki 305, Japan}
\author{Y.J.~Kim$^{\dag}$}~\affiliation{Center for High Energy Physics: Kyungpook National University, Daegu 702-701, Korea; Seoul National University, Seoul 151-742, Korea; Sungkyunkwan University, Suwon 440-746, Korea; Korea Institute of Science and Technology Information, Daejeon 305-806, Korea; Chonnam National University, Gwangju 500-757, Korea; Chonbuk National University, Jeonju 561-756, Korea; Ewha Womans University, Seoul, 120-750, Korea}
\author{Y.K.~Kim$^{\dag}$}~\affiliation{Enrico Fermi Institute, University of Chicago, Chicago, Illinois 60637, USA}
\author{N.~Kimura$^{\dag}$}~\affiliation{Waseda University, Tokyo 169, Japan}
\author{M.~Kirby$^{\dag}$}~\affiliation{Fermi National Accelerator Laboratory, Batavia, Illinois 60510, USA}
\author{I.~Kiselevich$^{\ddag}$}~\affiliation{Institution for Theoretical and Experimental Physics, ITEP, Moscow 117259, Russia}
\author{K.~Knoepfel$^{\dag}$}~\affiliation{Fermi National Accelerator Laboratory, Batavia, Illinois 60510, USA}
\author{J.M.~Kohli$^{\ddag}$}~\affiliation{Panjab University, Chandigarh, India}
\author{K.~Kondo\footnote{Deceased}$^{\dag}$}~\affiliation{Waseda University, Tokyo 169, Japan}
\author{D.J.~Kong$^{\dag}$}~\affiliation{Center for High Energy Physics: Kyungpook National University, Daegu 702-701, Korea; Seoul National University, Seoul 151-742, Korea; Sungkyunkwan University, Suwon 440-746, Korea; Korea Institute of Science and Technology Information, Daejeon 305-806, Korea; Chonnam National University, Gwangju 500-757, Korea; Chonbuk National University, Jeonju 561-756, Korea; Ewha Womans University, Seoul, 120-750, Korea}
\author{J.~Konigsberg$^{\dag}$}~\affiliation{University of Florida, Gainesville, Florida 32611, USA}
\author{A.V.~Kotwal$^{\dag}$}~\affiliation{Duke University, Durham, North Carolina 27708, USA}
\author{A.V.~Kozelov$^{\ddag}$}~\affiliation{Institute for High Energy Physics, Protvino, Russia}
\author{J.~Kraus$^{\ddag}$}~\affiliation{University of Mississippi, University, Mississippi 38677, USA}
\author{M.~Kreps$^{\dag}$}~\affiliation{Institut f\"{u}r Experimentelle Kernphysik, Karlsruhe Institute of Technology, D-76131 Karlsruhe, Germany}
\author{J.~Kroll$^{\dag}$}~\affiliation{University of Pennsylvania, Philadelphia, Pennsylvania 19104, USA}
\author{M.~Kruse$^{\dag}$}~\affiliation{Duke University, Durham, North Carolina 27708, USA}
\author{T.~Kuhr$^{\dag}$}~\affiliation{Institut f\"{u}r Experimentelle Kernphysik, Karlsruhe Institute of Technology, D-76131 Karlsruhe, Germany}
\author{A.~Kumar$^{\ddag}$}~\affiliation{State University of New York, Buffalo, New York 14260, USA}
\author{A.~Kupco$^{\ddag}$}~\affiliation{Center for Particle Physics, Institute of Physics, Academy of Sciences of the Czech Republic, Prague, Czech Republic}
\author{M.~Kurata$^{\dag}$}~\affiliation{University of Tsukuba, Tsukuba, Ibaraki 305, Japan}
\author{T.~Kur\v{c}a$^{\ddag}$}~\affiliation{IPNL, Universit\'e Lyon 1, CNRS/IN2P3, Villeurbanne, France and Universit\'e de Lyon, Lyon, France}
\author{V.A.~Kuzmin$^{\ddag}$}~\affiliation{Moscow State University, Moscow, Russia}
\author{A.T.~Laasanen$^{\dag}$}~\affiliation{Purdue University, West Lafayette, Indiana 47907, USA}
\author{S.~Lammel$^{\dag}$}~\affiliation{Fermi National Accelerator Laboratory, Batavia, Illinois 60510, USA}
\author{S.~Lammers$^{\ddag}$}~\affiliation{Indiana University, Bloomington, Indiana 47405, USA}
\author{M.~Lancaster$^{\dag}$}~\affiliation{University College London, London WC1E 6BT, United Kingdom}
\author{K.~Lannon$^{\dag n}$}~\affiliation{The Ohio State University, Columbus, Ohio 43210, USA}
\author{G.~Latino$^{\dag\sharp{d}}$}~\affiliation{Istituto Nazionale di Fisica Nucleare Pisa, $^{\sharp{c}}$University of Pisa, $^{\sharp{d}}$University of Siena and $^{\sharp{e}}$Scuola Normale Superiore, I-56127 Pisa, Italy, $^{\sharp{h}}$INFN Pavia and University of Pavia, I-27100 Pavia, Italy}
\author{P.~Lebrun$^{\ddag}$}~\affiliation{IPNL, Universit\'e Lyon 1, CNRS/IN2P3, Villeurbanne, France and Universit\'e de Lyon, Lyon, France}
\author{H.S.~Lee$^{\ddag}$}~\affiliation{Korea Detector Laboratory, Korea University, Seoul, Korea}
\author{H.S.~Lee$^{\dag}$}~\affiliation{Center for High Energy Physics: Kyungpook National University, Daegu 702-701, Korea; Seoul National University, Seoul 151-742, Korea; Sungkyunkwan University, Suwon 440-746, Korea; Korea Institute of Science and Technology Information, Daejeon 305-806, Korea; Chonnam National University, Gwangju 500-757, Korea; Chonbuk National University, Jeonju 561-756, Korea; Ewha Womans University, Seoul, 120-750, Korea}
\author{J.S.~Lee$^{\dag}$}~\affiliation{Center for High Energy Physics: Kyungpook National University, Daegu 702-701, Korea; Seoul National University, Seoul 151-742, Korea; Sungkyunkwan University, Suwon 440-746, Korea; Korea Institute of Science and Technology Information, Daejeon 305-806, Korea; Chonnam National University, Gwangju 500-757, Korea; Chonbuk National University, Jeonju 561-756, Korea; Ewha Womans University, Seoul, 120-750, Korea}
\author{S.W.~Lee$^{\ddag}$}~\affiliation{Iowa State University, Ames, Iowa 50011, USA}
\author{W.M.~Lee$^{\ddag}$}~\affiliation{Florida State University, Tallahassee, Florida 32306, USA}
\author{X.~Lei$^{\ddag}$}~\affiliation{University of Arizona, Tucson, Arizona 85721, USA}
\author{J.~Lellouch$^{\ddag}$}~\affiliation{LPNHE, Universit\'es Paris VI and VII, CNRS/IN2P3, Paris, France}
\author{S.~Leo$^{\dag}$}~\affiliation{Istituto Nazionale di Fisica Nucleare Pisa, $^{\sharp{c}}$University of Pisa, $^{\sharp{d}}$University of Siena and $^{\sharp{e}}$Scuola Normale Superiore, I-56127 Pisa, Italy, $^{\sharp{h}}$INFN Pavia and University of Pavia, I-27100 Pavia, Italy}
\author{S.~Leone$^{\dag}$}~\affiliation{Istituto Nazionale di Fisica Nucleare Pisa, $^{\sharp{c}}$University of Pisa, $^{\sharp{d}}$University of Siena and $^{\sharp{e}}$Scuola Normale Superiore, I-56127 Pisa, Italy, $^{\sharp{h}}$INFN Pavia and University of Pavia, I-27100 Pavia, Italy}
\author{J.D.~Lewis$^{\dag}$}~\affiliation{Fermi National Accelerator Laboratory, Batavia, Illinois 60510, USA}
\author{D.~Li$^{\ddag}$}~\affiliation{LPNHE, Universit\'es Paris VI and VII, CNRS/IN2P3, Paris, France}
\author{D.~Li$^{\ddag}$}~\affiliation{LPNHE, Universit\'es Paris VI and VII, CNRS/IN2P3, Paris, France}
\author{H.~Li$^{\ddag}$}~\affiliation{University of Virginia, Charlottesville, Virginia 22904, USA}
\author{L.~Li$^{\ddag}$}~\affiliation{University of California Riverside, Riverside, California 92521, USA}
\author{Q.Z.~Li$^{\ddag}$}~\affiliation{Fermi National Accelerator Laboratory, Batavia, Illinois 60510, USA}
\author{J.K.~Lim$^{\ddag}$}~\affiliation{Korea Detector Laboratory, Korea University, Seoul, Korea}
\author{A.~Limosani$^{\dag q}$}~\affiliation{Duke University, Durham, North Carolina 27708, USA}
\author{D.~Lincoln$^{\ddag}$}~\affiliation{Fermi National Accelerator Laboratory, Batavia, Illinois 60510, USA}
\author{J.~Linnemann$^{\ddag}$}~\affiliation{Michigan State University, East Lansing, Michigan 48824, USA}
\author{V.V.~Lipaev$^{\ddag}$}~\affiliation{Institute for High Energy Physics, Protvino, Russia}
\author{E.~Lipeles$^{\dag}$}~\affiliation{University of Pennsylvania, Philadelphia, Pennsylvania 19104, USA}
\author{R.~Lipton$^{\ddag}$}~\affiliation{Fermi National Accelerator Laboratory, Batavia, Illinois 60510, USA}
\author{A.~Lister$^{\dag ee}$}~\affiliation{University of Geneva, CH-1211 Geneva 4, Switzerland}
\author{H.~Liu$^{\dag}$}~\affiliation{University of Virginia, Charlottesville, Virginia 22904, USA}
\author{H.~Liu$^{\ddag}$}~\affiliation{Southern Methodist University, Dallas, Texas 75275, USA}
\author{Q.~Liu$^{\dag}$}~\affiliation{Purdue University, West Lafayette, Indiana 47907, USA}
\author{T.~Liu$^{\dag}$}~\affiliation{Fermi National Accelerator Laboratory, Batavia, Illinois 60510, USA}
\author{Y.~Liu$^{\ddag}$}~\affiliation{University of Science and Technology of China, Hefei, People's Republic of China}
\author{A.~Lobodenko$^{\ddag}$}~\affiliation{Petersburg Nuclear Physics Institute, St. Petersburg, Russia}
\author{S.~Lockwitz$^{\dag}$}~\affiliation{Yale University, New Haven, Connecticut 06520, USA}
\author{A.~Loginov$^{\dag}$}~\affiliation{Yale University, New Haven, Connecticut 06520, USA}
\author{M.~Lokajicek$^{\ddag}$}~\affiliation{Center for Particle Physics, Institute of Physics, Academy of Sciences of the Czech Republic, Prague, Czech Republic}
\author{R.~Lopes~de~Sa$^{\ddag}$}~\affiliation{State University of New York, Stony Brook, New York 11794, USA}
\author{D.~Lucchesi$^{\dag\sharp{b}}$}~\affiliation{Istituto Nazionale di Fisica Nucleare, Sezione di Padova-Trento, $^{\sharp{b}}$University of Padova, I-35131 Padova, Italy}
\author{J.~Lueck$^{\dag}$}~\affiliation{Institut f\"{u}r Experimentelle Kernphysik, Karlsruhe Institute of Technology, D-76131 Karlsruhe, Germany}
\author{P.~Lujan$^{\dag}$}~\affiliation{Ernest Orlando Lawrence Berkeley National Laboratory, Berkeley, California 94720, USA}
\author{P.~Lukens$^{\dag}$}~\affiliation{Fermi National Accelerator Laboratory, Batavia, Illinois 60510, USA}
\author{R.~Luna-Garcia$^{\ddag f}$}~\affiliation{CINVESTAV, Mexico City, Mexico}
\author{G.~Lungu$^{\dag}$}~\affiliation{The Rockefeller University, New York, New York 10065, USA}
\author{A.L.~Lyon$^{\ddag}$}~\affiliation{Fermi National Accelerator Laboratory, Batavia, Illinois 60510, USA}
\author{J.~Lys$^{\dag}$}~\affiliation{Ernest Orlando Lawrence Berkeley National Laboratory, Berkeley, California 94720, USA}
\author{R.~Lysak$^{\dag r}$}~\affiliation{Comenius University, 842 48 Bratislava, Slovakia; Institute of Experimental Physics, 040 01 Kosice, Slovakia}
\author{A.K.A.~Maciel$^{\ddag}$}~\affiliation{LAFEX, Centro Brasileiro de Pesquisas F\'{i}sicas, Rio de Janeiro, Brazil}
\author{R.~Madar$^{\ddag}$}~\affiliation{Physikalisches Institut, Universit\"at Freiburg, Freiburg, Germany}
\author{R.~Madrak$^{\dag}$}~\affiliation{Fermi National Accelerator Laboratory, Batavia, Illinois 60510, USA}
\author{P.~Maestro$^{\dag\sharp{d}}$}~\affiliation{Istituto Nazionale di Fisica Nucleare Pisa, $^{\sharp{c}}$University of Pisa, $^{\sharp{d}}$University of Siena and $^{\sharp{e}}$Scuola Normale Superiore, I-56127 Pisa, Italy, $^{\sharp{h}}$INFN Pavia and University of Pavia, I-27100 Pavia, Italy}
\author{R.~Maga\~na-Villalba$^{\ddag}$}~\affiliation{CINVESTAV, Mexico City, Mexico}
\author{S.~Malik$^{\dag}$}~\affiliation{The Rockefeller University, New York, New York 10065, USA}
\author{S.~Malik$^{\ddag}$}~\affiliation{University of Nebraska, Lincoln, Nebraska 68588, USA}
\author{V.L.~Malyshev$^{\ddag}$}~\affiliation{Joint Institute for Nuclear Research, Dubna, Russia}
\author{G.~Manca$^{\dag s}$}~\affiliation{University of Liverpool, Liverpool L69 7ZE, United Kingdom}
\author{A.~Manousakis-Katsikakis$^{\dag}$}~\affiliation{University of Athens, 157 71 Athens, Greece}
\author{J.~Mansour$^{\ddag}$}~\affiliation{II. Physikalisches Institut, Georg-August-Universit\"at G\"ottingen, G\"ottingen, Germany}
\author{F.~Margaroli$^{\dag}$}~\affiliation{Istituto Nazionale di Fisica Nucleare, Sezione di Roma 1, $^{\sharp{f}}$Sapienza Universit\`{a} di Roma, I-00185 Roma, Italy}
\author{P.~Marino$^{\dag \sharp{e}}$}~\affiliation{Istituto Nazionale di Fisica Nucleare Pisa, $^{\sharp{c}}$University of Pisa, $^{\sharp{d}}$University of Siena and $^{\sharp{e}}$Scuola Normale Superiore, I-56127 Pisa, Italy, $^{\sharp{h}}$INFN Pavia and University of Pavia, I-27100 Pavia, Italy}
\author{M.~Mart\'{\i}nez$^{\dag}$}~\affiliation{Institut de Fisica d'Altes Energies, ICREA, Universitat Autonoma de Barcelona, E-08193, Bellaterra (Barcelona), Spain}
\author{J.~Mart\'{\i}nez-Ortega$^{\ddag}$}~\affiliation{CINVESTAV, Mexico City, Mexico}
\author{K.~Matera$^{\dag}$}~\affiliation{University of Illinois, Urbana, Illinois 61801, USA}
\author{M.E.~Mattson$^{\dag}$}~\affiliation{Wayne State University, Detroit, Michigan 48201, USA}
\author{A.~Mazzacane$^{\dag}$}~\affiliation{Fermi National Accelerator Laboratory, Batavia, Illinois 60510, USA}
\author{P.~Mazzanti$^{\dag}$}~\affiliation{Istituto Nazionale di Fisica Nucleare Bologna, $^{\sharp{a}}$University of Bologna, I-40127 Bologna, Italy}
\author{R.~McCarthy$^{\ddag}$}~\affiliation{State University of New York, Stony Brook, New York 11794, USA}
\author{C.L.~McGivern$^{\ddag}$}~\affiliation{The University of Manchester, Manchester M13 9PL, United Kingdom}
\author{R.~McNulty$^{\dag t}$}~\affiliation{University of Liverpool, Liverpool L69 7ZE, United Kingdom}
\author{A.~Mehta$^{\dag}$}~\affiliation{University of Liverpool, Liverpool L69 7ZE, United Kingdom}
\author{P.~Mehtala$^{\dag}$}~\affiliation{Division of High Energy Physics, Department of Physics, University of Helsinki and Helsinki Institute of Physics, FIN-00014, Helsinki, Finland}
\author{M.M.~Meijer$^{\ddag}$}~\affiliation{Nikhef, Science Park, Amsterdam, the Netherlands}~\affiliation{Radboud University Nijmegen, Nijmegen, the Netherlands}
\author{A.~Melnitchouk$^{\ddag}$}~\affiliation{Fermi National Accelerator Laboratory, Batavia, Illinois 60510, USA}
\author{D.~Menezes$^{\ddag}$}~\affiliation{Northern Illinois University, DeKalb, Illinois 60115, USA}
\author{P.G.~Mercadante$^{\ddag}$}~\affiliation{Universidade Federal do ABC, Santo Andr\'e, Brazil}
\author{M.~Merkin$^{\ddag}$}~\affiliation{Moscow State University, Moscow, Russia}
\author{C.~Mesropian$^{\dag}$}~\affiliation{The Rockefeller University, New York, New York 10065, USA}
\author{A.~Meyer$^{\ddag} j$}~\affiliation{III. Physikalisches Institut A, RWTH Aachen University, Aachen, Germany}
\author{J.~Meyer$^{\ddag}$}~\affiliation{II. Physikalisches Institut, Georg-August-Universit\"at G\"ottingen, G\"ottingen, Germany}
\author{T.~Miao$^{\dag}$}~\affiliation{Fermi National Accelerator Laboratory, Batavia, Illinois 60510, USA}
\author{F.~Miconi$^{\ddag}$}~\affiliation{IPHC, Universit\'e de Strasbourg, CNRS/IN2P3, Strasbourg, France}
\author{D.~Mietlicki$^{\dag}$}~\affiliation{University of Michigan, Ann Arbor, Michigan 48109, USA}
\author{A.~Mitra$^{\dag}$}~\affiliation{Institute of Physics, Academia Sinica, Taipei, Taiwan 11529, Republic of China}
\author{H.~Miyake$^{\dag}$}~\affiliation{University of Tsukuba, Tsukuba, Ibaraki 305, Japan}
\author{S.~Moed$^{\dag}$}~\affiliation{Fermi National Accelerator Laboratory, Batavia, Illinois 60510, USA}
\author{N.~Moggi$^{\dag}$}~\affiliation{Istituto Nazionale di Fisica Nucleare Bologna, $^{\sharp{a}}$University of Bologna, I-40127 Bologna, Italy}
\author{N.K.~Mondal$^{\ddag}$}~\affiliation{Tata Institute of Fundamental Research, Mumbai, India}
\author{C.S.~Moon$^{\dag z}$}~\affiliation{Fermi National Accelerator Laboratory, Batavia, Illinois 60510, USA}
\author{R.~Moore$^{\dag tt}$}~\affiliation{Fermi National Accelerator Laboratory, Batavia, Illinois 60510, USA}
\author{M.J.~Morello$^{\dag\sharp{e}}$}~\affiliation{Istituto Nazionale di Fisica Nucleare Pisa, $^{\sharp{c}}$University of Pisa, $^{\sharp{d}}$University of Siena and $^{\sharp{e}}$Scuola Normale Superiore, I-56127 Pisa, Italy, $^{\sharp{h}}$INFN Pavia and University of Pavia, I-27100 Pavia, Italy}
\author{A.~Mukherjee$^{\dag}$}~\affiliation{Fermi National Accelerator Laboratory, Batavia, Illinois 60510, USA}
\author{M.~Mulhearn$^{\ddag}$}~\affiliation{University of Virginia, Charlottesville, Virginia 22904, USA}
\author{Th.~Muller$^{\dag}$}~\affiliation{Institut f\"{u}r Experimentelle Kernphysik, Karlsruhe Institute of Technology, D-76131 Karlsruhe, Germany}
\author{P.~Murat$^{\dag}$}~\affiliation{Fermi National Accelerator Laboratory, Batavia, Illinois 60510, USA}
\author{M.~Mussini$^{\dag\sharp{a}}$}~\affiliation{Istituto Nazionale di Fisica Nucleare Bologna, $^{\sharp{a}}$University of Bologna, I-40127 Bologna, Italy}
\author{J.~Nachtman$^{\dag u}$}~\affiliation{Fermi National Accelerator Laboratory, Batavia, Illinois 60510, USA}
\author{Y.~Nagai$^{\dag}$}~\affiliation{University of Tsukuba, Tsukuba, Ibaraki 305, Japan}
\author{J.~Naganoma$^{\dag}$}~\affiliation{Waseda University, Tokyo 169, Japan}
\author{E.~Nagy$^{\ddag}$}~\affiliation{CPPM, Aix-Marseille Universit\'e, CNRS/IN2P3, Marseille, France}
\author{M.~Naimuddin$^{\ddag}$}~\affiliation{Delhi University, Delhi, India}
\author{I.~Nakano$^{\dag}$}~\affiliation{Okayama University, Okayama 700-8530, Japan}
\author{A.~Napier$^{\dag}$}~\affiliation{Tufts University, Medford, Massachusetts 02155, USA}
\author{M.~Narain$^{\ddag}$}~\affiliation{Brown University, Providence, Rhode Island 02912, USA}
\author{R.~Nayyar$^{\ddag}$}~\affiliation{University of Arizona, Tucson, Arizona 85721, USA}
\author{H.A.~Neal$^{\ddag}$}~\affiliation{University of Michigan, Ann Arbor, Michigan 48109, USA}
\author{J.P.~Negret$^{\ddag}$}~\affiliation{Universidad de los Andes, Bogot\'a, Colombia}
\author{J.~Nett$^{\dag}$}~\affiliation{Mitchell Institute for Fundamental Physics and Astronomy, Texas A\&M University, College Station, Texas 77843, USA}
\author{C.~Neu$^{\dag}$}~\affiliation{University of Virginia, Charlottesville, Virginia 22904, USA}
\author{P.~Neustroev$^{\ddag}$}~\affiliation{Petersburg Nuclear Physics Institute, St. Petersburg, Russia}
\author{H.T.~Nguyen$^{\ddag}$}~\affiliation{University of Virginia, Charlottesville, Virginia 22904, USA}
\author{T.~Nigmanov$^{\dag}$}~\affiliation{University of Pittsburgh, Pittsburgh, Pennsylvania 15260, USA}
\author{L.~Nodulman$^{\dag}$}~\affiliation{Argonne National Laboratory, Argonne, Illinois 60439, USA}
\author{S.Y.~Noh$^{\dag}$}~\affiliation{Center for High Energy Physics: Kyungpook National University, Daegu 702-701, Korea; Seoul National University, Seoul 151-742, Korea; Sungkyunkwan University, Suwon 440-746, Korea; Korea Institute of Science and Technology Information, Daejeon 305-806, Korea; Chonnam National University, Gwangju 500-757, Korea; Chonbuk National University, Jeonju 561-756, Korea; Ewha Womans University, Seoul, 120-750, Korea}
\author{O.~Norniella$^{\dag}$}~\affiliation{University of Illinois, Urbana, Illinois 61801, USA}
\author{T.~Nunnemann$^{\ddag}$}~\affiliation{Ludwig-Maximilians-Universit\"at M\"unchen, M\"unchen, Germany}
\author{L.~Oakes$^{\dag}$}~\affiliation{University of Oxford, Oxford OX1 3RH, United Kingdom}
\author{S.H.~Oh$^{\dag}$}~\affiliation{Duke University, Durham, North Carolina 27708, USA}
\author{Y.D.~Oh$^{\dag}$}~\affiliation{Center for High Energy Physics: Kyungpook National University, Daegu 702-701, Korea; Seoul National University, Seoul 151-742, Korea; Sungkyunkwan University, Suwon 440-746, Korea; Korea Institute of Science and Technology Information, Daejeon 305-806, Korea; Chonnam National University, Gwangju 500-757, Korea; Chonbuk National University, Jeonju 561-756, Korea; Ewha Womans University, Seoul, 120-750, Korea}
\author{I.~Oksuzian$^{\dag}$}~\affiliation{University of Virginia, Charlottesville, Virginia 22904, USA}
\author{T.~Okusawa$^{\dag}$}~\affiliation{Osaka City University, Osaka 588, Japan}
\author{R.~Orava$^{\dag}$}~\affiliation{Division of High Energy Physics, Department of Physics, University of Helsinki and Helsinki Institute of Physics, FIN-00014, Helsinki, Finland}
\author{J.~Orduna$^{\ddag}$}~\affiliation{Rice University, Houston, Texas 77005, USA}
\author{L.~Ortolan$^{\dag}$}~\affiliation{Institut de Fisica d'Altes Energies, ICREA, Universitat Autonoma de Barcelona, E-08193, Bellaterra (Barcelona), Spain}
\author{N.~Osman$^{\ddag}$}~\affiliation{CPPM, Aix-Marseille Universit\'e, CNRS/IN2P3, Marseille, France}
\author{J.~Osta$^{\ddag}$}~\affiliation{University of Notre Dame, Notre Dame, Indiana 46556, USA}
\author{M.~Padilla$^{\ddag}$}~\affiliation{University of California Riverside, Riverside, California 92521, USA}
\author{C.~Pagliarone$^{\dag}$}~\affiliation{Istituto Nazionale di Fisica Nucleare Trieste/Udine; $^{\sharp{i}}$University of Trieste, I-34127 Trieste, Italy; $^{\sharp{g}}$University of Udine, I-33100 Udine, Italy}
\author{A.~Pal$^{\ddag}$}~\affiliation{University of Texas, Arlington, Texas 76019, USA}
\author{E.~Palencia$^{\dag e}$}~\affiliation{Instituto de Fisica de Cantabria, CSIC-University of Cantabria, 39005 Santander, Spain}
\author{P.~Palni$^{\dag}$}~\affiliation{University of New Mexico, Albuquerque, New Mexico 87131, USA}
\author{V.~Papadimitriou$^{\dag}$}~\affiliation{Fermi National Accelerator Laboratory, Batavia, Illinois 60510, USA}
\author{N.~Parashar$^{\ddag}$}~\affiliation{Purdue University Calumet, Hammond, Indiana 46323, USA}
\author{V.~Parihar$^{\ddag}$}~\affiliation{Brown University, Providence, Rhode Island 02912, USA}
\author{S.K.~Park$^{\ddag}$}~\affiliation{Korea Detector Laboratory, Korea University, Seoul, Korea}
\author{W.~Parker$^{\dag}$}~\affiliation{University of Wisconsin, Madison, Wisconsin 53706, USA}
\author{R.~Partridge$^{\ddag g}$}~\affiliation{Brown University, Providence, Rhode Island 02912, USA}
\author{N.~Parua$^{\ddag}$}~\affiliation{Indiana University, Bloomington, Indiana 47405, USA}
\author{A.~Patwa$^{\ddag k}$}~\affiliation{Brookhaven National Laboratory, Upton, New York 11973, USA}
\author{G.~Pauletta$^{\dag\sharp{g}}$}~\affiliation{Istituto Nazionale di Fisica Nucleare Trieste/Udine; $^{\sharp{i}}$University of Trieste, I-34127 Trieste, Italy; $^{\sharp{g}}$University of Udine, I-33100 Udine, Italy}
\author{M.~Paulini$^{\dag}$}~\affiliation{Carnegie Mellon University, Pittsburgh, Pennsylvania 15213, USA}
\author{C.~Paus$^{\dag}$}~\affiliation{Massachusetts Institute of Technology, Cambridge, Massachusetts 02139, USA}
\author{B.~Penning$^{\ddag}$}~\affiliation{Fermi National Accelerator Laboratory, Batavia, Illinois 60510, USA}
\author{M.~Perfilov$^{\ddag}$}~\affiliation{Moscow State University, Moscow, Russia}
\author{Y.~Peters$^{\ddag}$}~\affiliation{II. Physikalisches Institut, Georg-August-Universit\"at G\"ottingen, G\"ottingen, Germany}
\author{K.~Petridis$^{\ddag}$}~\affiliation{The University of Manchester, Manchester M13 9PL, United Kingdom}
\author{G.~Petrillo$^{\ddag}$}~\affiliation{University of Rochester, Rochester, New York 14627, USA}
\author{P.~P\'etroff$^{\ddag}$}~\affiliation{LAL, Universit\'e Paris-Sud, CNRS/IN2P3, Orsay, France}
\author{T.J.~Phillips$^{\dag}$}~\affiliation{Duke University, Durham, North Carolina 27708, USA}
\author{G.~Piacentino$^{\dag}$}~\affiliation{Istituto Nazionale di Fisica Nucleare Pisa, $^{\sharp{c}}$University of Pisa, $^{\sharp{d}}$University of Siena and $^{\sharp{e}}$Scuola Normale Superiore, I-56127 Pisa, Italy, $^{\sharp{h}}$INFN Pavia and University of Pavia, I-27100 Pavia, Italy}
\author{E.~Pianori$^{\dag}$}~\affiliation{University of Pennsylvania, Philadelphia, Pennsylvania 19104, USA}
\author{J.~Pilot$^{\dag}$}~\affiliation{The Ohio State University, Columbus, Ohio 43210, USA}
\author{K.~Pitts$^{\dag}$}~\affiliation{University of Illinois, Urbana, Illinois 61801, USA}
\author{C.~Plager$^{\dag}$}~\affiliation{University of California, Los Angeles, Los Angeles, California 90024, USA}
\author{M.-A.~Pleier$^{\ddag}$}~\affiliation{Brookhaven National Laboratory, Upton, New York 11973, USA}
\author{P.L.M.~Podesta-Lerma$^{\ddag h}$}~\affiliation{CINVESTAV, Mexico City, Mexico}
\author{V.M.~Podstavkov$^{\ddag}$}~\affiliation{Fermi National Accelerator Laboratory, Batavia, Illinois 60510, USA}
\author{L.~Pondrom$^{\dag}$}~\affiliation{University of Wisconsin, Madison, Wisconsin 53706, USA}
\author{A.V.~Popov$^{\ddag}$}~\affiliation{Institute for High Energy Physics, Protvino, Russia}
\author{S.~Poprocki$^{\dag j}$}~\affiliation{Fermi National Accelerator Laboratory, Batavia, Illinois 60510, USA}
\author{K.~Potamianos$^{\dag}$}~\affiliation{Ernest Orlando Lawrence Berkeley National Laboratory, Berkeley, California 94720, USA}
\author{A.~Pranko$^{\dag}$}~\affiliation{Ernest Orlando Lawrence Berkeley National Laboratory, Berkeley, California 94720, USA}
\author{M.~Prewitt$^{\ddag}$}~\affiliation{Rice University, Houston, Texas 77005, USA}
\author{D.~Price$^{\ddag}$}~\affiliation{Indiana University, Bloomington, Indiana 47405, USA}
\author{N.~Prokopenko$^{\ddag}$}~\affiliation{Institute for High Energy Physics, Protvino, Russia}
\author{F.~Prokoshin$^{\dag w}$}~\affiliation{Joint Institute for Nuclear Research, Dubna, Russia}
\author{F.~Ptohos$^{\dag x}$}~\affiliation{Laboratori Nazionali di Frascati, Istituto Nazionale di Fisica Nucleare, I-00044 Frascati, Italy}
\author{G.~Punzi$^{\dag\sharp{c}}$}~\affiliation{Istituto Nazionale di Fisica Nucleare Pisa, $^{\sharp{c}}$University of Pisa, $^{\sharp{d}}$University of Siena and $^{\sharp{e}}$Scuola Normale Superiore, I-56127 Pisa, Italy, $^{\sharp{h}}$INFN Pavia and University of Pavia, I-27100 Pavia, Italy}
\author{J.~Qian$^{\ddag}$}~\affiliation{University of Michigan, Ann Arbor, Michigan 48109, USA}
\author{A.~Quadt$^{\ddag}$}~\affiliation{II. Physikalisches Institut, Georg-August-Universit\"at G\"ottingen, G\"ottingen, Germany}
\author{B.~Quinn$^{\ddag}$}~\affiliation{University of Mississippi, University, Mississippi 38677, USA}
\author{M.S.~Rangel$^{\ddag}$}~\affiliation{LAFEX, Centro Brasileiro de Pesquisas F\'{i}sicas, Rio de Janeiro, Brazil}
\author{N.~Ranjan$^{\dag}$}~\affiliation{Purdue University, West Lafayette, Indiana 47907, USA}
\author{P.N.~Ratoff$^{\ddag}$}~\affiliation{Lancaster University, Lancaster LA1 4YB, United Kingdom}
\author{I.~Razumov$^{\ddag}$}~\affiliation{Institute for High Energy Physics, Protvino, Russia}
\author{I.~Redondo~Fern\'{a}ndez$^{\dag}$}~\affiliation{Centro de Investigaciones Energeticas Medioambientales y Tecnologicas, E-28040 Madrid, Spain}
\author{P.~Renton$^{\dag}$}~\affiliation{University of Oxford, Oxford OX1 3RH, United Kingdom}
\author{M.~Rescigno$^{\dag}$}~\affiliation{Istituto Nazionale di Fisica Nucleare, Sezione di Roma 1, $^{\sharp{f}}$Sapienza Universit\`{a} di Roma, I-00185 Roma, Italy}
\author{F.~Rimondi$^{*}$}~\affiliation{Istituto Nazionale di Fisica Nucleare Bologna, $^{\sharp{a}}$University of Bologna, I-40127 Bologna, Italy}
\author{I.~Ripp-Baudot$^{\ddag}$}~\affiliation{IPHC, Universit\'e de Strasbourg, CNRS/IN2P3, Strasbourg, France}
\author{L.~Ristori$^{\dag}$}\affiliation{Istituto Nazionale di Fisica Nucleare Pisa, $^{\sharp{c}}$University of Pisa, $^{\sharp{d}}$University of Siena and $^{\sharp{e}}$Scuola Normale Superiore, I-56127 Pisa, Italy, $^{\sharp{h}}$INFN Pavia and University of Pavia, I-27100 Pavia, Italy}~\affiliation{Fermi National Accelerator Laboratory, Batavia, Illinois 60510, USA}
\author{F.~Rizatdinova$^{\ddag}$}~\affiliation{Oklahoma State University, Stillwater, Oklahoma 74078, USA}
\author{A.~Robson$^{\dag}$}~\affiliation{Glasgow University, Glasgow G12 8QQ, United Kingdom}
\author{T.~Rodriguez$^{\dag}$}~\affiliation{University of Pennsylvania, Philadelphia, Pennsylvania 19104, USA}
\author{S.~Rolli$^{\dag y}$}~\affiliation{Tufts University, Medford, Massachusetts 02155, USA}
\author{M.~Rominsky$^{\ddag}$}~\affiliation{Fermi National Accelerator Laboratory, Batavia, Illinois 60510, USA}
\author{M.~Ronzani$^{\dag \sharp{c}}$}~\affiliation{Istituto Nazionale di Fisica Nucleare Pisa, $^{\sharp{c}}$University of Pisa, $^{\sharp{d}}$University of Siena and $^{\sharp{e}}$Scuola Normale Superiore, I-56127 Pisa, Italy, $^{\sharp{h}}$INFN Pavia and University of Pavia, I-27100 Pavia, Italy}
\author{R.~Roser$^{\dag}$}~\affiliation{Fermi National Accelerator Laboratory, Batavia, Illinois 60510, USA}
\author{J.L.~Rosner$^{\dag}$}~\affiliation{Enrico Fermi Institute, University of Chicago, Chicago, Illinois 60637, USA}
\author{A.~Ross$^{\ddag}$}~\affiliation{Lancaster University, Lancaster LA1 4YB, United Kingdom}
\author{C.~Royon$^{\ddag}$}~\affiliation{CEA, Irfu, SPP, Saclay, France}
\author{P.~Rubinov$^{\ddag}$}~\affiliation{Fermi National Accelerator Laboratory, Batavia, Illinois 60510, USA}
\author{R.~Ruchti$^{\ddag}$}~\affiliation{University of Notre Dame, Notre Dame, Indiana 46556, USA}
\author{F.~Ruffini$^{\dag\sharp{d}}$}~\affiliation{Istituto Nazionale di Fisica Nucleare Pisa, $^{\sharp{c}}$University of Pisa, $^{\sharp{d}}$University of Siena and $^{\sharp{e}}$Scuola Normale Superiore, I-56127 Pisa, Italy, $^{\sharp{h}}$INFN Pavia and University of Pavia, I-27100 Pavia, Italy}
\author{A.~Ruiz$^{\dag}$}~\affiliation{Instituto de Fisica de Cantabria, CSIC-University of Cantabria, 39005 Santander, Spain}
\author{J.~Russ$^{\dag}$}~\affiliation{Carnegie Mellon University, Pittsburgh, Pennsylvania 15213, USA}
\author{V.~Rusu$^{\dag}$}~\affiliation{Fermi National Accelerator Laboratory, Batavia, Illinois 60510, USA}
\author{G.~Sajot$^{\ddag}$}~\affiliation{LPSC, Universit\'e Joseph Fourier Grenoble 1, CNRS/IN2P3, Institut National Polytechnique de Grenoble, Grenoble, France}
\author{W.K.~Sakumoto$^{\dag}$}~\affiliation{University of Rochester, Rochester, New York 14627, USA}
\author{Y.~Sakurai$^{\dag}$}~\affiliation{Waseda University, Tokyo 169, Japan}
\author{P.~Salcido$^{\ddag}$}~\affiliation{Northern Illinois University, DeKalb, Illinois 60115, USA}
\author{A.~S\'anchez-Hern\'andez$^{\ddag}$}~\affiliation{CINVESTAV, Mexico City, Mexico}
\author{M.P.~Sanders$^{\ddag}$}~\affiliation{Ludwig-Maximilians-Universit\"at M\"unchen, M\"unchen, Germany}
\author{L.~Santi$^{\dag\sharp{g}}$}~\affiliation{Istituto Nazionale di Fisica Nucleare Trieste/Udine; $^{\sharp{i}}$University of Trieste, I-34127 Trieste, Italy; $^{\sharp{g}}$University of Udine, I-33100 Udine, Italy}
\author{A.S.~Santos$^{\ddag i}$}~\affiliation{LAFEX, Centro Brasileiro de Pesquisas F\'{i}sicas, Rio de Janeiro, Brazil}
\author{K.~Sato$^{\dag}$}~\affiliation{University of Tsukuba, Tsukuba, Ibaraki 305, Japan}
\author{G.~Savage$^{\ddag}$}~\affiliation{Fermi National Accelerator Laboratory, Batavia, Illinois 60510, USA}
\author{V.~Saveliev$^{\dag g}$}~\affiliation{Fermi National Accelerator Laboratory, Batavia, Illinois 60510, USA}
\author{A.~Savoy-Navarro$^{\dag z}$}~\affiliation{Fermi National Accelerator Laboratory, Batavia, Illinois 60510, USA}
\author{L.~Sawyer$^{\ddag}$}~\affiliation{Louisiana Tech University, Ruston, Louisiana 71272, USA}
\author{T.~Scanlon$^{\ddag}$}~\affiliation{Imperial College London, London SW7 2AZ, United Kingdom}
\author{R.D.~Schamberger$^{\ddag}$}~\affiliation{State University of New York, Stony Brook, New York 11794, USA}
\author{Y.~Scheglov$^{\ddag}$}~\affiliation{Petersburg Nuclear Physics Institute, St. Petersburg, Russia}
\author{H.~Schellman$^{\ddag}$}~\affiliation{Northwestern University, Evanston, Illinois 60208, USA}
\author{P.~Schlabach$^{\dag}$}~\affiliation{Fermi National Accelerator Laboratory, Batavia, Illinois 60510, USA}
\author{E.E.~Schmidt$^{\dag}$}~\affiliation{Fermi National Accelerator Laboratory, Batavia, Illinois 60510, USA}
\author{C.~Schwanenberger$^{\ddag}$}~\affiliation{The University of Manchester, Manchester M13 9PL, United Kingdom}
\author{T.~Schwarz$^{\dag}$}~\affiliation{University of Michigan, Ann Arbor, Michigan 48109, USA}
\author{R.~Schwienhorst$^{\ddag}$}~\affiliation{Michigan State University, East Lansing, Michigan 48824, USA}
\author{L.~Scodellaro$^{\dag}$}~\affiliation{Instituto de Fisica de Cantabria, CSIC-University of Cantabria, 39005 Santander, Spain}
\author{F.~Scuri$^{\dag}$}~\affiliation{Istituto Nazionale di Fisica Nucleare Pisa, $^{\sharp{c}}$University of Pisa, $^{\sharp{d}}$University of Siena and $^{\sharp{e}}$Scuola Normale Superiore, I-56127 Pisa, Italy, $^{\sharp{h}}$INFN Pavia and University of Pavia, I-27100 Pavia, Italy}
\author{S.~Seidel$^{\dag}$}~\affiliation{University of New Mexico, Albuquerque, New Mexico 87131, USA}
\author{Y.~Seiya$^{\dag}$}~\affiliation{Osaka City University, Osaka 588, Japan}
\author{J.~Sekaric$^{\ddag}$}~\affiliation{University of Kansas, Lawrence, Kansas 66045, USA}
\author{A.~Semenov$^{\dag}$}~\affiliation{Joint Institute for Nuclear Research, Dubna, Russia}
\author{H.~Severini$^{\ddag}$}~\affiliation{University of Oklahoma, Norman, Oklahoma 73019, USA}
\author{F.~Sforza$^{\dag\sharp{c}}$}~\affiliation{Istituto Nazionale di Fisica Nucleare Pisa, $^{\sharp{c}}$University of Pisa, $^{\sharp{d}}$University of Siena and $^{\sharp{e}}$Scuola Normale Superiore, I-56127 Pisa, Italy, $^{\sharp{h}}$INFN Pavia and University of Pavia, I-27100 Pavia, Italy}
\author{E.~Shabalina$^{\ddag}$}~\affiliation{II. Physikalisches Institut, Georg-August-Universit\"at G\"ottingen, G\"ottingen, Germany}
\author{S.Z.~Shalhout$^{\dag}$}~\affiliation{University of California, Davis, Davis, California 95616, USA}
\author{V.~Shary$^{\ddag}$}~\affiliation{CEA, Irfu, SPP, Saclay, France}
\author{S.~Shaw$^{\ddag}$}~\affiliation{Michigan State University, East Lansing, Michigan 48824, USA}
\author{A.A.~Shchukin$^{\ddag}$}~\affiliation{Institute for High Energy Physics, Protvino, Russia}
\author{T.~Shears$^{\dag}$}~\affiliation{University of Liverpool, Liverpool L69 7ZE, United Kingdom}
\author{P.F.~Shepard$^{\dag}$}~\affiliation{University of Pittsburgh, Pittsburgh, Pennsylvania 15260, USA}
\author{M.~Shimojima$^{\dag aa}$}~\affiliation{University of Tsukuba, Tsukuba, Ibaraki 305, Japan}
\author{R.K.~Shivpuri$^{\ddag}$}~\affiliation{Delhi University, Delhi, India}
\author{M.~Shochet$^{\dag}$}~\affiliation{Enrico Fermi Institute, University of Chicago, Chicago, Illinois 60637, USA}
\author{I.~Shreyber-Tecker$^{\dag}$}~\affiliation{Institution for Theoretical and Experimental Physics, ITEP, Moscow 117259, Russia}
\author{V.~Simak$^{\ddag}$}~\affiliation{Czech Technical University in Prague, Prague, Czech Republic}
\author{A.~Simonenko$^{\dag}$}~\affiliation{Joint Institute for Nuclear Research, Dubna, Russia}
\author{P.~Sinervo$^{\dag}$}~\affiliation{Institute of Particle Physics: McGill University, Montr\'{e}al, Qu\'{e}bec, Canada H3A~2T8; Simon Fraser University, Burnaby, British Columbia, Canada V5A~1S6; University of Toronto, Toronto, Ontario, Canada M5S~1A7; and TRIUMF, Vancouver, British Columbia, V6T~2A3, Canada}
\author{P.~Skubic$^{\ddag}$}~\affiliation{University of Oklahoma, Norman, Oklahoma 73019, USA}
\author{P.~Slattery$^{\ddag}$}~\affiliation{University of Rochester, Rochester, New York 14627, USA}
\author{K.~Sliwa$^{\dag}$}~\affiliation{Tufts University, Medford, Massachusetts 02155, USA}
\author{D.~Smirnov$^{\ddag}$}~\affiliation{University of Notre Dame, Notre Dame, Indiana 46556, USA}
\author{J.R.~Smith$^{\dag}$}~\affiliation{University of California, Davis, Davis, California 95616, USA}
\author{K.J.~Smith$^{\ddag}$}~\affiliation{State University of New York, Buffalo, New York 14260, USA}
\author{F.D.~Snider$^{\dag}$}~\affiliation{Fermi National Accelerator Laboratory, Batavia, Illinois 60510, USA}
\author{G.R.~Snow$^{\ddag}$}~\affiliation{University of Nebraska, Lincoln, Nebraska 68588, USA}
\author{J.~Snow$^{\ddag}$}~\affiliation{Langston University, Langston, Oklahoma 73050, USA}
\author{S.~Snyder$^{\ddag}$}~\affiliation{Brookhaven National Laboratory, Upton, New York 11973, USA}
\author{S.~S{\"o}ldner-Rembold$^{\ddag}$}~\affiliation{The University of Manchester, Manchester M13 9PL, United Kingdom}
\author{H.~Song$^{\dag}$}~\affiliation{University of Pittsburgh, Pittsburgh, Pennsylvania 15260, USA}
\author{L.~Sonnenschein$^{\ddag}$}~\affiliation{III. Physikalisches Institut A, RWTH Aachen University, Aachen, Germany}
\author{V.~Sorin$^{\dag}$}~\affiliation{Institut de Fisica d'Altes Energies, ICREA, Universitat Autonoma de Barcelona, E-08193, Bellaterra (Barcelona), Spain}
\author{K.~Soustruznik$^{\ddag}$}~\affiliation{Charles University, Faculty of Mathematics and Physics, Center for Particle Physics, Prague, Czech Republic}
\author{M.~Stancari$^{\dag}$}~\affiliation{Fermi National Accelerator Laboratory, Batavia, Illinois 60510, USA}
\author{R.~St.~Denis$^{\dag}$}~\affiliation{Glasgow University, Glasgow G12 8QQ, United Kingdom}
\author{J.~Stark$^{\ddag}$}~\affiliation{LPSC, Universit\'e Joseph Fourier Grenoble 1, CNRS/IN2P3, Institut National Polytechnique de Grenoble, Grenoble, France}
\author{B.~Stelzer$^{\dag}$}~\affiliation{Institute of Particle Physics: McGill University, Montr\'{e}al, Qu\'{e}bec, Canada H3A~2T8; Simon Fraser University, Burnaby, British Columbia, Canada V5A~1S6; University of Toronto, Toronto, Ontario, Canada M5S~1A7; and TRIUMF, Vancouver, British Columbia, V6T~2A3, Canada}
\author{O.~Stelzer-Chilton$^{\dag}$}~\affiliation{Institute of Particle Physics: McGill University, Montr\'{e}al, Qu\'{e}bec, Canada H3A~2T8; Simon Fraser University, Burnaby, British Columbia, Canada V5A~1S6; University of Toronto, Toronto, Ontario, Canada M5S~1A7; and TRIUMF, Vancouver, British Columbia, V6T~2A3, Canada}
\author{D.~Stentz$^{\dag b}$}~\affiliation{Fermi National Accelerator Laboratory, Batavia, Illinois 60510, USA}
\author{D.A.~Stoyanova$^{\ddag}$}~\affiliation{Institute for High Energy Physics, Protvino, Russia}
\author{M.~Strauss$^{\ddag}$}~\affiliation{University of Oklahoma, Norman, Oklahoma 73019, USA}
\author{J.~Strologas$^{\dag}$}~\affiliation{University of New Mexico, Albuquerque, New Mexico 87131, USA}
\author{Y.~Sudo$^{\dag}$}~\affiliation{University of Tsukuba, Tsukuba, Ibaraki 305, Japan}
\author{A.~Sukhanov$^{\dag}$}~\affiliation{Fermi National Accelerator Laboratory, Batavia, Illinois 60510, USA}
\author{I.~Suslov$^{\dag}$}~\affiliation{Joint Institute for Nuclear Research, Dubna, Russia}
\author{L.~Suter$^{\ddag}$}~\affiliation{The University of Manchester, Manchester M13 9PL, United Kingdom}
\author{P.~Svoisky$^{\ddag}$}~\affiliation{University of Oklahoma, Norman, Oklahoma 73019, USA}
\author{K.~Takemasa$^{\dag}$}~\affiliation{University of Tsukuba, Tsukuba, Ibaraki 305, Japan}
\author{Y.~Takeuchi$^{\dag}$}~\affiliation{University of Tsukuba, Tsukuba, Ibaraki 305, Japan}
\author{J.~Tang$^{\dag}$}~\affiliation{Enrico Fermi Institute, University of Chicago, Chicago, Illinois 60637, USA}
\author{M.~Tecchio$^{\dag}$}~\affiliation{University of Michigan, Ann Arbor, Michigan 48109, USA}
\author{P.K.~Teng$^{\dag}$}~\affiliation{Institute of Physics, Academia Sinica, Taipei, Taiwan 11529, Republic of China}
\author{J.~Thom$^{\dag j}$}~\affiliation{Fermi National Accelerator Laboratory, Batavia, Illinois 60510, USA}
\author{E.~Thomson$^{\dag}$}~\affiliation{University of Pennsylvania, Philadelphia, Pennsylvania 19104, USA}
\author{V.~Thukral$^{\dag}$}~\affiliation{Mitchell Institute for Fundamental Physics and Astronomy, Texas A\&M University, College Station, Texas 77843, USA}
\author{M.~Titov$^{\ddag}$}~\affiliation{CEA, Irfu, SPP, Saclay, France}
\author{D.~Toback$^{\dag}$}~\affiliation{Mitchell Institute for Fundamental Physics and Astronomy, Texas A\&M University, College Station, Texas 77843, USA}
\author{S.~Tokar$^{\dag}$}~\affiliation{Comenius University, 842 48 Bratislava, Slovakia; Institute of Experimental Physics, 040 01 Kosice, Slovakia}
\author{V.V.~Tokmenin$^{\ddag}$}~\affiliation{Joint Institute for Nuclear Research, Dubna, Russia}
\author{K.~Tollefson$^{\dag}$}~\affiliation{Michigan State University, East Lansing, Michigan 48824, USA}
\author{T.~Tomura$^{\dag}$}~\affiliation{University of Tsukuba, Tsukuba, Ibaraki 305, Japan}
\author{D.~Tonelli$^{\dag e}$}~\affiliation{Fermi National Accelerator Laboratory, Batavia, Illinois 60510, USA}
\author{S.~Torre$^{\dag}$}~\affiliation{Laboratori Nazionali di Frascati, Istituto Nazionale di Fisica Nucleare, I-00044 Frascati, Italy}
\author{D.~Torretta$^{\dag}$}~\affiliation{Fermi National Accelerator Laboratory, Batavia, Illinois 60510, USA}
\author{P.~Totaro$^{\dag}$}~\affiliation{Istituto Nazionale di Fisica Nucleare, Sezione di Padova-Trento, $^{\sharp{b}}$University of Padova, I-35131 Padova, Italy}
\author{M.~Trovato$^{\dag\sharp{e}}$}~\affiliation{Istituto Nazionale di Fisica Nucleare Pisa, $^{\sharp{c}}$University of Pisa, $^{\sharp{d}}$University of Siena and $^{\sharp{e}}$Scuola Normale Superiore, I-56127 Pisa, Italy, $^{\sharp{h}}$INFN Pavia and University of Pavia, I-27100 Pavia, Italy}
\author{Y.-T.~Tsai$^{\ddag}$}~\affiliation{University of Rochester, Rochester, New York 14627, USA}
\author{D.~Tsybychev$^{\ddag}$}~\affiliation{State University of New York, Stony Brook, New York 11794, USA}
\author{B.~Tuchming$^{\ddag}$}~\affiliation{CEA, Irfu, SPP, Saclay, France}
\author{C.~Tully$^{\ddag}$}~\affiliation{Princeton University, Princeton, New Jersey 08544, USA}
\author{F.~Ukegawa$^{\dag}$}~\affiliation{University of Tsukuba, Tsukuba, Ibaraki 305, Japan}
\author{S.~Uozumi$^{\dag}$}~\affiliation{Center for High Energy Physics: Kyungpook National University, Daegu 702-701, Korea; Seoul National University, Seoul 151-742, Korea; Sungkyunkwan University, Suwon 440-746, Korea; Korea Institute of Science and Technology Information, Daejeon 305-806, Korea; Chonnam National University, Gwangju 500-757, Korea; Chonbuk National University, Jeonju 561-756, Korea; Ewha Womans University, Seoul, 120-750, Korea}
\author{L.~Uvarov$^{\ddag}$}~\affiliation{Petersburg Nuclear Physics Institute, St. Petersburg, Russia}
\author{S.~Uvarov$^{\ddag}$}~\affiliation{Petersburg Nuclear Physics Institute, St. Petersburg, Russia}
\author{S.~Uzunyan$^{\ddag}$}~\affiliation{Northern Illinois University, DeKalb, Illinois 60115, USA}
\author{R.~Van~Kooten$^{\ddag}$}~\affiliation{Indiana University, Bloomington, Indiana 47405, USA}
\author{W.M.~van~Leeuwen$^{\ddag}$}~\affiliation{Nikhef, Science Park, Amsterdam, the Netherlands}
\author{N.~Varelas$^{\ddag}$}~\affiliation{University of Illinois at Chicago, Chicago, Illinois 60607, USA}
\author{E.W.~Varnes$^{\ddag}$}~\affiliation{University of Arizona, Tucson, Arizona 85721, USA}
\author{I.A.~Vasilyev$^{\ddag}$}~\affiliation{Institute for High Energy Physics, Protvino, Russia}
\author{F.~V\'{a}zquez$^{\dag c}$}~\affiliation{University of Florida, Gainesville, Florida 32611, USA}
\author{G.~Velev$^{\dag}$}~\affiliation{Fermi National Accelerator Laboratory, Batavia, Illinois 60510, USA}
\author{C.~Vellidis$^{\dag}$}~\affiliation{Fermi National Accelerator Laboratory, Batavia, Illinois 60510, USA}
\author{A.Y.~Verkheev$^{\ddag}$}~\affiliation{Joint Institute for Nuclear Research, Dubna, Russia}
\author{C.~Vernieri$^{\dag{\sharp{e}}}$}~\affiliation{Istituto Nazionale di Fisica Nucleare Pisa, $^{\sharp{c}}$University of Pisa, $^{\sharp{d}}$University of Siena and $^{\sharp{e}}$Scuola Normale Superiore, I-56127 Pisa, Italy, $^{\sharp{h}}$INFN Pavia and University of Pavia, I-27100 Pavia, Italy}
\author{L.S.~Vertogradov$^{\ddag}$}~\affiliation{Joint Institute for Nuclear Research, Dubna, Russia}
\author{M.~Verzocchi$^{\ddag}$}~\affiliation{Fermi National Accelerator Laboratory, Batavia, Illinois 60510, USA}
\author{M.~Vesterinen$^{\ddag}$}~\affiliation{The University of Manchester, Manchester M13 9PL, United Kingdom}
\author{M.~Vidal$^{\dag}$}~\affiliation{Purdue University, West Lafayette, Indiana 47907, USA}
\author{D.~Vilanova$^{\ddag}$}~\affiliation{CEA, Irfu, SPP, Saclay, France}
\author{R.~Vilar$^{\dag}$}~\affiliation{Instituto de Fisica de Cantabria, CSIC-University of Cantabria, 39005 Santander, Spain}
\author{J.~Viz\'{a}n$^{\dag uu}$}~\affiliation{Instituto de Fisica de Cantabria, CSIC-University of Cantabria, 39005 Santander, Spain}
\author{M.~Vogel$^{\dag}$}~\affiliation{University of New Mexico, Albuquerque, New Mexico 87131, USA}
\author{P.~Vokac$^{\ddag}$}~\affiliation{Czech Technical University in Prague, Prague, Czech Republic}
\author{G.~Volpi$^{\dag}$}~\affiliation{Laboratori Nazionali di Frascati, Istituto Nazionale di Fisica Nucleare, I-00044 Frascati, Italy}
\author{P.~Wagner$^{\dag}$}~\affiliation{University of Pennsylvania, Philadelphia, Pennsylvania 19104, USA}
\author{H.D.~Wahl$^{\ddag}$}~\affiliation{Florida State University, Tallahassee, Florida 32306, USA}
\author{R.~Wallny$^{\dag}$}~\affiliation{University of California, Los Angeles, Los Angeles, California 90024, USA}
\author{S.M.~Wang$^{\dag}$}~\affiliation{Institute of Physics, Academia Sinica, Taipei, Taiwan 11529, Republic of China}
\author{M.H.L.S.~Wang$^{\ddag}$}~\affiliation{Fermi National Accelerator Laboratory, Batavia, Illinois 60510, USA}
\author{R.-J.~Wang$^{\ddag}$}~\affiliation{Northeastern University, Boston, Massachusetts 02115, USA}
\author{A.~Warburton$^{\dag}$}~\affiliation{Institute of Particle Physics: McGill University, Montr\'{e}al, Qu\'{e}bec, Canada H3A~2T8; Simon Fraser University, Burnaby, British Columbia, Canada V5A~1S6; University of Toronto, Toronto, Ontario, Canada M5S~1A7; and TRIUMF, Vancouver, British Columbia, V6T~2A3, Canada}
\author{J.~Warchol$^{\ddag}$}~\affiliation{University of Notre Dame, Notre Dame, Indiana 46556, USA}
\author{D.~Waters$^{\dag}$}~\affiliation{University College London, London WC1E 6BT, United Kingdom}
\author{G.~Watts$^{\ddag}$}~\affiliation{University of Washington, Seattle, Washington 98195, USA}
\author{M.~Wayne$^{\ddag}$}~\affiliation{University of Notre Dame, Notre Dame, Indiana 46556, USA}
\author{J.~Weichert$^{\ddag}$}~\affiliation{Institut f\"ur Physik, Universit\"at Mainz, Mainz, Germany}
\author{L.~Welty-Rieger$^{\ddag}$}~\affiliation{Northwestern University, Evanston, Illinois 60208, USA}
\author{W.C.~Wester~III$^{\dag}$}~\affiliation{Fermi National Accelerator Laboratory, Batavia, Illinois 60510, USA}
\author{A.~White$^{\ddag}$}~\affiliation{University of Texas, Arlington, Texas 76019, USA}
\author{D.~Whiteson$^{\dag bb}$}~\affiliation{University of Pennsylvania, Philadelphia, Pennsylvania 19104, USA}
\author{D.~Wicke$^{\ddag}$}~\affiliation{Fachbereich Physik, Bergische Universit\"at Wuppertal, Wuppertal, Germany}
\author{A.B.~Wicklund$^{\dag}$}~\affiliation{Argonne National Laboratory, Argonne, Illinois 60439, USA}
\author{S.~Wilbur$^{\dag}$}~\affiliation{Enrico Fermi Institute, University of Chicago, Chicago, Illinois 60637, USA}
\author{H.H.~Williams$^{\dag}$}~\affiliation{University of Pennsylvania, Philadelphia, Pennsylvania 19104, USA}
\author{M.R.J.~Williams$^{\ddag}$}~\affiliation{Lancaster University, Lancaster LA1 4YB, United Kingdom}
\author{G.W.~Wilson$^{\ddag}$}~\affiliation{University of Kansas, Lawrence, Kansas 66045, USA}
\author{J.S.~Wilson$^{\dag}$}~\affiliation{University of Michigan, Ann Arbor, Michigan 48109, USA}
\author{P.~Wilson$^{\dag}$}~\affiliation{Fermi National Accelerator Laboratory, Batavia, Illinois 60510, USA}
\author{B.L.~Winer$^{\dag}$}~\affiliation{The Ohio State University, Columbus, Ohio 43210, USA}
\author{P.~Wittich$^{\dag j}$}~\affiliation{Fermi National Accelerator Laboratory, Batavia, Illinois 60510, USA}
\author{M.~Wobisch$^{\ddag}$}~\affiliation{Louisiana Tech University, Ruston, Louisiana 71272, USA}
\author{S.~Wolbers$^{\dag}$}~\affiliation{Fermi National Accelerator Laboratory, Batavia, Illinois 60510, USA}
\author{H.~Wolfe$^{\dag}$}~\affiliation{The Ohio State University, Columbus, Ohio 43210, USA}
\author{D.R.~Wood$^{\ddag}$}~\affiliation{Northeastern University, Boston, Massachusetts 02115, USA}
\author{T.~Wright$^{\dag}$}~\affiliation{University of Michigan, Ann Arbor, Michigan 48109, USA}
\author{X.~Wu$^{\dag}$}~\affiliation{University of Geneva, CH-1211 Geneva 4, Switzerland}
\author{Z.~Wu$^{\dag}$}~\affiliation{Baylor University, Waco, Texas 76798, USA}
\author{T.R.~Wyatt$^{\ddag}$}~\affiliation{The University of Manchester, Manchester M13 9PL, United Kingdom}
\author{Y.~Xie$^{\ddag}$}~\affiliation{Fermi National Accelerator Laboratory, Batavia, Illinois 60510, USA}
\author{R.~Yamada$^{\ddag}$}~\affiliation{Fermi National Accelerator Laboratory, Batavia, Illinois 60510, USA}
\author{K.~Yamamoto$^{\dag}$}~\affiliation{Osaka City University, Osaka 588, Japan}
\author{D.~Yamato$^{\dag}$}~\affiliation{Osaka City University, Osaka 588, Japan}
\author{S.~Yang$^{\ddag}$}~\affiliation{University of Science and Technology of China, Hefei, People's Republic of China}
\author{T.~Yang$^{\dag}$}~\affiliation{Fermi National Accelerator Laboratory, Batavia, Illinois 60510, USA}
\author{U.K.~Yang$^{\dag cc}$}~\affiliation{Enrico Fermi Institute, University of Chicago, Chicago, Illinois 60637, USA}
\author{Y.C.~Yang$^{\dag}$}~\affiliation{Center for High Energy Physics: Kyungpook National University, Daegu 702-701, Korea; Seoul National University, Seoul 151-742, Korea; Sungkyunkwan University, Suwon 440-746, Korea; Korea Institute of Science and Technology Information, Daejeon 305-806, Korea; Chonnam National University, Gwangju 500-757, Korea; Chonbuk National University, Jeonju 561-756, Korea; Ewha Womans University, Seoul, 120-750, Korea}
\author{W.-M.~Yao$^{\dag}$}~\affiliation{Ernest Orlando Lawrence Berkeley National Laboratory, Berkeley, California 94720, USA}
\author{T.~Yasuda$^{\ddag}$}~\affiliation{Fermi National Accelerator Laboratory, Batavia, Illinois 60510, USA}
\author{Y.A.~Yatsunenko$^{\ddag}$}~\affiliation{Joint Institute for Nuclear Research, Dubna, Russia}
\author{W.~Ye$^{\ddag}$}~\affiliation{State University of New York, Stony Brook, New York 11794, USA}
\author{Z.~Ye$^{\ddag}$}~\affiliation{Fermi National Accelerator Laboratory, Batavia, Illinois 60510, USA}
\author{G.P.~Yeh$^{\dag}$}~\affiliation{Fermi National Accelerator Laboratory, Batavia, Illinois 60510, USA}
\author{K.~Yi$^{\dag u}$}~\affiliation{Fermi National Accelerator Laboratory, Batavia, Illinois 60510, USA}
\author{H.~Yin$^{\ddag}$}~\affiliation{Fermi National Accelerator Laboratory, Batavia, Illinois 60510, USA}
\author{K.~Yip$^{\ddag}$}~\affiliation{Brookhaven National Laboratory, Upton, New York 11973, USA}
\author{J.~Yoh$^{\dag}$}~\affiliation{Fermi National Accelerator Laboratory, Batavia, Illinois 60510, USA}
\author{K.~Yorita$^{\dag}$}~\affiliation{Waseda University, Tokyo 169, Japan}
\author{T.~Yoshida$^{\dag dd}$}~\affiliation{Osaka City University, Osaka 588, Japan}
\author{S.W.~Youn$^{\ddag}$}~\affiliation{Fermi National Accelerator Laboratory, Batavia, Illinois 60510, USA}
\author{G.B.~Yu$^{\dag}$}~\affiliation{Duke University, Durham, North Carolina 27708, USA}
\author{I.~Yu$^{\dag}$}~\affiliation{Center for High Energy Physics: Kyungpook National University, Daegu 702-701, Korea; Seoul National University, Seoul 151-742, Korea; Sungkyunkwan University, Suwon 440-746, Korea; Korea Institute of Science and Technology Information, Daejeon 305-806, Korea; Chonnam National University, Gwangju 500-757, Korea; Chonbuk National University, Jeonju 561-756, Korea; Ewha Womans University, Seoul, 120-750, Korea}
\author{J.M.~Yu$^{\ddag}$}~\affiliation{University of Michigan, Ann Arbor, Michigan 48109, USA}
\author{A.~Zanetti$^{\dag}$}~\affiliation{Istituto Nazionale di Fisica Nucleare Trieste/Udine; $^{\sharp{i}}$University of Trieste, I-34127 Trieste, Italy; $^{\sharp{g}}$University of Udine, I-33100 Udine, Italy}
\author{Y.~Zeng$^{\dag}$}~\affiliation{Duke University, Durham, North Carolina 27708, USA}
\author{J.~Zennamo$^{\ddag}$}~\affiliation{State University of New York, Buffalo, New York 14260, USA}
\author{T.G.~Zhao$^{\ddag}$}~\affiliation{The University of Manchester, Manchester M13 9PL, United Kingdom}
\author{B.~Zhou$^{\ddag}$}~\affiliation{University of Michigan, Ann Arbor, Michigan 48109, USA}
\author{C.~Zhou$^{\dag}$}~\affiliation{Duke University, Durham, North Carolina 27708, USA}
\author{J.~Zhu$^{\ddag}$}~\affiliation{University of Michigan, Ann Arbor, Michigan 48109, USA}
\author{M.~Zielinski$^{\ddag}$}~\affiliation{University of Rochester, Rochester, New York 14627, USA}
\author{D.~Zieminska$^{\ddag}$}~\affiliation{Indiana University, Bloomington, Indiana 47405, USA}
\author{L.~Zivkovic$^{\ddag}$}~\affiliation{LPNHE, Universit\'es Paris VI and VII, CNRS/IN2P3, Paris, France}
\author{S.~Zucchelli$^{\dag\sharp{a}}$}~\affiliation{Istituto Nazionale di Fisica Nucleare Bologna, $^{\sharp{a}}$University of Bologna, I-40127 Bologna, Italy}
\collaboration{CDF\footnote{
With CDF visitors from
$^{{\dag}a}$Universidad de Oviedo, E-33007 Oviedo, Spain,
$^{{\dag}b}$Northwestern University, Evanston, IL 60208, USA,
$^{{\dag}c}$Universidad Iberoamericana, Mexico D.F., Mexico,
$^{{\dag}d}$ETH, 8092 Z\"urich, Switzerland,
$^{{\dag}e}$CERN, CH-1211 Geneva, Switzerland,
$^{{\dag}f}$Queen Mary, University of London, London, E1 4NS, United Kingdom,
$^{{\dag}g}$National Research Nuclear University, Moscow, Russia,
$^{{\dag}h}$Yarmouk University, Irbid 211-63, Jordan,
$^{{\dag}i}$Muons, Inc., Batavia, IL 60510, USA,
$^{{\dag}j}$Cornell University, Ithaca, NY 14853, USA,
$^{{\dag}k}$Kansas State University, Manhattan, KS 66506, USA,
$^{{\dag}l}$Kinki University, Higashi-Osaka City, Japan 577-8502,
$^{{\dag}n}$University of Notre Dame, Notre Dame, IN 46556, USA,
$^{{\dag}q}$University of Melbourne, Victoria 3010, Australia,
$^{{\dag}r}$Institute of Physics, Academy of Sciences of the Czech Republic, Czech Republic,
$^{{\dag}s}$Istituto Nazionale di Fisica Nucleare, Sezione di Cagliari, 09042 Monserrato (Cagliari), Italy,
$^{{\dag}t}$University College Dublin, Dublin 4, Ireland,
$^{{\dag}u}$University of Iowa, Iowa City, IA 52242, USA,
$^{{\dag}w}$Universidad Tecnica Federico Santa Maria, 110v Valparaiso, Chile,
$^{{\dag}x}$University of Cyprus, Nicosia CY-1678, Cyprus,
$^{{\dag}y}$Office of Science, U.S. Department of Energy, Washington, DC 20585, USA,
$^{{\dag}z}$CNRS-IN2P3, Paris, F-75205 France,
$^{{\dag}aa}$Nagasaki Institute of Applied Science, Nagasaki, Japan,
$^{{\dag}bb}$University of California Irvine, Irvine, CA 92697, USA,
$^{{\dag}cc}$University of Manchester, Manchester M13 9PL, United Kingdom,
$^{{\dag}dd}$University of Fukui, Fukui City, Fukui Prefecture, Japan 910-0017,
$^{{\dag}ee}$University of British Columbia, Vancouver, BC V6T 1Z1, Canada,
$^{{\dag}oo}$University of Z\"urich, 8006 Z\"urich, Switzerland,
$^{{\dag}qq}$Hampton University, Hampton, VA 23668, USA,
$^{{\dag}rr}$Brookhaven National Laboratory, Upton, NY 11973, USA,
$^{{\dag}ss}$Los Alamos National Laboratory, Los Alamos, NM 87544, USA,
$^{{\dag}tt}$Massachusetts General Hospital and Harvard Medical School, Boston, MA 02114 USA,
and
$^{{\dag}uu}$Universite catholique de Louvain, 1348 Louvain-La-Neuve, Belgium.
} and D0\footnote{
and D0 visitors from
$^{{\ddag}a}$Augustana College, Sioux Falls, SD, USA,
$^{{\ddag}b}$The University of Liverpool, Liverpool, UK,
$^{{\ddag}c}$UPIITA-IPN, Mexico City, Mexico,
$^{{\ddag}d}$DESY, Hamburg, Germany,
$^{{\ddag}e}$University College London, London, UK,
$^{{\ddag}f}$Centro de Investigacion en Computacion - IPN, Mexico City, Mexico,
$^{{\ddag}g}$SLAC, Menlo Park, CA, USA,
$^{{\ddag}h}$ECFM, Universidad Autonoma de Sinaloa, Culiac\'an, Mexico,
$^{{\ddag}i}$Universidade Estadual Paulista, S\~ao Paulo, Brazil,
$^{{\ddag}j}$Karlsruher Institut f\"ur Technologie (KIT) - Steinbuch Centre for Computing (SCC)
and
$^{{\ddag}k}$Office of Science, U.S. Department of Energy, Washington, D.C. 20585, USA.
} Collaborations}
\noaffiliation

\vskip 0.25cm

\date{March 25th, 2013}

\begin{abstract}

We combine searches by the CDF and D0 Collaborations for the standard model Higgs boson 
with mass in the range 90--200~GeV$/c^2$ produced in the gluon-gluon fusion, $WH$, $ZH$, 
$t{\bar{t}}H$, and vector boson fusion processes, and decaying in the $H\rightarrow 
b{\bar{b}}$, $H\rightarrow W^+W^-$, $H\rightarrow ZZ$, $H\rightarrow\tau^+\tau^-$, and 
$H\rightarrow \gamma\gamma$ modes.  The data correspond to integrated luminosities 
of up to 10 fb$^{-1}$ and were collected at the Fermilab Tevatron in $p{\bar{p}}$ 
collisions at $\sqrt{s}=1.96$~TeV. The searches are also interpreted in the context 
of fermiophobic and fourth generation models.
We observe a significant excess of events in the mass range between 115 and 140~GeV/$c^2$. 
The local significance corresponds to 3.0 standard deviations at $m_H=125$~GeV/$c^2$, consistent 
with the mass of the Higgs boson observed at the LHC, and we expect a local 
significance of 1.9 standard deviations.  We separately 
combine searches for $H \to b\bar{b}$, $H \to W^+W^-$, $H\rightarrow\tau^+\tau^-$, and 
$H\rightarrow\gamma\gamma$. The observed signal strengths in all channels are consistent 
with the presence of a standard model Higgs boson with a mass of 125~GeV/$c^2$. 

\end{abstract}

\pacs{13.85.Rm, 14.80.Bn}
\maketitle

\section{\label{sec:intro}Introduction}

Within the standard model (SM)~\cite{gws}, spontaneous breaking of electroweak symmetry gives mass to the $W$ and
$Z$ bosons~\cite{higgs}, and to the fundamental fermions via their Yukawa interactions with the Higgs field.
In the SM, the symmetry-breaking mechanism predicts the existence of one neutral
scalar particle, the Higgs boson, whose mass ($m_{H}$) is a free parameter.

Precision electroweak data, including the recently updated measurements of the $W$-boson and top-quark masses from
the CDF and \DZ Collaborations~\cite{topmass,wmass}, yield an
indirect constraint on the allowed mass of the Higgs boson, $m_H < 152$~GeV/$c^2$~\cite{lepewwg}, at the 
95\% confidence level (C.L.)~\cite{clnote}.  Direct searches at LEP2 
exclude SM Higgs boson masses below 114.4~GeV/$c^2$~\cite{sm-lep}.
The ATLAS and CMS Collaborations at the Large Hadron Collider (LHC) have recently reported the observation of 
a new boson with mass of around 125~GeV/$c^2$~\cite{atlas,cms}.  
Much of the sensitivity of the LHC searches comes from gluon-gluon fusion ($gg\rightarrow H$) production and Higgs 
boson decays to $\gamma\gamma$,  $ZZ$, and $W^+W^-$.
Published searches for associated production $VH\rightarrow Vb{\bar{b}}$ at the LHC, where $V=W$ or $Z$~\cite{atlasbb,cmsbb}, 
have not yet reached sensitivity to SM Higgs boson production.  The CDF and D0 Collaborations have recently reported
combined evidence for a particle, with a mass consistent with that of the new boson observed at LHC, produced in 
association with a $W$ or $Z$ boson and decaying to a bottom-antibottom quark pair~\cite{tevhbbprl}.

In this article, we combine the most recent results of SM Higgs boson
searches in \pp~collisions at~\tevE\  using
the full Tevatron Run~II integrated luminosity of up to 10~fb$^{-1}$ per experiment.
The analyses combined here seek signals of Higgs bosons
in the mass range
90--200 GeV/$c^2$, produced in association with
a vector boson ($q\bar{q}\rightarrow VH$), in association with top quarks, through gluon-gluon
fusion, and through vector boson fusion (VBF)
($q\bar{q}\rightarrow q^{\prime}\bar{q}^{\prime}H$). The Higgs boson decay modes studied are
$H\rightarrow b{\bar{b}}$, $H\rightarrow W^+W^-$, $H\rightarrow ZZ$,
$H\rightarrow\tau^+\tau^-$, and $H\rightarrow \gamma\gamma$.
For Higgs boson masses greater than 130 GeV$/c^2$, searches for $H\rightarrow W^+W^-$ decays with subsequent
leptonic $W$ decays provide the greatest sensitivity. Below 130 GeV$/c^2$, sensitivity comes
mainly from associated $VH$ production, with the $H$ boson decaying to $b\bar{b}$ and
the $W$ or $Z$ boson decaying leptonically.   While we present our results in the full mass range,
we also focus specifically on the mass hypothesis $m_H=125$~GeV/$c^2$, due to the recent LHC findings.
Specifically, we show the sensitivity of the searches over the full mass range to a SM Higgs boson 
signal with $m_H=125$~GeV/$c^2$.
Previous Tevatron SM combinations, focused respectively on
the $H\rightarrow b{\bar{b}}$ and $H\rightarrow W^+W^-$ decay modes, are published in Refs.~\cite{tevhbbprl,tevwwprl}. 
The results presented here are based on the combinations of the searches from each experiment as published in Refs.~\cite{cdfprd,d0prd}.

This article is structured as follows.  
Section~\ref{sec:simulation} discusses the simulation methods used to predict the yields from the signal and SM background processes.
Section~\ref{sec:detector} briefly describes the CDF and D0 detectors. 
Section~\ref{sec:selection} describes the event selections used by the various analyses and 
Section~\ref{sec:candidates} presents the data.
Section~\ref{sec:stats} provides a brief introduction to the statistical procedures used and 
Section~\ref{sec:syst} discusses the different sources of systematic uncertainties and how they are controlled. 
Sections~\ref{sec:result} and~\ref{sec:bsm} present the results in the contexts of the SM and extensions to it.
Section~\ref{sec:conclusion} summarizes the article.
\section{\label{sec:simulation}Event simulation}

Higgs boson signal events are simulated using the leading-order (LO) calculation from 
\PYTHIA~\cite{pythia},
with CTEQ5L (CDF) and CTEQ6L1 (D0)~\cite{cteq} parton distribution functions (PDFs).
The normalization of
these Monte Carlo (MC) samples is obtained using the highest-order cross-section
calculation available for the corresponding production process.  The
cross section for the gluon-gluon fusion process is calculated at
next-to-next-to-leading order (NNLO) in quantum chromodynamics (QCD) with
soft gluon resummation to next-to-next-to-leading-log (NNLL)
accuracy~\cite{anastasiou,grazzinideflorian}.  These calculations include two-loop
electroweak corrections, and also three-loop ${\cal{O}}(\alpha\alpha_s)$ corrections.
The {\it WH} and {\it ZH} cross-section 
calculations are performed at NNLO precision in QCD and
next-to-leading-order (NLO) precision in the electroweak corrections~\cite{vhtheory}. 
The VBF cross section is computed at NNLO in QCD~\cite{vbfnnlo}, and the
electroweak corrections are computed with the {\sc hawk} program~\cite{hawk}.
The $t{\bar{t}}H$ production cross sections are taken from Ref.~\cite{tth}.
The signal production cross sections are
computed using the MSTW2008 PDF set~\cite{mstw08}, except for the $t{\bar{t}}H$ production
cross section which uses the CTEQ6M~\cite{cteq} PDF set.  The Higgs boson decay branching fractions
are from Ref.~\cite{lhcdifferential} and rely on calculations using \HDECAY~\cite{hdecay}
and {\sc prophecy4f}~\cite{prophecy4f}. The distribution of the transverse momentum
($p_{T}$) of the Higgs boson in the {\sc pythia}-generated gluon-fusion sample is 
reweighted to match the $p_{T}$ as calculated by {\sc hqt}~\cite{Higgs_pT}, at NNLL and NNLO 
accuracy.

We model SM and instrumental background processes using a mixture of MC and data-driven 
methods.  In the CDF analyses, backgrounds from SM processes
with electroweak gauge bosons or top quarks are modeled using \PYTHIA,
\ALPGEN~\cite{Mangano:2002ea}, {\sc{mc@nlo}}~\cite{MC@NLO}, and \HERWIG~\cite{herwig}.  
For D0, these backgrounds are modeled using
\PYTHIA, \ALPGEN, and {\sc{singletop}}~\cite{comphep}.  An
interface to \PYTHIA\ provides parton showering and hadronization for
generators without this functionality.

Diboson ({\it{WW}}, {\it{WZ}}, {\it{ZZ}}) MC samples are normalized using
the NLO calculations from \MCFM~\cite{mcfm}.  For top-quark-pair production 
($t{\bar{t}}$), we use a production cross section of $7.04\pm 0.49$~pb~\cite{mochuwer}, 
which is based on a top-quark mass of 173~GeV/$c^2$~\cite{topmass} and MSTW 2008 
PDFs~\cite{mstw08}.  The single-top-quark production cross section is
taken to be $3.15\pm 0.31$~pb~\cite{kidonakis_st}. For many analyses, the
{\it V}+jet processes are normalized using the NNLO cross section calculations 
of Ref.~\cite{v-xs}, though in some cases data-driven techniques are used.
Likewise, the normalization of the instrumental, multijet and, for the CDF searches, 
the {\it V}+heavy-flavor jet backgrounds~\cite{hfjet}
are constrained from data samples where the expected signal-to-background ratio 
is several orders of magnitude smaller than in the search samples. For the D0 
searches, the {\it V}+light-flavor is normalized to data in a control region, and 
the {\it V}+heavy-flavor normalization, relative to the {\it V}+light-flavor, is 
taken from \MCFM.  In addition, for the D0 searches, prior to $b$-tagging~\cite{btag} 
{\it V}+jets samples are compared to data and corrections applied to mitigate any 
discrepancies in kinematic distributions.

All MC samples are processed through a {\sc{geant}}~\cite{geant} simulation of the 
detector, and reconstructed in the same way as data. The effects of instrumental noise 
and additional $p{\bar{p}}$ interactions are modeled using MC in the CDF analyses,
while recorded data from randomly selected beam crossings with the same instantaneous luminosity 
profile as data are overlaid on to the MC events in the D0 analyses.
In the entire Run~II data sample, the average number of reconstructed primary vertices is 
approximately 3 -- including the hard scatter.

For the \hww\ analyses, the dominant irreducible background process 
is diboson production, while the dominant reducible backgrounds are
$Z/\gamma^{*}+$jets, $t\bar{t}$, $W$+$\gamma$, $W$+jets, and multijet
production where in the latter three cases photons or jets can be 
misidentified as leptons. For the analyses targeting \hbb\ the main 
backgrounds originate from $V$+heavy-flavor-jets
and $t\bar{t}$ production. 

\section{Detectors and object reconstruction}\label{sec:detector}

The CDF and D0 detectors 
have central trackers
 surrounded by
hermetic calorimeters and muon detectors and are designed to study the products
of 1.96 TeV proton-antiproton collisions~\cite{cdfdetector,d0detector}.
Most searches combined here use the complete Tevatron data sample,
which corresponds to up to 10~\ifb\ depending on the experiment and the 
search channel, after data-quality requirements.
The online event selections (triggers) rely on fast reconstruction
of combinations of high-$p_T$ lepton candidates, jets, and missing transverse energy (\met{}), 
defined below.  To maximize sensitivity, all events satisfying any trigger
requirement from the complete suite of triggers used for data taking
are considered whenever possible. For instance, while most of the \hww\ candidate
events are selected by single-lepton and dilepton triggers, a gain in
efficiency of up to 20\%, depending on the channel, is achieved by
including events that pass lepton+jets and lepton$+\etmiss$ triggers.

High-quality electron candidates are identified by associating charged-particle 
tracks with deposits of energy in the electromagnetic calorimeters when both 
measurements are available.  High-quality muon candidates are identified by 
associating tracks with hits in the muon detectors surrounding the calorimeters 
in the CDF and D0 detectors.  Lepton candidates are categorized based on the 
quality of the contributing measurements.  Tight selection requirements yield 
samples of leptons with low background rates from hadrons or jets of hadrons 
misidentified as leptons. Looser requirements are designed to increase the 
acceptance for lepton candidates with poorly measured or partially missing 
information, with resulting higher rates for backgrounds.  To optimize the 
sensitivity of the combined results, events that are selected with high-quality 
leptons are analyzed separately from those with low-quality leptons.

Jets are clustered from energy deposits in the electromagnetic
and hadronic calorimeters and, in some analyses, combine information 
from charged particle tracks to improve purity or energy resolution.  The transverse energy vector 
${\vec{E}_T}$ of a calorimeter energy deposit is $E\sin\theta {\hat n}$, where 
$E$ is the measured energy, $\theta$ is the angle with respect to the proton beam 
axis of a line drawn from the collision point to the energy deposit, and ${\hat n}$ 
is a unit vector in the plane perpendicular to the beam pointing along that line.  
The missing transverse energy \met{} is the magnitude of the vector opposite to 
the sum of the ${\vec{E}_T}$ vectors measured in the calorimeter, after propagation of all corrections to the calorimetric objects
and for identified muons (which deposit only small amounts of energy in the calorimeters) contributing to the signal topology.
  Further details of the object reconstruction
algorithms used in the Higgs boson searches can be found in the references for the 
individual analyses (see Tables~\ref{tab:cdfacc} and~\ref{tab:dzacc}).

\section{Event selection}\label{sec:selection}

Event selections are similar in the CDF and D0 analyses, typically
consisting of a preselection based on event topology and kinematics. 
Multivariate analysis (MVA) techniques~\cite{mva} are 
used to combine several discriminating variables into a single final
discriminant that is used in the statistical interpretation to compute upper limits,
$p$-values, and fitted cross sections. 
Each channel is divided into exclusive sub-channels according to various
lepton, jet multiplicity, and $b$-tagging characterization criteria.
This procedure groups events with similar signal-to-background ratio to optimize the overall sensitivity.
Such subdivision allows, for example, the efficient use of poorly reconstructed
leptons or those in the forward region, the exploitation of the different dominant
signal and backgrounds when training the MVAs separately in each 
sub-channel, or reduction of the impact of systematic uncertainties.  The MVAs are
trained separately at each value of $m_H$ in their respective mass ranges, in 5~GeV/$c^2$ steps.

For the analyses exploiting the \hbb\ decay, $b$-tagging and dijet mass resolution are of great importance.
Both collaborations have developed multivariate approaches to maximize the performance of the $b$-tagging 
algorithms. The CDF $b$-tagging algorithm is based on an MVA~\cite{tagging}, and depending on the chosen 
operating point provides $b$-tagging efficiencies of 50\%--70\% with misidentification rates for light 
($u$, $d$, $s$, and gluon) jets of 0.5\%--6\%.  In the D0 analyses, the MVA builds and improves upon the previous neural network 
$b$-tagger~\cite{Abazov:2010ab,dzZHv2} and achieves identification efficiencies of about 80\% (50\%) for 
$b$ jets for a light jet misidentification rate of about 10\% (0.5\%).

The decay width of the SM Higgs boson is predicted to be much smaller than the 
experimental dijet mass resolution, which is typically 15\% of the mean reconstructed 
mass.  A SM Higgs boson signal would appear as a broad enhancement in the reconstructed 
di-$b$-jet mass distribution.  The CDF and D0 Collaborations search for $H\rightarrow 
b{\bar{b}}$ produced in association with a leptonically decaying
$W$ boson, or a leptonically or invisibly decaying $Z$ boson.  CDF also contributes 
searches for $WH+ZH\rightarrow jjb{\bar{b}}$ and $t{\bar{t}}H\rightarrow t{\bar{t}}b{\bar{b}}$, 
where in the latter case one of the top quarks decays to a leptonically decaying $W$ boson.  

Both collaborations search for the \hww\ signal in which both $W$ bosons decay 
leptonically by selecting events with large missing transverse energy and
two oppositely-charged, isolated leptons. The presence of neutrinos in
the final state prevents reconstruction of the Higgs boson mass.   
Other observables are used for separating the signal from  background.  
For example, the azimuthal angle between the leptons in signal events is 
smaller on average than that in background events due to the scalar nature 
of the Higgs boson and parity violation in $W^\pm$ decays. Furthermore, the 
missing transverse momentum is larger and the total transverse energy of 
the jets is lower than they are typically in background events. The D0 Collaboration also includes
channels in which one of the $W$ bosons in the \hww\ process decays leptonically and the other hadronically.

Although the primary sensitivity at low mass ($m_H\leq 130$~GeV/$c^2$) is provided by the \hbb{} analyses
and at high mass ($m_H>130$~GeV/$c^2$) by the \hww{} analyses,
significant additional sensitivity is achieved by the inclusion of other channels.
Both collaborations contribute analyses searching for Higgs bosons decaying into 
tri-lepton final states, tau-lepton pairs and diphoton pairs. The full list of channels 
included is shown in Tables~\ref{tab:cdfacc} and~\ref{tab:dzacc} which summarize, 
for the CDF and D0 analyses respectively, the integrated luminosities, the Higgs boson mass
ranges over which the searches are performed, and references to further details 
for each analysis.

\begin{table*}[!]
\caption{\label{tab:cdfacc}Luminosities, explored mass ranges, and references
for the different processes and final states ($\ell$ = $e$ or $\mu$, and 
$\tau_{\rm{had}}$ denotes a hadronic tau-lepton decay) for the CDF analyses.  The 
generic labels ``$1\times$'', ``$2\times$'', ``$3\times$'', and ``$4\times$'' 
refer to separations based on lepton or photon categories.   The analyses are 
grouped in five categories, corresponding to the Higgs boson decay mode to which 
the analysis is most sensitive: $H\to b\bar{b}$, $H\to W^+W^-$, $H\to\tau^+\tau^-$, 
$H\to\gamma\gamma$, and $H\to ZZ$.}
\begin{ruledtabular}
\begin{tabular}{lcccc} \\
Channel & & Luminosity  & $m_H$ range & Reference \\
        & & (fb$^{-1}$) & (GeV/$c^2$) &           \\ \hline
$WH\rightarrow \ell\nu b\bar{b}$ 2-jet channels \ \ \ 4$\times$(5 $b$-tag categories)                                 &     & 9.45 & 90--150 & \cite{cdfWH} \\
$WH\rightarrow \ell\nu b\bar{b}$ 3-jet channels \ \ \ 3$\times$(2 $b$-tag categories)                                 &     & 9.45 & 90--150 & \cite{cdfWH} \\
$ZH\rightarrow \nu\bar{\nu} b\bar{b}$ \ \ \ (3 $b$-tag categories)                                                    &     & 9.45 & 90--150 & \cite{cdfmetbb} \\
$ZH\rightarrow \ell^+\ell^- b\bar{b}$ 2-jet channels \ \ \ 2$\times$(4 $b$-tag categories)                            & $H\rightarrow b\bar{b}$ & 9.45 & 90--150 & \cite{cdfZH} \\
$ZH\rightarrow \ell^+\ell^- b\bar{b}$ 3-jet channels \ \ \ 2$\times$(4 $b$-tag categories)                            &     & 9.45 & 90--150 & \cite{cdfZH} \\
$WH+ZH\rightarrow jjb{\bar{b}}$ \ \ \  (2 $b$-tag categories)                                                         &     & 9.45 & 100--150 & \cite{cdfjjbb} \\
$t\bar{t}H \rightarrow W^+ b W^- \bar{b} b\bar{b}$ \ \ \  (4 jets,5 jets,$\ge$6 jets)$\times$(5 $b$-tag categories)   &     & 9.45 & 100-150 & \cite{cdfttHLep} \\ \hline
$H\rightarrow W^+ W^-$ \ \ \ 2$\times$(0 jets)+2$\times$(1 jet)+1$\times$($\ge$2 jets)+1$\times$(low-$m_{\ell\ell}$)  &     & 9.7  & 110--200 & \cite{cdfHWW} \\ 
$H\rightarrow W^+ W^-$ \ \ \ ($e$-$\tau_{\rm{had}}$)+($\mu$-$\tau_{\rm{had}}$)                                        &     & 9.7  & 130--200 & \cite{cdfHWW} \\
$WH\rightarrow WW^+ W^-$ \ \ \ (same-sign leptons)+(tri-leptons)                                                      & $H\rightarrow W^+W^-$ & 9.7  & 110--200 & \cite{cdfHWW} \\
$WH\rightarrow WW^+ W^-$ \ \ \ (tri-leptons with 1 $\tau_{\rm{had}}$)                                                 &     & 9.7  & 130--200 & \cite{cdfHWW} \\
$ZH\rightarrow ZW^+ W^-$ \ \ \ (tri-leptons with 1 jet,$\ge$2 jets)                                                   &     & 9.7  & 110--200 & \cite{cdfHWW} \\ \hline
$H\rightarrow \tau^+ \tau^-$ \ \ \ (1 jet)+($\ge$2 jets)                                                              & $H\rightarrow \tau^+\tau^-$ & 6.0  & 100--150 & \cite{cdfHtt} \\ \hline
$H \rightarrow \gamma \gamma$ \ \ \  1$\times$(0 jet)+1$\times$($\ge$1 jet)+3$\times$(all jets)                       & $H\rightarrow\gamma\gamma$ & 10.0 & 100--150 & \cite{cdfHgg} \\ \hline
$H\rightarrow ZZ$ \ \ \ (four leptons)                                                                                & $H\rightarrow ZZ$ & 9.7  & 120--200 & \cite{cdfHZZ} \\
\end{tabular}
\end{ruledtabular}
\end{table*}

\vglue 0.5cm

\begin{table*}[!]
\caption{\label{tab:dzacc}Luminosities, explored mass ranges, and references
for the different processes and final states ($\ell=e$ or $\mu$, and $\tau_{\rm{had}}$ 
denotes a hadronic tau-lepton decay) for the D0 analyses. The 
generic labels ``$1\times$'', ``$2\times$'', ``$3\times$'', and ``$4\times$'' 
refer to separations based on lepton, photon or background characterization categories. The analyses are grouped in 
four categories, corresponding to the Higgs boson decay mode to which the analysis is most 
sensitive: $H\to b\bar{b}$, $H\to W^+W^-$, $H\to\tau^+\tau^-$, and $H\to\gamma\gamma$.}
\begin{ruledtabular}
\begin{tabular}{lcccc} \\
Channel && Luminosity  & $m_H$ range & Reference \\
        && (fb$^{-1}$) & (GeV/$c^2$) &           \\ \hline
%
$WH\rightarrow \ell\nu b\bar{b}$ 2-jet channels \ \ \ 2$\times$(4 $b$-tag categories) & & 9.7  & 90--150 & \cite{dzWHl-jul12,dzWHl} \\
$WH\rightarrow \ell\nu b\bar{b}$ 3-jet channels \ \ \ 2$\times$(4 $b$-tag categories) & & 9.7  & 90--150 & \cite{dzWHl-jul12,dzWHl} \\
$ZH\rightarrow \nu\bar{\nu} b\bar{b}$ \ \ \ (2 $b$-tag categories)   & $H\to b\bar{b}$ & 9.5  & 100--150 & \cite{dzZHv2} \\
$ZH\rightarrow \ell^+\ell^- b\bar{b}$  \ \ \ 2$\times$(2 $b$-tag)$\times$(4 lepton categories) & & 9.7  & 90--150 & \cite{dzZHll1-jul12,dzZHll1} \\
\hline
$H\rightarrow W^+ W^- \rightarrow \ell^\pm\nu \ell^\mp\nu$ \ \ \ 2$\times$(0 jets,1 jet,$\ge$2 jets) &  & 9.7  & 115--200 & \cite{dzHWW}\\
$H~+~X\rightarrow W^+ W^- \rightarrow \mu^\mp \nu \tau_{\rm{had}}^\pm \nu$ \ \ \ (3 $\tau$ categories) & & 7.3  & 115--200 & \cite{dzVHt1}\\
$H\rightarrow W^+ W^- \rightarrow \ell\bar{\nu} jj$ \ \ \ 2$\times$(2 $b$-tag categories)$\times$(2 jets, 3 jets) &\multirow{2}{*}{$H\to W^+W^-$}    & 9.7  & 100--200 & \cite{dzWHl}\\
$VH \rightarrow e^\pm \mu^\pm + X $ \ \ \ &  & 9.7  & 100--200 & \cite{dzWWW2} \\
$VH \rightarrow \ell\ell\ell + X $ 	($\mu\mu e$, 3 $\times$ $e\mu\mu$) &   & 9.7  & 100--200 & \cite{dzWWW2} \\
$VH\rightarrow  \ell\bar{\nu} jjjj$ \ \ \ 2$\times$($\ge$4 jets)    & & 9.7  & 100-200 & \cite{dzWHl}\\
\hline
$VH \rightarrow \tau_{\rm{had}} \tau_{\rm{had}} \mu + X $ 	\ \ \ (3 $\tau$ categories)			&\multirow{2}{*}{$H\rightarrow \tau^+\tau^-$}		 & 8.6  & 100-150 & \cite{dzWWW2} \\
$H$+$X$$\rightarrow$$ \ell^\pm \tau^{\mp}_{\rm{had}}jj$  \ \ \ 2$\times$(3 $\tau$ categories)    & & 9.7  & 105--150 & \cite{dzVHt2} \\
\hline
$H \rightarrow \gamma \gamma$ \ \ \ (4 categories)    & $H\rightarrow \gamma\gamma$     & 9.6  & 100-150 & \cite{dzHgg} \\
\end{tabular}
\end{ruledtabular}
\end{table*}

\section{\label{sec:candidates}Candidate distribution}

The number of contributing channels is large, and several different kinds of discriminating varibles are
used.  Visual comparison of the observed 
data with the predictions is challenging in some of the sub-channels due to low data counts.
For a more robust comparison, we display the data from all the sub-channels together, aggregating bins with 
similar signal to background ratios ($s/b$) from all contributing sub-channels. 
We collect 
the signal predictions, the background predictions, and the data in narrow bins of $s/b$, 
summing the contributions from bins in the final discriminant histograms in the sub-channels.  
A fit of the background model (see Section~\ref{sec:stats}) to the data is performed before 
this aggregation procedure, in order to provide the best prediction for the background model 
in bins with the highest sensitivity.  The classification of analysis events according to 
their $s/b$ preserves the importance of each of the events in the histogram, to the extent 
that they are not added to other events that are selected with different $s/b$.   This 
representation of the data is not used to compute the final results, since the distribution 
indiscriminately sums unrelated backgrounds which are fit separately.  It does, however, 
provide a guide to how much individual events contribute to the results and how well the 
signal is separated from backgrounds in the combined search.  The resulting distribution 
of $\log_{10}(s/b)$ is shown for $m_H=125$~GeV/$c^2$ in Fig.~\ref{fig:lnsb}, demonstrating 
agreement with background over five orders of magnitude.

\begin{figure}[!]
\begin{center}
\begin{tabular}{cc}
\includegraphics[width=1.0\columnwidth]{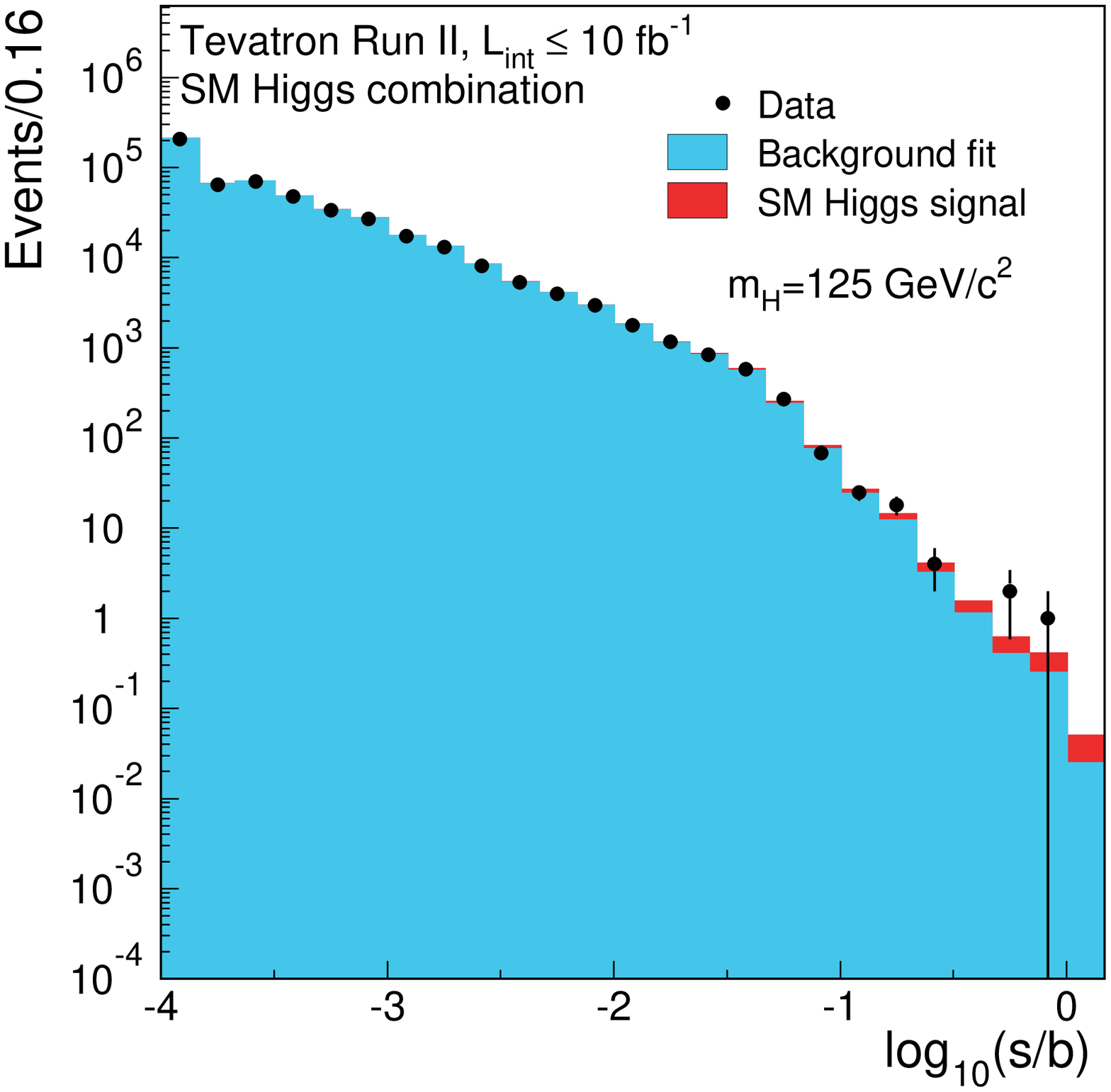}
\end{tabular}
\end{center}
\caption{ \label{fig:lnsb} (color online).  Distribution of $\log_{10}(s/b)$, for the data 
from all contributing Higgs boson search channels from CDF and D0, for $m_H=125$~GeV/$c^2$.  
The data are shown with points, and the expected signal is shown stacked on top of the 
backgrounds, which are fit to the data within their systematic uncertainties.  The error bars shown
on the data correspond in each bin to the square root of the observed data count.  Underflows 
and overflows are collected into the leftmost and rightmost bins, respectively.}
\end{figure}

\section{STATISTICAL TECHNIQUES}\label{sec:stats}

The results are interpreted using both Bayesian and modified frequentist techniques, 
separately at each value of $m_H$, as was done previously~\cite{tevwwprl,tevhbbprl,pdgstats}.
The two methods yield results that are numerically consistent; 
limits on the Higgs boson production rate typically agree within 5\% at each
value of $m_H$, and with a 1\% deviation when averaged over all positive and negative departures. For simplicity, when
summarizing the results, we quote one set of values as the default,
and the {\it a priori} decision made for the earlier Tevatron
combinations to use the Bayesian method is retained here. Both methods
use the distributions of the final discriminants, and not only the
total event counts passing selection requirements.

Each of the techniques is built on a combined likelihood (including prior probability 
densities on systematic uncertainties,
$\pi({\vec{\theta}})$) based on the product of likelihoods for the
individual channels, each of which is a product over histogram bins,
\begin{equation}
{\cal{L}}(R,{\vec{s}},{\vec{b}}|{\vec{n}},{\vec{\theta}})\!\times\!\pi({\vec{\theta}})
= \prod_{i=1}^{N_{\rm C}}\prod_{j=1}^{N_{\rm bins}} \mu_{ij}^{n_{ij}} \frac{e^{-\mu_{ij}}}{n_{ij}!}
\times\prod_{k=1}^{n_{\rm sys}}e^{-\theta_k^2/2},
\end{equation}
\noindent where the first product is over the number of channels
($N_{\rm C}$) and the second product is over histogram bins containing 
$n_{ij}$ events, binned in ranges of the final discriminants used for 
the individual analyses.  The predictions for the bin contents are 
$\mu_{ij} =R \times s_{ij}({\vec{\theta}}) + b_{ij}({\vec{\theta}})$ 
for channel $i$ and histogram bin $j$, where $s_{ij}$ and $b_{ij}$ 
represent the expected SM signal and background in the bin, and $R$ 
is a scaling factor applied to the signal.  By scaling all signal 
contributions by the same factor we assume that the 
relative contributions of the different processes at each $m_H$ are 
as predicted by the SM.  Systematic uncertainties are parametrized by 
the dependence of $s_{ij}$ and $b_{ij}$ on ${\vec\theta}$.  Each of 
the $n_{\rm sys}$ components of ${\vec\theta}$, $\theta_k$, corresponds 
to a single independent source of systematic uncertainty scaled by its 
standard deviation, and each parameter may affect the predictions of 
several sources of signal and background in different channels, thus 
accounting for correlations. Gaussian prior densities are assumed for the 
nuisance parameters, truncated to ensure that no prediction is negative.

In the Bayesian calculation, we assume a uniform prior 
probability density for non-negative values of 
$R$ and integrate the likelihood function multiplied by prior densities 
for the nuisance parameters to obtain the posterior density for $R$.  The 
observed 95\% credibility level upper limit on $R$, $R_{95}^{\rm{obs}}$, 
is the value of $R$ such that the integral of the posterior density of 
$R$ from zero to $R_{95}^{\rm{obs}}$ corresponds to 95\% of the integral
of $R$ from zero to infinity.  The expected distribution of $R_{95}$ is 
computed in an ensemble of simulated experimental outcomes assuming no 
signal is present.  In each simulated outcome, random values of the 
nuisance parameters are drawn from their prior densities.  A combined 
measurement of the cross section for Higgs boson production times the 
branching fraction ${\cal B}(H\rightarrow XX)$, in units of the SM 
production rate, is given by $R^{\rm{fit}}$, which is the value of $R$ 
that maximizes the posterior density. The 68\% credibility interval, 
which corresponds to one standard deviation (s.d.), is quoted as the 
smallest interval containing 68\% of the integral of the posterior.

We also perform calculations with the modified frequentist technique
${\rm CL}_{\rm s}$~\cite{pdgstats}, using a log-likelihood ratio (LLR) 
as the test statistic:
LLR$=-2\ln\frac{p({\mathrm{data}}|s+b)}{p({\mathrm{data}}|b)}$,
where $p({\mathrm{data}}|s+b)$ and $p({\mathrm{data}}|b)$ are the
probabilities that the data (either simulated or experimental data)
are drawn from distributions predicted under the signal-plus-background 
and background-only hypotheses, respectively. The probabilities $p$ 
are computed using the best-fit values of the parameters $\theta_k$, 
separately for each of the two hypotheses~\cite{pflh}.  The use of 
these fits extends the procedure used at LEP~\cite{lep}, improving 
the sensitivity when the expected signals are small and the uncertainties 
on the backgrounds are large.  The  ${\rm CL}_{\rm s}$ technique involves 
computing two $p$-values,
${\rm CL}_{\rm b} = p$(LLR $\ge$ LLR$_{\mathrm{obs}} | b$),
where LLR$_{\mathrm{obs}}$ is the value of the test statistic computed for the
data, and
${\rm CL}_{\rm s+b} = p$(LLR$\ge$ LLR$_{\mathrm{obs}} | s+b$).
To compute limits, we use the ratio of $p$-values, ${\rm CL}_{\rm s}={\rm CL}_{\rm s+b}/{\rm CL}_{\rm b}$.
If ${\rm CL}_{\rm s}<0.05$ for a particular choice
of the signal-plus-background hypothesis, parametrized by the signal scale factor $R$,
that hypothesis is excluded at least at the 95\% C.L. The value of $R_{95}^{\rm{obs}}$ in the CL$_{\rm s}$ method is
the smallest value of $R$ excluded at the 95\% C.L. The expected limit is computed
using the median LLR value expected in the background-only hypothesis. Systematic
uncertainties are included by fluctuating around their Gaussian priors the predictions for $s_{ij}$
and $b_{ij}$ when generating the pseudoexperiments used to compute
CL$_{\mathrm{s+b}}$ and CL$_{\mathrm{b}}$.

In this framework, a second estimate of the signal rate, $R^{\rm{fit}}_{\rm{profile}}$ is computed, maximizing
the likelihood as a function of the unconstrained signal rate $R$ and the nuisance parameters $\theta_k$.  
This estimate of the combined signal rate may differ from the Bayesian calculation of $R^{\rm{fit}}$ when the
likelihood function deviates from a Gaussian form, since the best fit depends on the likelihood near the maximum
and the Bayesian calculation integrates over all values of the nuisance parameters  
which result in positive signal and background rates in all histogram bins.

\section{\label{sec:syst}Systematic Uncertainties}

Systematic uncertainties are evaluated for each final state,
background, and signal process.  Uncertainties that modify only the
normalization and uncertainties that change the shape of the final
discriminant distribution are included.  To study 
the shape uncertainties on the distributions of the final
discriminants, the relevant parameter is varied within one standard 
deviation of its uncertainty and the full analysis repeated using 
the modified distribution.  For example, for the jet energy scale 
and resolution, the parameters of the energy scale and resolution 
are varied within one s.d. of their uncertainties and the analysis 
carried out using the kinematic distributions of the modified jets, 
also including the changes in sample composition resulting from 
the change in the jet energy parameters.  No retraining of the MVAs is 
performed during the propagation of systematic uncertainties to the 
distributions of the discriminants. Correlations between signal and 
background, across different channels within an experiment and across 
the two experiments are taken into account.  Full details on the treatment
of the systematic uncertainties in the individual channels can be found
in the relevant references.

The uncertainties on the inclusive
signal production cross sections are estimated from the variations in the factorization
and renormalization scale, which include the impact of
uncalculated higher-order corrections, uncertainties due to PDFs, and
the dependence on the strong coupling constant, $\alpha_s$, as recommended
by the PDF4LHC working group~\cite{pdf_uncertainties,ggH01jetUncert}.
The resulting uncertainties on the inclusive {\it VH} and VBF
production rates are taken to be 7\% and 5\%,
respectively~\cite{vhtheory}. 
Uncertainties on the branching fractions
are taken from Ref.~\cite{HBR-err}.

For analyses focusing on $gg\rightarrow H$ production that divide events 
into categories based on the number of reconstructed jets, the uncertainties 
associated with the renormalization and factorization scale are estimated 
following Ref.~\cite{errMatr}. By propagating the uncorrelated uncertainties 
of the NNLL inclusive~\cite{anastasiou,grazzinideflorian}, NLO $\ge 1$ jet\,
\cite{ggH01jetUncert}, and NLO $\ge 2$ jets\,\cite{campbellh2j} cross 
sections to the exclusive ${gg\to H}+0$ jet, $\ge 1$ jet, and $\ge 2$ jets 
rates, an uncertainty matrix containing correlated and uncorrelated 
uncertainty contributions between exclusive jet categories is obtained. 
The total uncertainty on $gg\rightarrow H$ production originating from 
these contributions varies from 10\% to 35\% in individual channels depending 
on the number of jets in the final state.  The PDF uncertainties are evaluated 
following Refs.~\cite{anastasiou,ggH01jetUncert}.

Significant sources of uncertainty for all analyses are the integrated luminosities 
used to normalize the expected signal yield and MC-based backgrounds, and the cross
sections for the simulated backgrounds.  For the former, uncertainties  of 6\% (CDF) 
and 6.1\% (D0) are used, with 4\% arising from the inelastic $p{\bar{p}}$ cross section
which is taken to be 100\% correlated between CDF and D0. Cross-section uncertainties 
of 6\% and 7\% are used for diboson and $t\bar{t}$ production respectively.
The uncertainty on the expected multijet
background in each channel is dominated by the statistics
of the data sample from which it is estimated and varies from 10\% to 30\%.

Sources of systematic uncertainty that affect both the normalization
and the shape of the final discriminant distribution include jet energy
scale (1--4)\%, jet energy resolution (1--3)\%, lepton identification, 
trigger efficiencies, and $b$-tagging.  Uncertainties on lepton identification 
and trigger efficiencies range from 2\% to 6\% and are applied to both the signal 
and MC-based background predictions.  These uncertainties are estimated from 
data-based methods separately by CDF and D0, and differ based on lepton flavor and
identification category.  The $b$-tag efficiencies and mistag rates are similarly
constrained by auxiliary data samples, such as inclusive jet data or
$t{\bar{t}}$ events.  The uncertainty on the per-jet $b$-tag efficiency is
approximately 4\%, and the mistag uncertainties vary between 7\% and 15\%.

For the analyses targeting the \hbb\ decay, the largest sources of uncertainty on 
the dominant backgrounds are the rates of $V+$heavy flavor jets, which are 
typically 20--30\% of the predicted values.
Using constraints from the data, the uncertainties on 
these rates are typically 8\% or less.
The data samples in the $V+$jets selections prior to $b$-tagging are used as control 
samples to constrain systematic uncertainties in the MC modeling of the 
energies and angles of jets. Any residual discrepancy coming from the difference 
between light- and heavy-flavor components is shown to be smaller than the systematic 
uncertainties associated with the generator or the correction procedures themselves.

A total of 326 independent sources
of systematic uncertainty are included in the combination of the Higgs boson search
results at $m_H=125$~GeV/$c^2$, not including the independent uncertainties in each bin of each
template from limited Monte Carlo (or data) statistics.  The uncertainties that are considered correlated
between CDF and D0 are those on the 
differential and inclusive theoretical production cross section predictions for the 
Higgs boson signals (itemized by PDF$+\alpha_s$ and scales), the Higgs boson decay branching fractions, 
the $t{\bar{t}}$, single top, and diboson background processes, and the correlated part of the luminosity
estimate.  All other uncertainties are associated with parameters whose central values are estimated using
techniques specific to the experiments and the analysis channels.  We consider these uncorrelated so as
not to extrapolate fit information improperly from one channel or experiment to another where the central
value or the uncertainty scale may be different.

\section{\label{sec:result}Results - Standard model interpretation}

\subsection{Diboson Production}\label{sec:diboson}
To validate our background modeling and methodology,
independent measurements of SM diboson production in the same 
final states used for the SM Higgs searches are carried out. The high 
mass analyses measure $p\bar{p}\rightarrow VV^\prime$ cross sections, while 
the low mass analyses target $VZ(\to b\bar{b})$ production. The data 
sample, reconstruction, process modeling, uncertainties, and sub-channel 
divisions are identical to those of the SM Higgs boson searches.  However, 
discriminant functions are trained to distinguish the contributions of 
SM diboson production from those of other backgrounds, and potential 
contributions from Higgs boson production are not considered. By way of 
illustration, below, we focus on {\it VZ} production.

The NLO SM cross section for {\it VZ} production times the branching fraction 
of $Z\rar b\bar{b}$ is 0.68 $\pm$ 0.05~pb~\cite{mcfm,pdgbr}. This is about six 
times larger than the 0.12 $\pm$ 0.01~pb~\cite{vhtheory,lhcdifferential} cross 
section times branching fraction of $H(\to b\bar{b})V$ for a 125~GeV/$c^2$ SM 
Higgs boson, but the associated background is larger, due to the distribution of
the dijet invariant mass in the $V$+jets events. {\it WW} 
production is considered as background.  The measured cross section, using the 
MVA discriminants, for {\it VZ} is $3.0 \pm 0.6$~(stat)~$\pm 0.7$~(syst)~pb 
whereas the SM prediction is $4.4 \pm 0.3$~pb~\cite{mcfm}. The combined 
background-subtracted dijet-mass distribution for the {\it VZ} 
analysis is shown in Fig.~\ref{fig:bgsubmjj} for illustration.  
\begin{figure}[bth] \begin{centering}
\includegraphics[width=0.8\columnwidth]{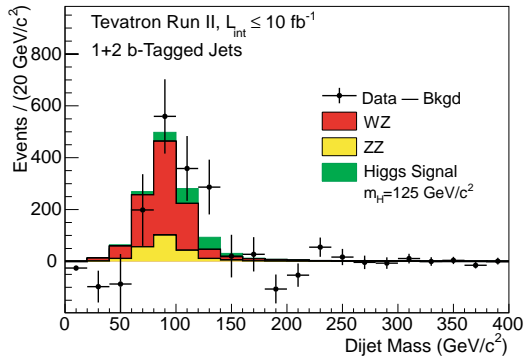}
\caption{
\label{fig:bgsubmjj} (color online). Background-subtracted distribution of the 
reconstructed dijet mass, summed over CDF and D0's channels contributing to the {\it VZ} 
analysis.  The {\it VZ} signal and the background contributions are fit to the 
data, and the fitted background is subtracted.  The fitted {\it VZ} and expected 
SM Higgs ($m_H=125$~GeV/$c^2$) contributions are shown with filled histograms.  
The error bars shown on the data points correspond in each bin to the square 
root of the sum of the expected signal and background yields.}
\end{centering}
\end{figure}
The {\it VZ} 
signal and the background contributions are fit to the data, and the fitted 
background is then subtracted.  Also shown is the contribution expected from 
a SM Higgs boson with $m_H=125$~GeV/$c^2$.  The {\it VV}$^\prime$ boson cross 
sections measured by the high mass analyses are likewise in good agreement 
with SM predictions~\cite{cdfww,d0prd}.
\subsection{Higgs boson combination using all decay modes}\label{sec:fullcomb}

For the search for the Higgs boson, the results produced by the multivariate 
analyses can be visualized by combining the histograms of the final 
discriminants, adding the contents of bins with similar signal-to-background 
ratio ($s/b$) as shown in Fig.~\ref{fig:lnsb}.  Figure~\ref{fig:bgsub} shows 
the signal expectation and the data with the background subtracted, as a 
function of the $s/b$ of the collected bins, for the combined search for a 
Higgs boson with mass $m_H=125$~GeV/$c^2$.  The background model is fit to 
the data, allowing the nuisance parameters to vary within their constraints. 
The uncertainties on the background predictions in each bin are those after 
the fit.  An excess of events in the highest $s/b$ bins relative to the 
background-only expectation is observed.

\begin{figure}[!]
\begin{center}
\begin{tabular}{cc}
\includegraphics[width=1.0\columnwidth]{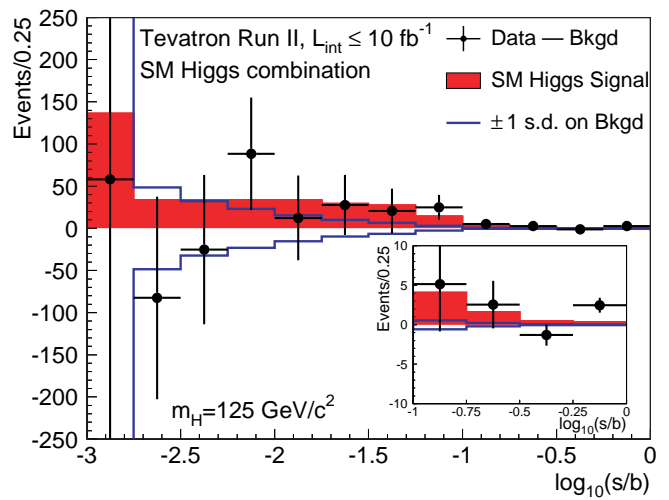}
\end{tabular}
\end{center}
\caption{ (color online). Background-subtracted distribution 
of the discriminant histograms, summed for bins with similar 
signal-to-background ratio ($s/b$) over all contributing Higgs
boson search channels from CDF and D0, for $m_H=125$~GeV/$c^2$. 
The background is fit to the data, and the uncertainty on the
background, shown with the unfilled histogram, is after the 
fit.  The signal model, scaled to the SM expectation, is shown 
with a filled histogram.  The error bars shown on the data 
points correspond in each bin to the square root of the sum 
of the expected signal and background yields.}
\label{fig:bgsub}
\end{figure}

\begin{figure}[htb]
\begin{centering}
\includegraphics[width=1.0\columnwidth]{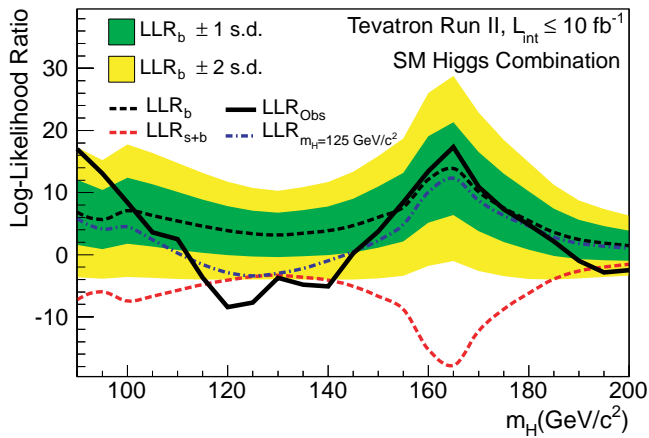}
\caption{
\label{fig:combollr}
(color online). The log-likelihood ratio LLR as a function of Higgs boson 
mass for all of CDF and D0's SM Higgs boson searches in all decay modes combined.  The solid line shows the 
observed LLR values, the dark long-dashed line shows the median expectation 
assuming no Higgs boson signal is present, and the dark- and light-shaded 
bands correspond, respectively, to the regions encompassing one and two 
s.d.~fluctuations around the background-only expectation.  The red long-dashed 
line shows the median expectation assuming a SM Higgs boson signal is present 
at each value of $m_H$ in turn.  The blue short-dashed line shows the median 
expected LLR assuming the SM Higgs boson is present at $m_H=125$~GeV/$c^2$.}
\end{centering}
\end{figure}

Figure~\ref{fig:combollr} displays the LLR distributions for the combined
analyses as functions of $m_{H}$. Included are the median of the LLR distributions for the
background-only hypothesis (LLR$_{b}$), the signal-plus-background hypothesis (LLR$_{s+b}$),
and the observed value for the data (LLR$_{\rm{obs}}$). For mass hypotheses of 95~GeV/$c^2$ 
and less, fewer channels are available for 
combination, giving rise to the behavior of the limits shown. The shaded bands represent the
one and two s.d.~departures for LLR$_{b}$ centered on the median.  These results are 
listed in Table~\ref{tab:smllrVals}.
The separation between the medians of the LLR$_{b}$ and LLR$_{s+b}$ distributions
provides a measure of the discriminating power of the search.  The widths
of the one- and two-s.d.~LLR$_{b}$ bands indicate the width of the LLR$_{b}$
distribution, assuming no signal and that fluctuations originate from statistical 
fluctuations and systematic effects only.  The value of LLR$_{\rm{obs}}$ relative 
to LLR$_{s+b}$ and LLR$_{b}$ indicates whether the data distribution more closely 
resembles the distributions expected if a signal is present (i.e., the LLR$_{s+b}$ 
distribution, which is negative by construction) or only background is present. 
The significance of departures of LLR$_{\rm{obs}}$ from LLR$_{b}$ can be evaluated 
by the width of the LLR$_{b}$ bands. The separation of the median signal-plus-background
and background-only hypotheses is about two s.d., or greater, for Higgs boson masses 
up to $\approx$~185 GeV/$c^2$.  The data are consistent with the background-only 
hypothesis (the black dashed line) at masses smaller than $\approx$~110 GeV/$c^2$ 
and above approximately 145 GeV/$c^2$. A slight excess is seen above approximately 
195 GeV/$c^2$, where our ability to separate the two hypotheses is limited. For 
$m_{H}$ from 115 to 140~GeV/$c^2$,
an excess above two s.d.~in the data with respect to the SM background expectation 
has an amplitude consistent with the expectation for a standard model Higgs boson 
(dashed red line).  Additionally, the LLR curve under the hypothesis that a SM Higgs 
boson is present with $m_H=125$~GeV/$c^2$ is shown. This signal-injected-LLR curve 
has a similar shape to the observed one. While the search for a 125~GeV/$c^2$ Higgs 
boson is optimized to find a Higgs boson of that mass, the excess of events over the 
SM background estimates also affects the results of Higgs boson searches at other 
masses.  Nearby masses are the most affected, but the expected presence of $H\to 
W^+W^-$ decays for a 125~GeV/$c^2$ Higgs boson implies a small expected excess in 
the $H\to W^+W^-$ searches at all masses due to the poor reconstructed mass resolution 
in this final state.

\begin{figure}[thb]
\begin{centering}
\includegraphics[width=1.0\columnwidth]{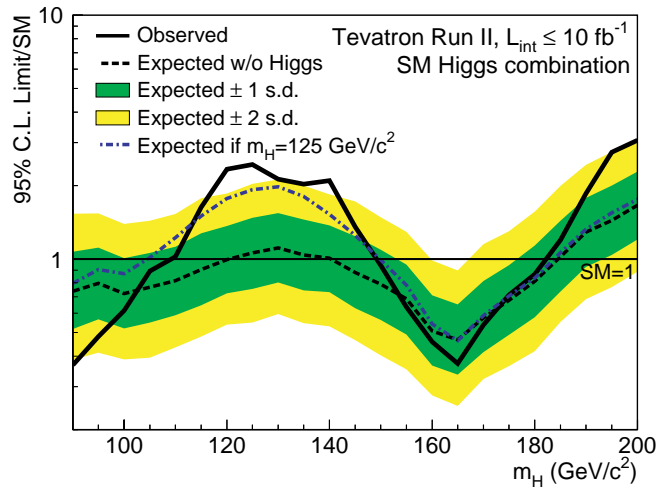}
\caption{
\label{fig:comboRatio}
(color online). Observed and median expected (for the background-only 
hypothesis) 95\% C.L. Bayesian upper production limits expressed as 
multiples of the SM cross section as a function of Higgs boson mass 
for the combined CDF and D0 searches in all decay modes.
The dark- and light-shaded bands indicate, respectively, the one and 
two s.d~probability regions in which the limits are expected to fluctuate in the 
absence of signal.  The blue short-dashed line shows median expected 
limits assuming the SM Higgs boson is present at $m_H=125$~GeV/$c^2$.}
\end{centering}
\end{figure}

\begin{table*}[htb]
\caption{\label{tab:smllrVals} Log-likelihood ratio (LLR) values obtained 
from the combination of all of CDF and D0's Higgs boson search channels using 
the ${\rm CL}_{\rm s}$ method.}
\begin{ruledtabular}
\begin{tabular}{lccccccc}\\
$m_{H}$ (GeV/$c^2$) &  LLR$_{obs}$ & LLR$_{s+b}$ &  LLR$_{b}^{-2\sigma}$ & LLR$_{b}^{-1\sigma}$ & LLR$_{b}$ &  LLR$_{b}^{+1\sigma}$ & LLR$_{b}^{+2\sigma}$ \\
\hline
90 & 17.02 & $-7.24$ & 17.31 & 12.08 & 6.84 & 1.61 & $-3.62$ \\ 
95 & 13.07 & $-5.96$ & 15.21 & 10.44 & 5.68 & 0.91 & $-3.85$ \\ 
100 & 8.39 & $-7.44$ & 17.73 & 12.40 & 7.08 & 1.76 & $-3.56$ \\ 
105 & 3.62 & $-6.69$ & 16.38 & 11.35 & 6.32 & 1.29 & $-3.74$ \\ 
110 & 2.53 & $-5.73$ & 14.79 & 10.12 & 5.45 & 0.78 & $-3.89$ \\ 
115 & $-3.67$ & $-4.81$ & 13.17 & 8.88 & 4.59 & 0.31 & $-3.98$ \\ 
120 & $-8.44$ & $-4.09$ & 11.76 & 7.82 & 3.88 & $-0.06$ & $-4.00$ \\ 
125 & $-7.72$ & $-3.52$ & 10.76 & 7.07 & 3.39 & $-0.29$ & $-3.97$ \\ 
130 & $-3.74$ & $-3.30$ & 10.31 & 6.74 & 3.18 & $-0.39$ & $-3.95$ \\ 
135 & $-4.81$ & $-3.64$ & 10.89 & 7.17 & 3.45 & $-0.26$ & $-3.98$ \\ 
140 & $-5.08$ & $-4.09$ & 11.72 & 7.79 & 3.86 & $-0.07$ & $-4.00$ \\ 
145 & 0.20 & $-5.07$ & 13.35 & 9.02 & 4.69 & 0.36 & $-3.97$ \\ 
150 & 3.72 & $-6.68$ & 15.87 & 10.95 & 6.04 & 1.12 & $-3.79$ \\ 
155 & 8.44 & $-8.80$ & 18.72 & 13.18 & 7.65 & 2.12 & $-3.41$ \\ 
160 & 13.45 & $-15.25$ & 26.04 & 19.08 & 12.12 & 5.15 & $-1.81$ \\ 
165 & 17.33 & $-17.81$ & 28.76 & 21.31 & 13.87 & 6.42 & $-1.03$ \\ 
170 & 10.93 & $-12.26$ & 22.87 & 16.50 & 10.13 & 3.77 & $-2.60$ \\ 
175 & 7.33 & $-8.77$ & 18.50 & 13.02 & 7.53 & 2.04 & $-3.45$ \\ 
180 & 4.86 & $-6.17$ & 14.87 & 10.18 & 5.50 & 0.81 & $-3.88$ \\ 
185 & 2.14 & $-3.92$ & 11.23 & 7.42 & 3.62 & $-0.19$ & $-3.99$ \\ 
190 & $-0.99$ & $-2.61$ & 8.73 & 5.60 & 2.46 & $-0.68$ & $-3.81$ \\ 
195 & $-2.83$ & $-1.98$ & 7.34 & 4.60 & 1.87 & $-0.87$ & $-3.60$ \\ 
200 & $-2.50$ & $-1.53$ & 6.29 & 3.88 & 1.46 & $-0.96$ & $-3.37$ \\ 
\end{tabular}
\end{ruledtabular}
\end{table*}

\begin{figure}[htb]
\begin{centering}
\includegraphics[width=1.0\columnwidth]{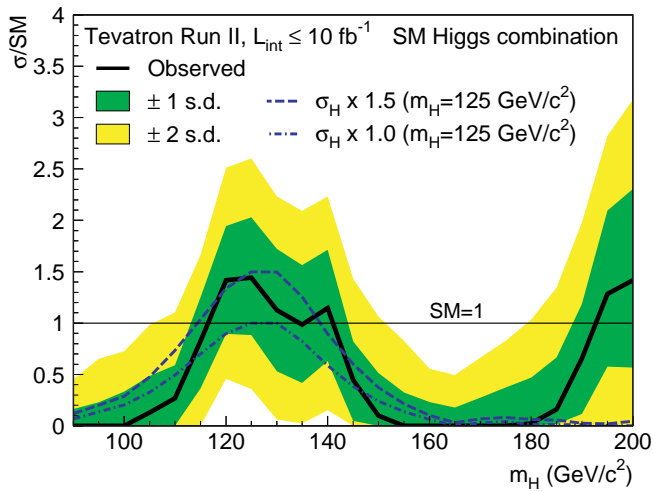}
\caption{
\label{fig:combBestFit}
(color online). The best-fit signal cross section expressed as a ratio to 
the SM cross section as a function of Higgs boson mass for all of CDF and D0's SM Higgs 
boson searches in all decay modes combined.  The dark- and light-shaded bands show the one 
and two s.d.~uncertainty ranges on the fitted signal, respectively.  Also 
shown with blue lines are the median fitted cross sections expected for a 
SM Higgs boson with $m_H=125$~GeV/$c^2$ at signal strengths of 1.0 times 
(short-dashed) and 1.5 times (long-dashed) the SM prediction.}
\end{centering}
\end{figure}
\begin{figure}[htb]
\begin{centering}
\includegraphics[width=1.0\columnwidth]{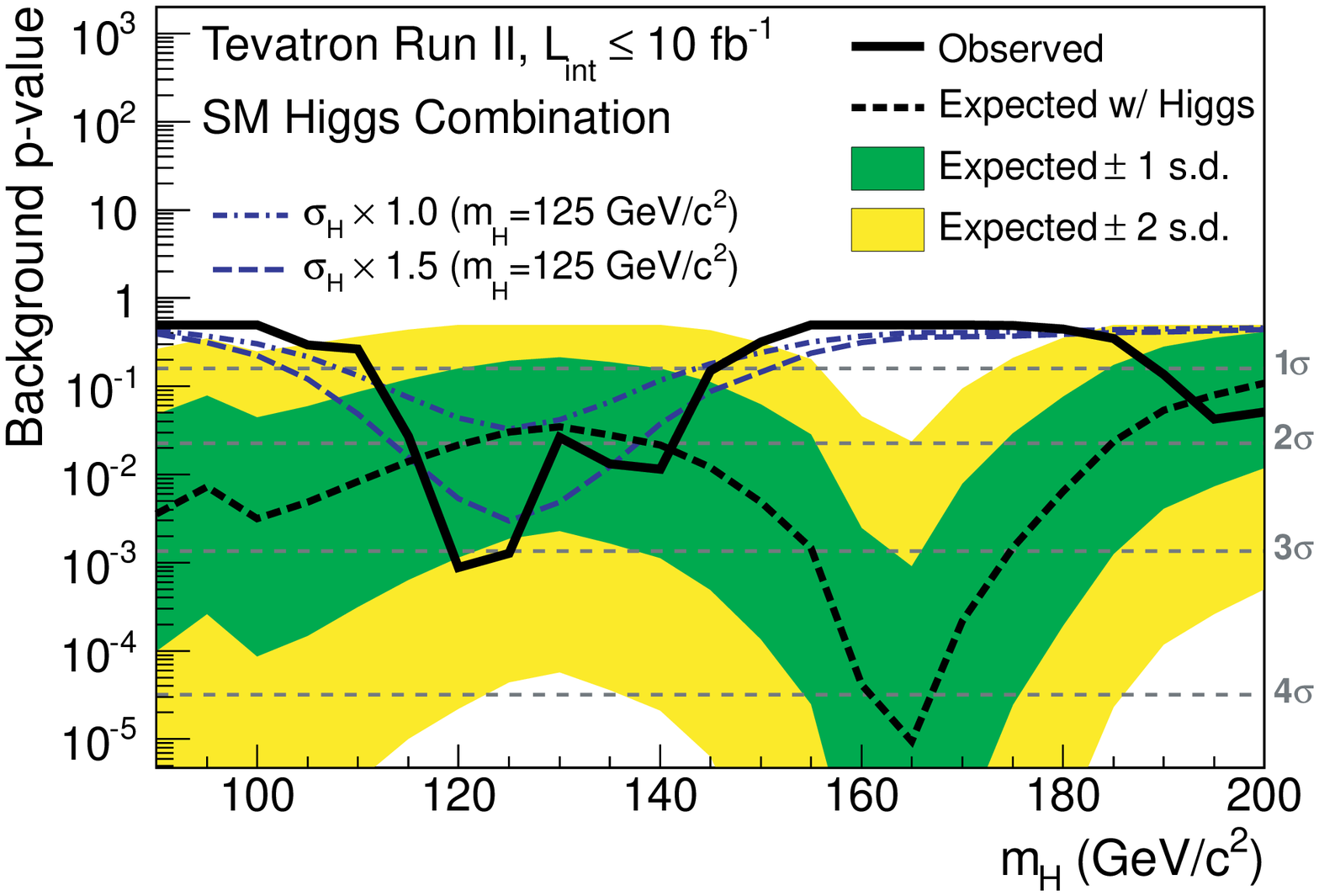} \\
\caption{
\label{fig:pvalue}
(color online). The solid black line shows the background $p$-value as a 
function of $m_H$ for all of CDF and D0's SM Higgs boson searches in all decay modes combined.  The dotted 
black line shows the median expected values assuming a SM signal is present, 
evaluated separately at each $m_H$.  The associated dark- and light-shaded 
bands indicate the one and two s.d.~fluctuations of possible experimental 
outcomes under this scenario.  The blue lines show the median expected 
$p$-values assuming the SM Higgs boson is present with $m_H$=125~GeV/$c^2$ 
at signal strengths of 1.0 times (short-dashed) and 1.5 times (long-dashed) 
the SM prediction.}
\end{centering}
\end{figure}

The upper limit on SM Higgs boson production as a function of $m_H$ is extracted in the range
90--200~GeV/$c^2$ in terms of $R_{95}^{\mathrm{obs}}$, the ratio of the observed limit to the 
predicted SM rate. 
The ratios of the 95\% C.L. expected and observed limit to the SM cross
section using the Bayesian method are shown in Fig.~\ref{fig:comboRatio} 
for the combined CDF and D0 analyses.  The observed and median-expected 
ratios are listed for the tested Higgs boson masses in Table~\ref{tab:ratios}, 
as obtained by the Bayesian and the ${\rm CL}_{\rm s}$ methods.

\begin{table}[htb]
\caption{\label{tab:ratios} Ratios of observed and median expected (for 
the background-only hypothesis) 95\% C.L. upper production limits to the 
SM cross section as a function of the Higgs boson mass for the combined 
CDF and D0 searches in all decay modes, obtained using the Bayesian and 
${\rm CL}_{\rm s}$ methods.}
\begin{ruledtabular}
\begin{tabular}{lcccc}
 & \multicolumn{2}{c}{Bayesian} & \multicolumn{2}{c}{${\rm CL}_{\rm s}$} \\ 
$m_H$ (GeV/$c^2$) & $R_{95}^{\mathrm{obs}}$ & $R_{95}^{\mathrm{exp}}$ & $R_{95}^{\mathrm{obs}}$ & $R_{95}^{\mathrm{exp}}$ \\ \hline
 90 &  0.37     &  0.74   &  0.39  &   0.74      \\
 95 &  0.48     &  0.80   &  0.49  &   0.81      \\
100 &  0.62     &  0.72   &  0.62  &   0.73      \\
105 &  0.89     &  0.77   &  0.93  &   0.77      \\
110 &  1.02     &  0.82   &  1.03  &   0.83      \\
115 &  1.63     &  0.90   &  1.67  &   0.91      \\
120 &  2.33     &  1.00   &  2.40  &   0.99      \\
125 &  2.44     &  1.06   &  2.62  &   1.07      \\
130 &  2.13     &  1.11   &  2.10  &   1.10      \\
135 &  2.03     &  1.04   &  2.12  &   1.06      \\
140 &  2.10     &  1.01   &  2.08  &   1.00      \\
145 &  1.35     &  0.88   &  1.29  &   0.90      \\
150 &  0.94     &  0.79   &  0.91  &   0.78      \\
155 &  0.64     &  0.69   &  0.62  &   0.68      \\
160 &  0.46     &  0.51   &  0.45  &   0.51      \\
165 &  0.37     &  0.47   &  0.36  &   0.47      \\
170 &  0.54     &  0.57   &  0.53  &   0.57      \\
175 &  0.71     &  0.68   &  0.68  &   0.68      \\
180 &  0.87     &  0.81   &  0.86  &   0.82      \\
185 &  1.20     &  1.02   &  1.18  &   1.04      \\
190 &  1.86     &  1.29   &  1.86  &   1.27      \\
195 &  2.74     &  1.44   &  2.64  &   1.48      \\
200 &  3.07     &  1.66   &  2.97  &   1.67      \\
\end{tabular}
\end{ruledtabular}
\end{table}

Intersections of piecewise linear interpolations of the observed and 
expected rate limits with the SM=1 line are used to quote ranges of Higgs 
boson masses that are excluded and that are expected to be excluded.
The regions of Higgs boson masses excluded at the 95\% C.L. are 
\SMLobslow\ $<m_H<$ \SMLobshigh~GeV/$c^{2}$ and \SMHobslow\ $<m_{H}<$ \SMHobshigh~GeV/$c^{2}$.  
The expected exclusion regions are
\SMLexplow\ $<m_H<$ \SMLexphigh~GeV/$c^{2}$ and \SMHexplow\ $<m_{H}<$ \SMHexphigh~GeV/$c^{2}$.

The observed excess for $m_{H}$ from 115 to 140~GeV/$c^2$ is driven by an excess of 
data events with respect to the background predictions in the most sensitive bins
of the discriminant distributions, favoring the hypothesis that a signal is present. 
To characterize the compatibility of this excess with the signal-plus-background 
hypothesis, the best-fit rate cross section, $R^{\rm{fit}}$, is computed using the Bayesian calculation, and shown 
in Fig.~\ref{fig:combBestFit}.  The measured signal strength is within 1~s.d.~of the 
expectation for a SM Higgs boson in the range $115<m_H<140$~GeV$/c^2$, with maximal 
strength between 120~GeV$/c^2$ and 125~GeV$/c^2$. At 125~GeV$/c^2$, 
$R^{\rm{fit}} = 1.44^{+0.49}_{-0.47}~({\rm{stat}})^{+0.33}_{-0.31}~({\rm{syst}}) \pm 0.10~({\rm{theory}})$.

The significance of the excess in the data over the background prediction is
computed at each hypothesized Higgs boson mass by calculating the local $p$-value
under the background-only hypothesis using $R^{\rm{fit}}_{\rm{profile}}$, chosen {\it a priori}, as the test statistic.
This $p$-value expresses the probability to obtain the value of $R^{\rm{fit}}_{\rm{profile}}$ 
observed in the data or larger, assuming a signal is absent.  
These $p$-values are shown in Fig.~\ref{fig:pvalue} along with the expected $p$-values assuming 
a SM signal is present, separately for each value of $m_H$. The median expected 
$p$-values assuming the SM Higgs boson is present with $m_H$=125~GeV/$c^2$ for 
signal strengths of 1.0 and 1.5 times the SM prediction are also shown.  The median
expected excess at $m_H=125$ GeV/$c^2$ corresponds to 1.9 standard deviations assuming
the SM Higgs boson is present at that mass.   The observed local significance at
$m_H=125$ GeV/$c^2$ corresponds to 3.0 standard deviations.
The maximum observed local significance is at $m_H=120$~GeV/$c^2$ and corresponds to 
3.1 standard deviations.  The fluctuations seen in the observed $p$-value as a 
function of the tested $m_H$ result from excesses seen in different search 
channels, as well as from point-to-point fluctuations due to the separate 
discriminants at each $m_H$, and are discussed in more detail below.  The width 
of the dip in the observed $p$-values from 115 to 140 GeV/$c^{2}$ is consistent 
with the resolution of the combination of the $H \to b\bar{b}$ and $H \to W^+W^-$ 
channels, as illustrated by the injected signal curves in Fig.~\ref{fig:pvalue}. 
The effective resolution of this search comes from two independent sources of 
information.  The reconstructed candidate masses help constrain $m_H$, but more 
importantly, the expected cross sections times the relevant branching ratios for 
the $H \to b\bar{b}$ and $H \to W^+W^-$ channels are strong functions of $m_H$ 
in the SM.  The observed excess in
the $H \to b\bar{b}$ channels coupled with the slight excess in the $H \to W^+W^-$ 
channels determine the shape of the observed $p$-value as a function of $m_H$.

\begin{figure}[htb]
\begin{centering}
\includegraphics[width=1.0\columnwidth]{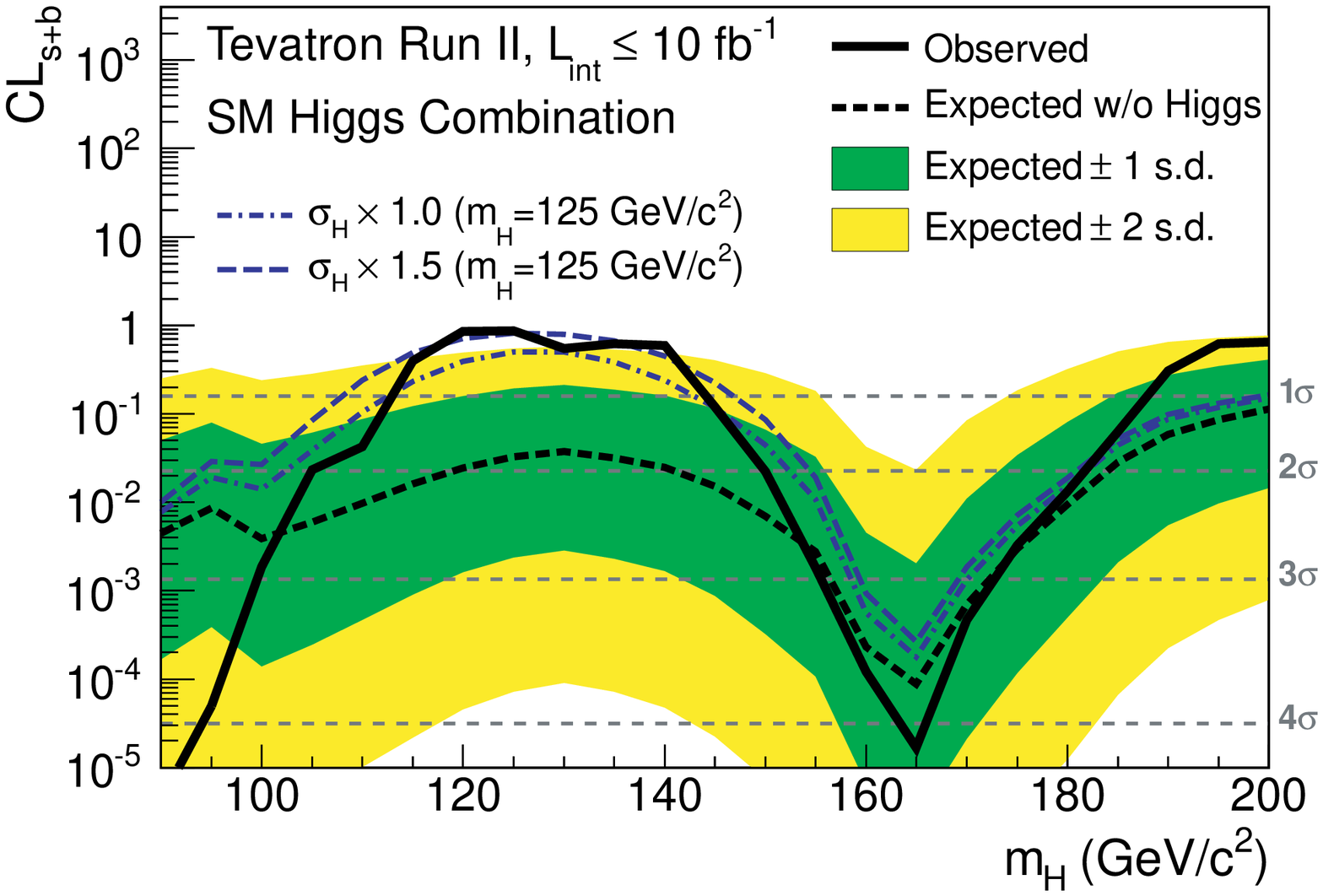} \\
\caption{
\label{fig:clsb}
(color online). The solid black line shows the signal-plus-background $p$-value 
as a function of $m_H$ for all of CDF and D0's SM Higgs boson searches in all decay modes combined.  The dotted 
black line shows the median expected values assuming no SM signal is present, 
evaluated separately at each $m_H$.  The associated dark and light-shaded bands 
indicate the one and two s.d.~fluctuations of possible experimental outcomes 
under this scenario.  The blue lines show the median expected $p$-values assuming 
the SM Higgs boson is present with $m_H$=125~GeV/$c^2$ at signal strengths of 
1.0 times (short-dashed) and 1.5 times (long-dashed) the SM prediction.}
\end{centering}
\end{figure}

Figure~\ref{fig:clsb} shows the quantity ${\rm CL}_{\rm s+b}$, corresponding to 
the $p$-value for the signal-plus-background hypothesis.
The observed value, along with the expected $p$-values assuming 
a signal is absent, are shown separately for each value of $m_H$. The median expected 
$p$-values assuming the SM Higgs boson is present with $m_H$=125~GeV/$c^2$ for 
signal strengths of 1.0 and 1.5 times the SM prediction are also shown.
In the mass region from 115 to 140 GeV/$c^{2}$ the observed values above 50\%
indicate a high level of consistency with the signal-plus-background hypothesis.

We also separate CDF and D0's searches into combinations focusing on the 
$H\to b\bar{b}$, $H \to W^+W^-$, $\ H \to \gamma \gamma$, and $\ H \to \tau^+ 
\tau^-$ decay modes, and these are discussed in the following sections.

\subsection{{\boldmath{$H\rightarrow b{\bar{b}}$}} Decay Mode}\label{sec:hbb}

Below 130~GeV/$c^2$, the $H\rightarrow b{\bar{b}}$ searches
contribute the majority of the search sensitivity.  The $WH\rightarrow \ell\nu b\bar{b}$, 
$ZH\rightarrow\nu\bar{\nu} b\bar{b}$, and $ZH\rightarrow \ell^+\ell^- b\bar{b}$ 
channels from both experiments are included in this sub-combination.  Two of the 
six contributing channels were updated for this sub-combination compared with 
that reported in Ref.~\cite{tevhbbprl}.  The CDF $ZH\rightarrow\nu\bar{\nu} 
b\bar{b}$~\cite{cdfmetbb-jul12} analysis was updated to use a more powerful MVA 
b-tagging algorithm~\cite{tagging} along with changes to the kinematic selections. 
The assignment of correlated systematic uncertainties between channels was updated 
in the D0 $WH\rightarrow \ell\nu b\bar{b}$ analysis~\cite{dzWHl-jul12}.  The 
observed LLR distribution is shown in Fig.~\ref{fig:hbbllr}, along with its 
expected values under the background-only and signal-plus-background hypotheses. 
The hypotheses that a SM Higgs boson is present with $m_H=125$~GeV/$c^2$ for signal 
strengths of 1.0 and 1.5 times the SM prediction are also given. The LLR values as
a function of Higgs boson mass are listed in Table~\ref{tab:hbbllrVals}.
\begin{figure}[htb]
\begin{centering}
\includegraphics[width=0.85\columnwidth]{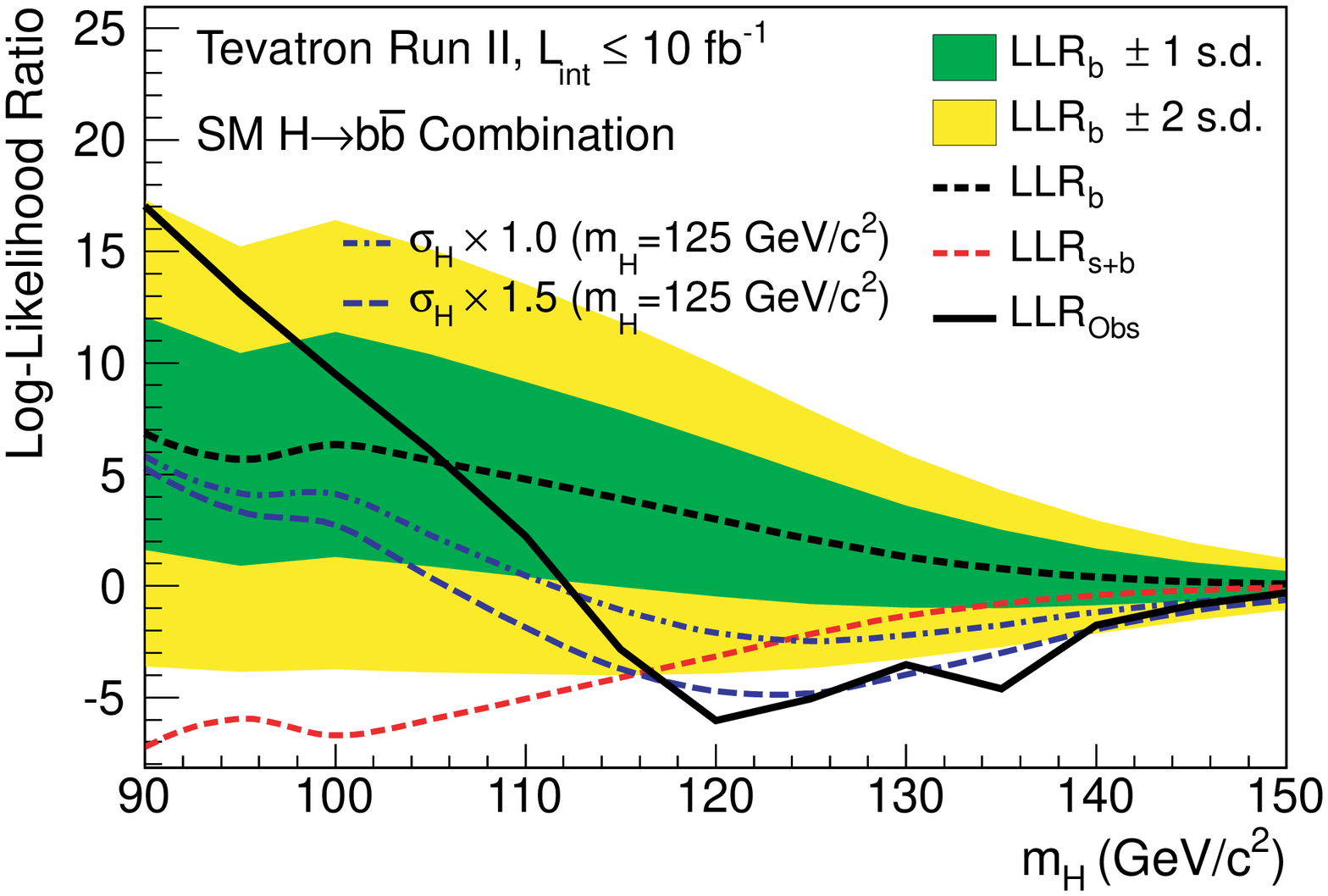}
\caption{
\label{fig:hbbllr}
(color online). The log-likelihood ratio LLR as a function of Higgs boson mass 
from the combination of CDF and D0's SM Higgs boson searches focusing on the $H\rightarrow b{\bar{b}}$ decay
mode.  The solid line shows the observed LLR values, the dark short-dashed line 
shows the median expectation assuming no Higgs boson signal is present, and the 
dark- and light-shaded bands correspond, respectively, to the regions encompassing 
one and two s.d.~fluctuations around the background-only expectation.  The red 
long-dashed line shows the median expectation assuming a SM Higgs boson signal 
is present at each value of $m_H$ in turn.  The blue lines show the median 
expected LLR assuming the SM Higgs boson is present at $m_H=125$~GeV/$c^2$ with 
signal strengths of 1.0 times (short-dashed) and 1.5 times (long-dashed) the SM
prediction.}
\end{centering}
\end{figure}
\begin{figure}[htb] \begin{centering}
\includegraphics[width=0.9\columnwidth]{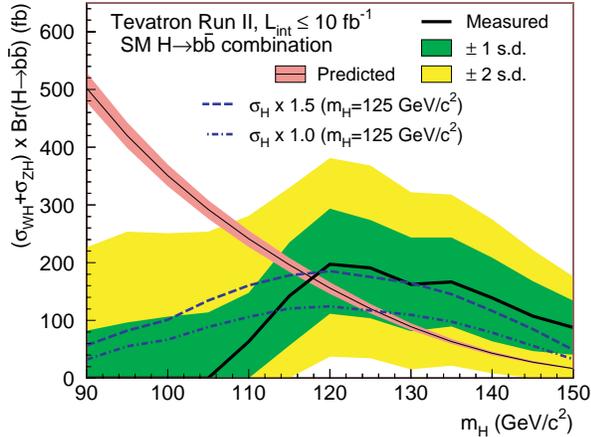}
\caption{
\label{fig:xsfit} (color online). The best-fit signal cross section times 
branching ratio $(\sigma_{WH}+\sigma_{ZH})\times \mathcal{B}(H\rightarrow
b{\bar{b}})$ as a function of Higgs boson mass from the combination of 
CDF and D0's SM Higgs boson searches focusing on the $H\rightarrow 
b{\bar{b}}$ decay mode.  The dark- and light-shaded bands show the one and 
two s.d.~uncertainty ranges on the fitted signal, respectively.  Also 
shown with blue lines are the median fitted cross sections expected for a 
SM Higgs boson with $m_H=125$~GeV/$c^2$ at signal strengths of 1.0 times 
(short-dashed) and 1.5 times (long-dashed) the SM prediction.  The SM 
prediction is shown as the smooth, falling curve where the narrow band 
indicates the theoretical uncertainty.}
\end{centering} 
\end{figure}

\begin{table*}[htb]
\caption{\label{tab:hbbllrVals} Log-likelihood ratio (LLR) values obtained 
from the combination of CDF and D0's Higgs boson search channels focusing on 
the $H\rightarrow b{\bar{b}}$ decay mode using the ${\rm CL}_{\rm s}$ method.}
\begin{ruledtabular}\begin{tabular}{lccccccc} \\
$m_{H}$ (GeV/$c^2$) &  LLR$_{\rm{obs}}$ & LLR$_{s+b}$ &  LLR$_{b}^{-2\sigma}$ & LLR$_{b}^{-1\sigma}$ & LLR$_{b}$ &  LLR$_{b}^{+1\sigma}$ & LLR$_{b}^{+2\sigma}$ \\
\hline
90 & 17.02 & $-7.24$ & 17.31 & 12.08 & 6.84 & 1.61 & $-3.62$ \\ 
95 & 13.07 & $-5.96$ & 15.21 & 10.44 & 5.68 & 0.91 & $-3.85$ \\ 
100 & 9.50 & $-6.71$ & 16.41 & 11.38 & 6.34 & 1.30 & $-3.73$ \\ 
105 & 6.09 & $-6.00$ & 15.12 & 10.38 & 5.63 & 0.88 & $-3.86$ \\ 
110 & 2.21 & $-5.06$ & 13.52 & 9.15 & 4.78 & 0.41 & $-3.97$ \\ 
115 & $-2.84$ & $-4.12$ & 11.83 & 7.87 & 3.91 & $-0.04$ & $-4.00$ \\ 
120 & $-6.05$ & $-3.16$ & 9.91 & 6.45 & 2.99 & $-0.47$ & $-3.93$ \\ 
125 & $-5.05$ & $-2.16$ & 7.88 & 4.99 & 2.09 & $-0.80$ & $-3.69$ \\ 
130 & $-3.53$ & $-1.34$ & 5.89 & 3.60 & 1.31 & $-0.98$ & $-3.27$ \\ 
135 & $-4.60$ & $-0.80$ & 4.28 & 2.52 & 0.77 & $-0.98$ & $-2.74$ \\ 
140 & $-1.77$ & $-0.41$ & 2.93 & 1.67 & 0.40 & $-0.87$ & $-2.13$ \\ 
145 & $-0.86$ & $-0.20$ & 1.95 & 1.07 & 0.19 & $-0.68$ & $-1.56$ \\ 
150 & $-0.31$ & $-0.08$ & 1.23 & 0.66 & 0.08 & $-0.49$ & $-1.07$ \\ 
\end{tabular}
\end{ruledtabular}
\end{table*}

\begin{figure}[htb]
\begin{centering}
\includegraphics[width=1.0\columnwidth]{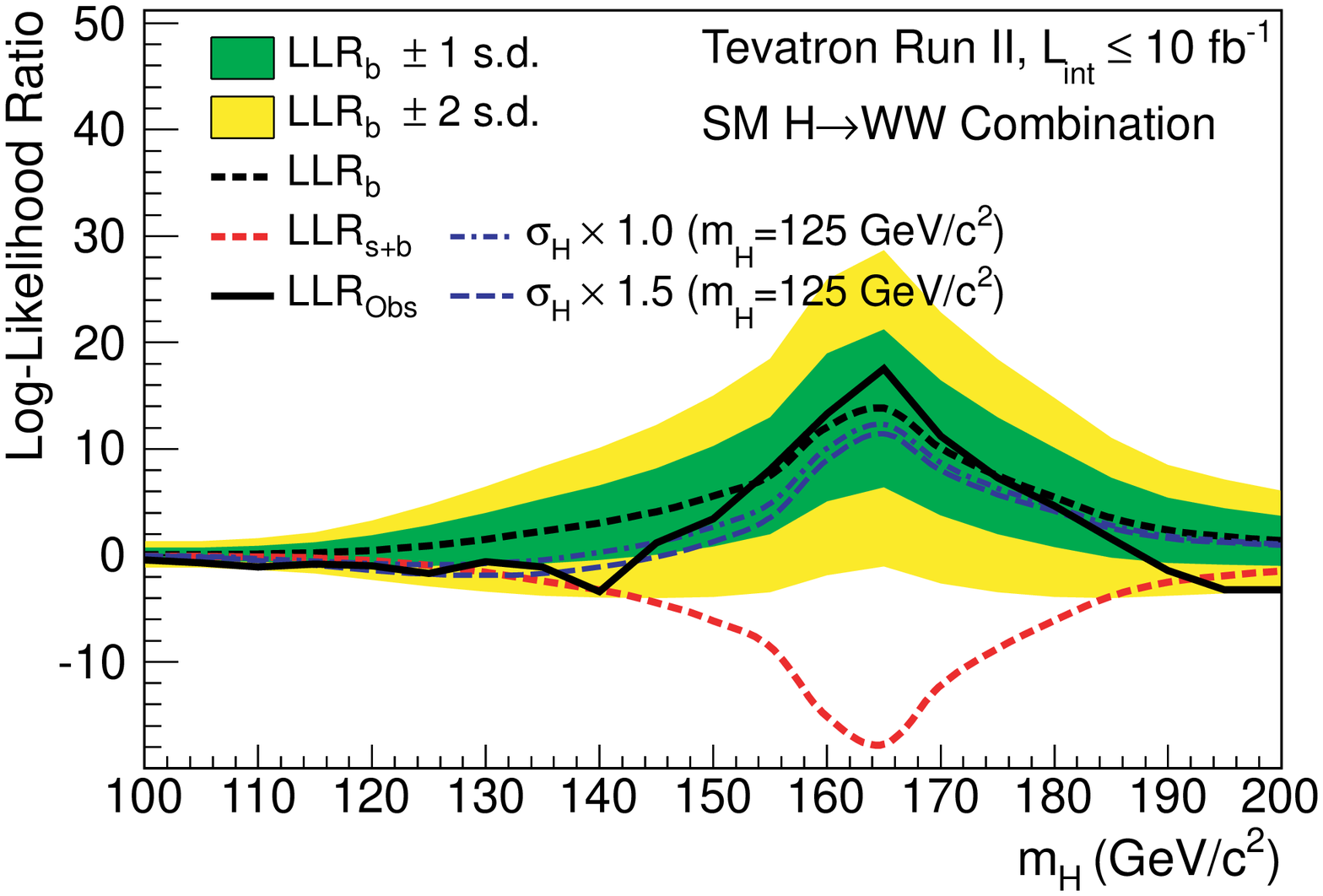}
\caption{
\label{fig:hwwLLR}
(color online). The log-likelihood ratio LLR as a function of Higgs boson mass 
from the combination of CDF and D0's SM Higgs boson searches focusing on the $H\rightarrow W^+W^-$ decay
mode.  The solid line shows the observed LLR values, the dark short-dashed line 
shows the median expectation assuming no Higgs boson signal is present, and the 
dark- and light-shaded bands correspond, respectively, to the regions encompassing 
one and two s.d.~fluctuations around the background-only expectation.  The red 
long-dashed line shows the median expectation assuming a SM Higgs boson signal 
is present at each value of $m_H$ in turn.  The blue lines show the median 
expected LLR assuming the SM Higgs boson is present at $m_H=125$~GeV/$c^2$ with 
signal strengths of 1.0 times (short-dashed) and 1.5 times (long-dashed) the SM
prediction.}
\end{centering}
\end{figure}

We multiply the best-fit rate cross section, $R^{\rm{fit}}$, for this sub-combination 
by the SM prediction for the associated-production cross section times the decay 
branching ratio ($\sigma_{WH}+\sigma_{ZH})\times \mathcal{B}(H\rightarrow b{\bar{b}})$, 
to obtain the observed value for this quantity.  We show the fitted 
($\sigma_{WH}+\sigma_{ZH})\times \mathcal{B}(H\rightarrow b{\bar{b}})$ as a function 
of $m_H$, along with the SM prediction, in Fig.~\ref{fig:xsfit}.  The figure also 
shows the expected cross section fits for each $m_H$, assuming that the SM Higgs boson 
with $m_H=125$~GeV/$c^2$ is present, both at the rate predicted by the SM, and also at
a multiple of 1.5 times that of the SM.   The best-fit rate corresponds to
$(\sigma_{WH}+\sigma_{ZH})\times
\mathcal{B}(H\rightarrow b{\bar{b}})= 0.19^{+0.08}_{-0.09}~(\mathrm{stat+syst})$~pb.
The shift in this result compared with the value of $0.23 \pm 0.09~(\mathrm{stat+syst})$~pb
obtained previously~\cite{tevhbbprl} is due to the updated $ZH\rightarrow \nu\bar{\nu} b\bar{b}$
analysis from CDF~\cite{cdfmetbb,cdfmetbb-jul12}, and corresponds to a change in the central value of 0.5~times
the total uncertainty.  For $m_H=125$~GeV/$c^2$, the SM predicts
($\sigma_{WH}+\sigma_{ZH})\times \mathcal{B}(H\rightarrow b{\bar{b}})=0.12  \pm$~0.01~pb.

\subsection{ {\boldmath{$H\rightarrow W^+W^-$}} Decay Mode}\label{sec:hww}

\begin{figure}[htb]
\begin{centering}
\includegraphics[width=0.9\columnwidth]{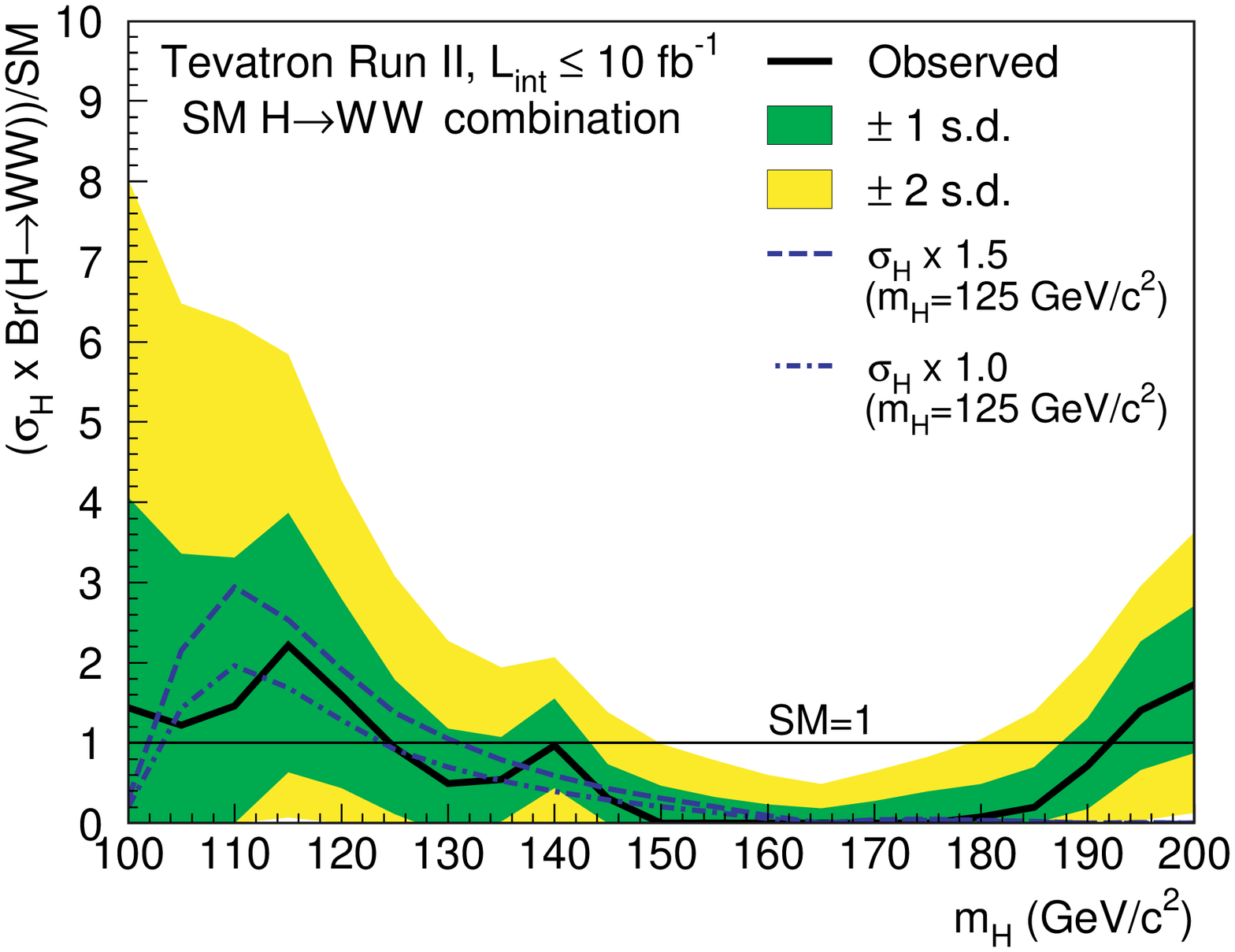}
\caption{
\label{fig:wwxs} (color online). The best-fit signal cross section expressed 
as a ratio to the SM cross section as a function of Higgs boson mass from 
the combination of CDF and D0's SM Higgs boson searches focusing on the 
$H\rightarrow W^+W^-$ decay mode.  The dark- and light-shaded bands show 
the one and two s.d.~uncertainty ranges on the fitted signal, respectively.  
Also shown with blue lines are the median fitted cross sections expected 
for a SM Higgs boson with $m_H=125$~GeV/$c^2$ at signal strengths of 1.0 
times (short-dashed) and 1.5 times (long-dashed) the SM prediction.}  
\end{centering}
\end{figure}

Above 130~GeV/$c^2$, the $H\rightarrow W^+W^-$ channels contribute the majority of the search 
sensitivity.  We combine all $H \to W^+W^-$ searches from CDF and D0, incorporating potential 
signal contributions from gluon-gluon fusion, $WH$, $ZH$, and vector boson fusion production. 
Approximately 75\% of the signal comes from the gluon-gluon fusion process, 20\% from associated 
production, and 5\% from the VBF process. The LLR distributions are shown in Fig.~\ref{fig:hwwLLR} 
and the values as a function of Higgs boson mass are listed in Table~\ref{tab:hwwllrVals}. 
The data present a one to two s.d.~excess in the region from 115 to 140 GeV/$c^2$ where there 
is some separation between the two hypotheses.  An excess is also seen in the searches for 
Higgs bosons with mass $m_H>195$~GeV/$c^2$, as mentioned in Section~\ref{sec:fullcomb}, but 
the sensitivity to the SM Higgs boson is not as large at these masses as it is at lower masses.  
Figure~\ref{fig:wwxs} shows the best-fit cross 
section for the combined $H\rightarrow W^+W^-$ searches, normalized to the SM prediction, 
as a function of $m_H$, along with the expectations assuming the SM Higgs boson is present 
at $m_H=125$~GeV/$c^2$ for signal strengths of 1.0 and 1.5 times the SM prediction.

\begin{table*}[htb]
\caption{\label{tab:hwwllrVals} Log-likelihood ratio (LLR) values obtained 
from the combination of CDF and D0's Higgs boson search channels focusing on 
the $H\rightarrow W^+W^-$ decay mode using the ${\rm CL}_{\rm s}$ method.}
\begin{ruledtabular}
\begin{tabular}{lccccccc} \\
$m_{H}$ (GeV/$c^2$) &  LLR$_{\rm{obs}}$ & LLR$_{s+b}$ &  LLR$_{b}^{-2\sigma}$ & LLR$_{b}^{-1\sigma}$ & LLR$_{b}$ &  LLR$_{b}^{+1\sigma}$ & LLR$_{b}^{+2\sigma}$ \\
\hline
100 & $-0.42$ & $-0.10$ & 1.36 & 0.73 & 0.10 & $-0.53$ & $-1.16$ \\ 
105 & $-0.72$ & $-0.10$ & 1.34 & 0.72 & 0.10 & $-0.53$ & $-1.15$ \\ 
110 & $-1.07$ & $-0.14$ & 1.62 & 0.88 & 0.14 & $-0.60$ & $-1.34$ \\ 
115 & $-0.80$ & $-0.24$ & 2.18 & 1.21 & 0.24 & $-0.74$ & $-1.71$ \\ 
120 & $-0.98$ & $-0.49$ & 3.26 & 1.87 & 0.48 & $-0.91$ & $-2.30$ \\ 
125 & $-1.69$ & $-0.94$ & 4.73 & 2.82 & 0.91 & $-1.00$ & $-2.91$ \\ 
130 & $-0.59$ & $-1.57$ & 6.44 & 3.98 & 1.52 & $-0.95$ & $-3.41$ \\ 
135 & $-1.11$ & $-2.40$ & 8.36 & 5.33 & 2.30 & $-0.73$ & $-3.77$ \\ 
140 & $-3.38$ & $-3.24$ & 10.09 & 6.58 & 3.08 & $-0.43$ & $-3.94$ \\ 
145 & 1.19 & $-4.42$ & 12.23 & 8.17 & 4.12 & 0.06 & $-4.00$ \\ 
150 & 3.43 & $-6.13$ & 15.01 & 10.29 & 5.57 & 0.85 & $-3.87$ \\ 
155 & 8.05 & $-8.59$ & 18.45 & 12.97 & 7.50 & 2.02 & $-3.46$ \\ 
160 & 13.27 & $-15.15$ & 25.92 & 18.98 & 12.04 & 5.10 & $-1.84$ \\ 
165 & 17.55 & $-17.75$ & 28.69 & 21.25 & 13.82 & 6.38 & $-1.05$ \\ 
170 & 11.19 & $-12.21$ & 22.80 & 16.45 & 10.09 & 3.74 & $-2.61$ \\ 
175 & 7.28 & $-8.72$ & 18.44 & 12.96 & 7.49 & 2.02 & $-3.46$ \\ 
180 & 4.63 & $-6.12$ & 14.80 & 10.13 & 5.46 & 0.78 & $-3.89$ \\ 
185 & 1.56 & $-3.83$ & 11.05 & 7.29 & 3.53 & $-0.23$ & $-3.99$ \\ 
190 & $-1.39$ & $-2.50$ & 8.51 & 5.44 & 2.36 & $-0.71$ & $-3.79$ \\ 
195 & $-3.24$ & $-1.88$ & 7.12 & 4.45 & 1.78 & $-0.89$ & $-3.56$ \\ 
200 & $-3.23$ & $-1.45$ & 6.08 & 3.73 & 1.38 & $-0.97$ & $-3.32$ \\ 
\end{tabular}
\end{ruledtabular}
\end{table*}

\begin{figure}[htb]
\begin{centering}
\includegraphics[width=1.0\columnwidth]{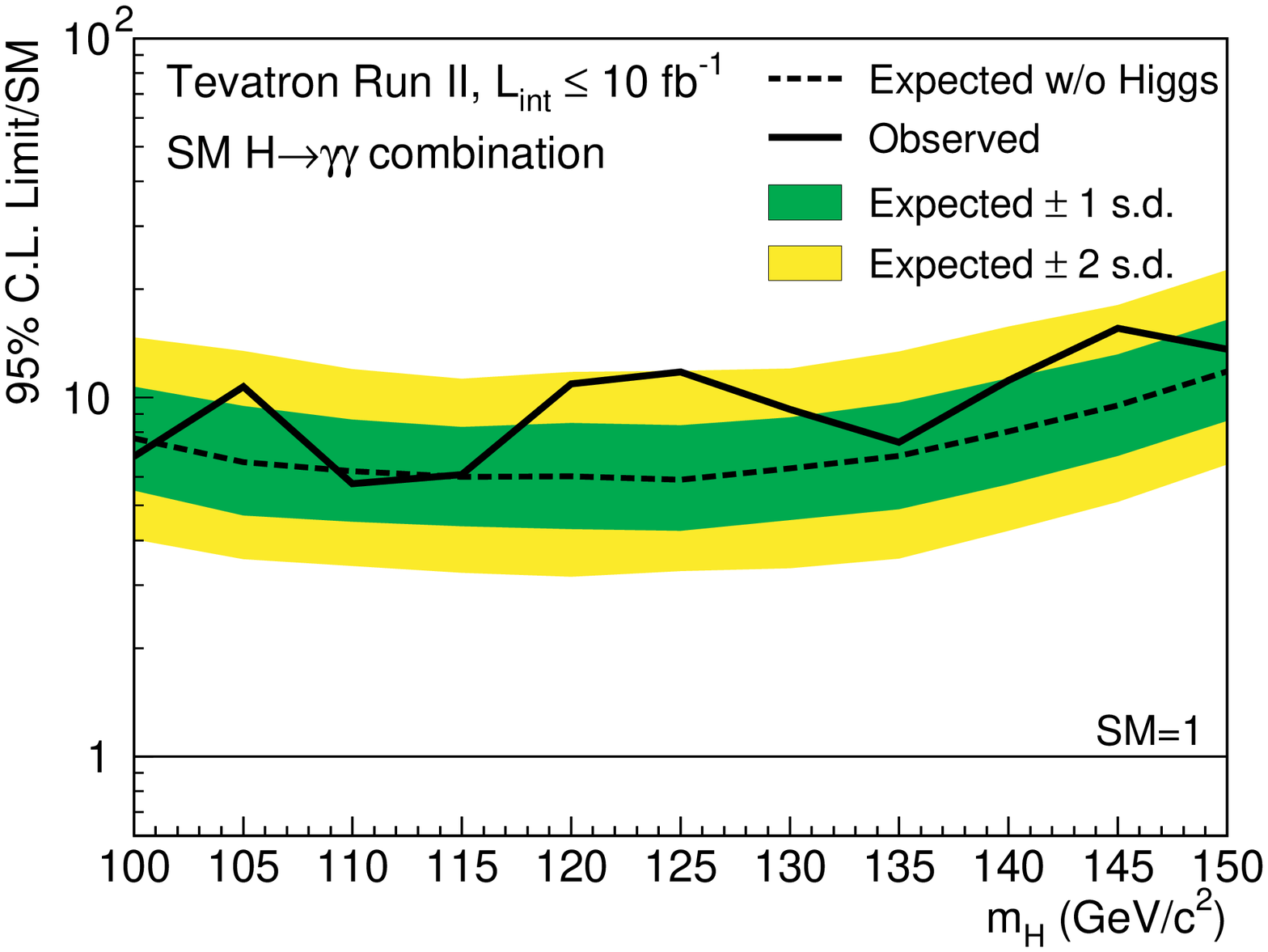}
\caption{\label{fig:comboRatiogamgam}
(color online). Observed and median expected (for the background-only 
hypothesis) 95\% C.L. Bayesian upper production limits expressed as 
multiples of the SM cross section as a function of Higgs boson mass 
from the combination of CDF and D0's SM Higgs boson searches focusing on the $H\rightarrow \gamma
\gamma$ decay mode.
The dark- and light-shaded bands indicate, respectively, the one and 
two s.d.~probability regions in which the limits are expected to fluctuate in the 
absence of signal.}
\end{centering}
\end{figure}

\subsection{ {\boldmath{$H\rightarrow\gamma\gamma$}} Decay Mode}\label{sec:hgg}

We also separately combine CDF and D0's searches focusing on the $H\rightarrow\gamma\gamma$ 
decay mode and display the resulting upper limits on the production cross section times 
the decay branching ratio normalized to the SM prediction in Fig.~\ref{fig:comboRatiogamgam}.  
An excess of approximately two s.d.~is seen in these searches at $m_H=125$~GeV/$c^2$, but 
its contributions to the fully combined SM cross section and limit are small due to the low 
expected signal yield in this channel.  However, the observed excess in the $H\rightarrow
\gamma\gamma$ search channel has a visible impact on Higgs boson coupling constraints as 
described in Section~\ref{sec:charact}.

\subsection{{\boldmath{$H\rightarrow\tau\tau$}} Decay Mode}\label{sec:htt}

We also separately combine CDF and D0's searches focusing on the  $H\rightarrow
\tau^+\tau^-$ decay mode and display the resulting upper limits on the production 
cross section times the decay branching ratio normalized to the SM prediction in 
Fig.~\ref{fig:comboRatiotautau}.

\begin{figure}[htb]
\begin{centering}
\includegraphics[width=1.0\columnwidth]{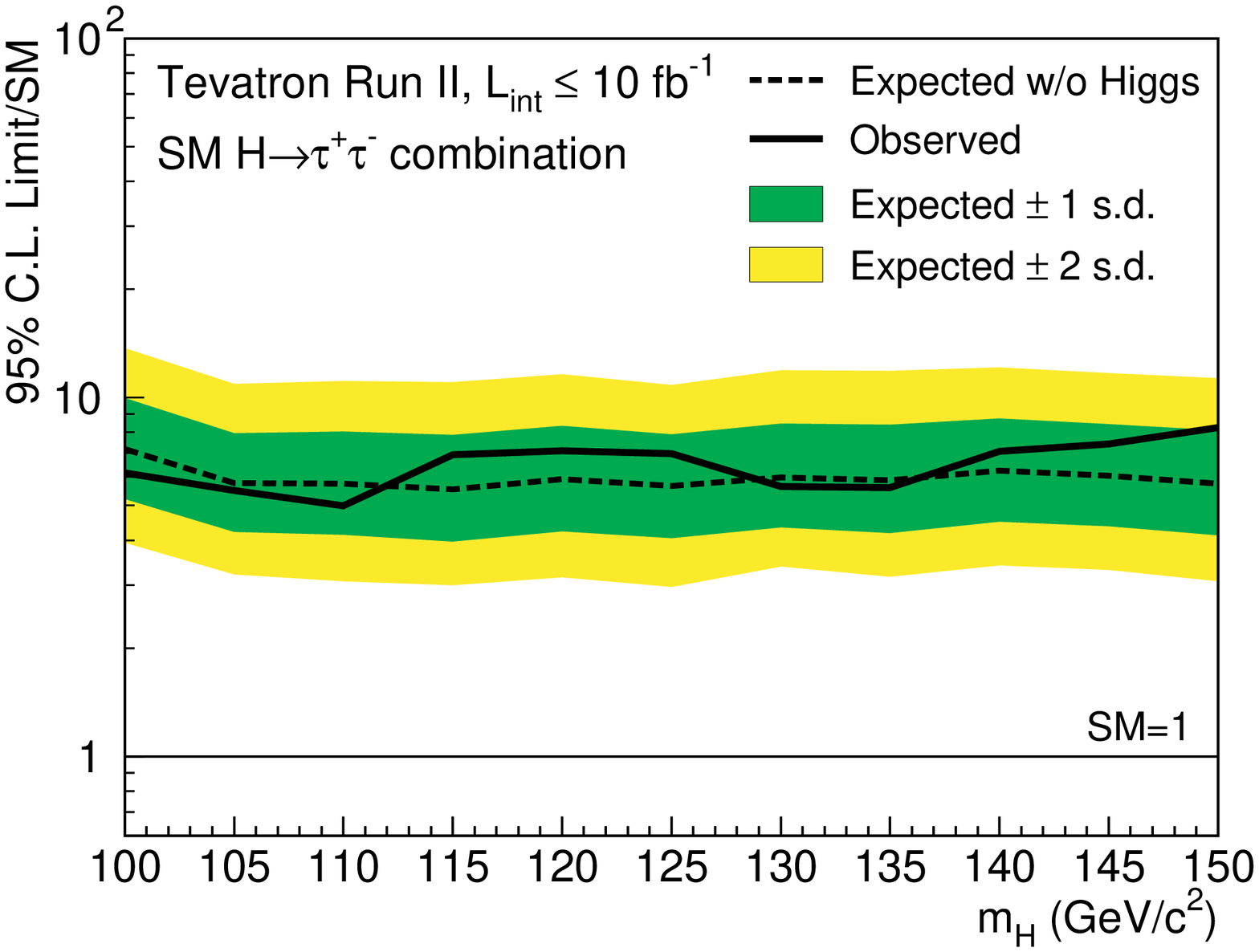}
\caption{
\label{fig:comboRatiotautau}
(color online). Observed and median expected (for the background-only 
hypothesis) 95\% C.L. Bayesian upper production limits expressed as 
multiples of the SM cross section as a function of Higgs boson mass 
from the combination of CDF and D0's SM Higgs boson searches focusing on the $H\rightarrow \tau^+\tau^-$ decay mode.
The dark- and light-shaded bands indicate, respectively, the one and 
two s.d.~probability regions in which the limits are expected to fluctuate in the 
absence of signal.}
\end{centering}
\end{figure}

\subsection{Compatibility of the Excess with the SM Higgs Boson Hypothesis}\label{sec:charact}

The best-fit rate parameters, $R^{\rm{fit}}$, for the full combination 
of all channels and the combinations of channels focusing on the 
$H\rightarrow W^+W^-$, $H\rightarrow b{\bar{b}}$, $H\rightarrow\gamma
\gamma$, and $H\rightarrow\tau^+\tau^-$ decay modes~\cite{mixture} are 
listed in Table~\ref{tab:xsmeas} as a function of Higgs boson mass over 
the range 115~$< m_H\ <$~140~GeV/$c^2$, where the combined result has 
sensitivity to a signal and a clear excess exists. For $m_H$ = 125~GeV/$c^2$, 
we obtain $R^{\rm{fit}} = 1.44^{+0.59}_{-0.56}$ using all decay modes.

\begin{table*}[htb]
\caption{\label{tab:xsmeas} Best-fit values of $R=(\sigma\times{\cal B})$/SM using 
the Bayesian method for all SM Higgs boson decay modes combined and the combinations 
of CDF and D0's Higgs boson search channels focusing on the $H\rightarrow W^+W^-$, 
$H\rightarrow b{\bar{b}}$, $H\rightarrow\gamma\gamma$, and $H\rightarrow\tau^+\tau^-$ 
decay modes as a function of Higgs boson mass over the range 115~$< m_H < 140$~GeV/$c^2$.  
The quoted uncertainties bound the smallest interval containing 68\% of the integral 
of the posterior probability density.}
\begin{ruledtabular}
\begin{tabular}{lcccccccc}\\
$m_H$ (GeV/$c^2$) &  115 &  120 &  125 & 130 & 135 & 140 \\ \hline
\rule[-2mm]{0mm}{7mm}$R_{\rm{fit}}$(SM)                          & $0.82^{+0.43}_{-0.46}$ & $1.42^{+0.53}_{-0.52}$ & $1.44^{+0.59}_{-0.56}$ & $1.13^{+0.60}_{-0.60}$ & $0.99^{+0.58}_{-0.57}$ & $1.15^{+0.57}_{-0.52}$ \\
\rule[-2mm]{0mm}{7mm}$R_{\rm{fit}}$($H\rightarrow W^+W^-$)       & $2.22^{+1.65}_{-1.59}$ & $1.59^{+1.20}_{-1.15}$ & $0.94^{+0.85}_{-0.83}$ & $0.49^{+0.69}_{-0.63}$ & $0.54^{+0.53}_{-0.52}$ & $0.97^{+0.58}_{-0.53}$ \\
\rule[-2mm]{0mm}{7mm}$R_{\rm{fit}}$($H\rightarrow b{\bar{b}}$)   & $0.72^{+0.47}_{-0.44}$ & $1.26^{+0.62}_{-0.55}$ & $1.59^{+0.69}_{-0.72}$ & $1.82^{+0.91}_{-0.91}$ & $2.62^{+1.22}_{-1.21}$ & $3.23^{+1.61}_{-1.74}$ \\
\rule[-2mm]{0mm}{7mm}$R_{\rm{fit}}$($H\rightarrow \gamma\gamma$) & $0.65^{+2.66}_{-0.54}$ & $5.34^{+3.20}_{-2.76}$ & $5.97^{+3.39}_{-3.12}$ & $3.17^{+2.69}_{-2.81}$ & $0.00^{+4.04}_{-0.00}$ & $3.31^{+3.30}_{-3.13}$ \\
\rule[-2mm]{0mm}{7mm}$R_{\rm{fit}}$($H\rightarrow \tau^+\tau^-$) & $1.70^{+2.20}_{-1.70}$ & $2.00^{+2.22}_{-1.90}$ & $1.68^{+2.28}_{-1.68}$ & $0.00^{+2.88}_{-0.00}$ & $0.00^{+2.83}_{-0.00}$ & $1.25^{+2.62}_{-1.15}$ \\
\end{tabular}
\end{ruledtabular}
\end{table*}


Figure~\ref{fig:xsectbychannel} shows the contribution of the four combinations
for the different decay modes to the best-fit signal cross section for 
$m_H=125$~GeV/$c^2$.  The results are consistent with each other, with the full 
combination, and with the production of the SM Higgs boson at that mass.  
Figure~\ref{fig:xsposterior125} shows the posterior probability densities 
obtained for the cross section scale factors from the $H\rightarrow b{\bar{b}}$, 
$H\rightarrow W^+W^-$, $H\rightarrow\gamma\gamma$, and $H\rightarrow\tau^+\tau^-$
combinations.

\begin{figure}[htb]
\begin{centering}
\includegraphics[width=1.0\columnwidth]{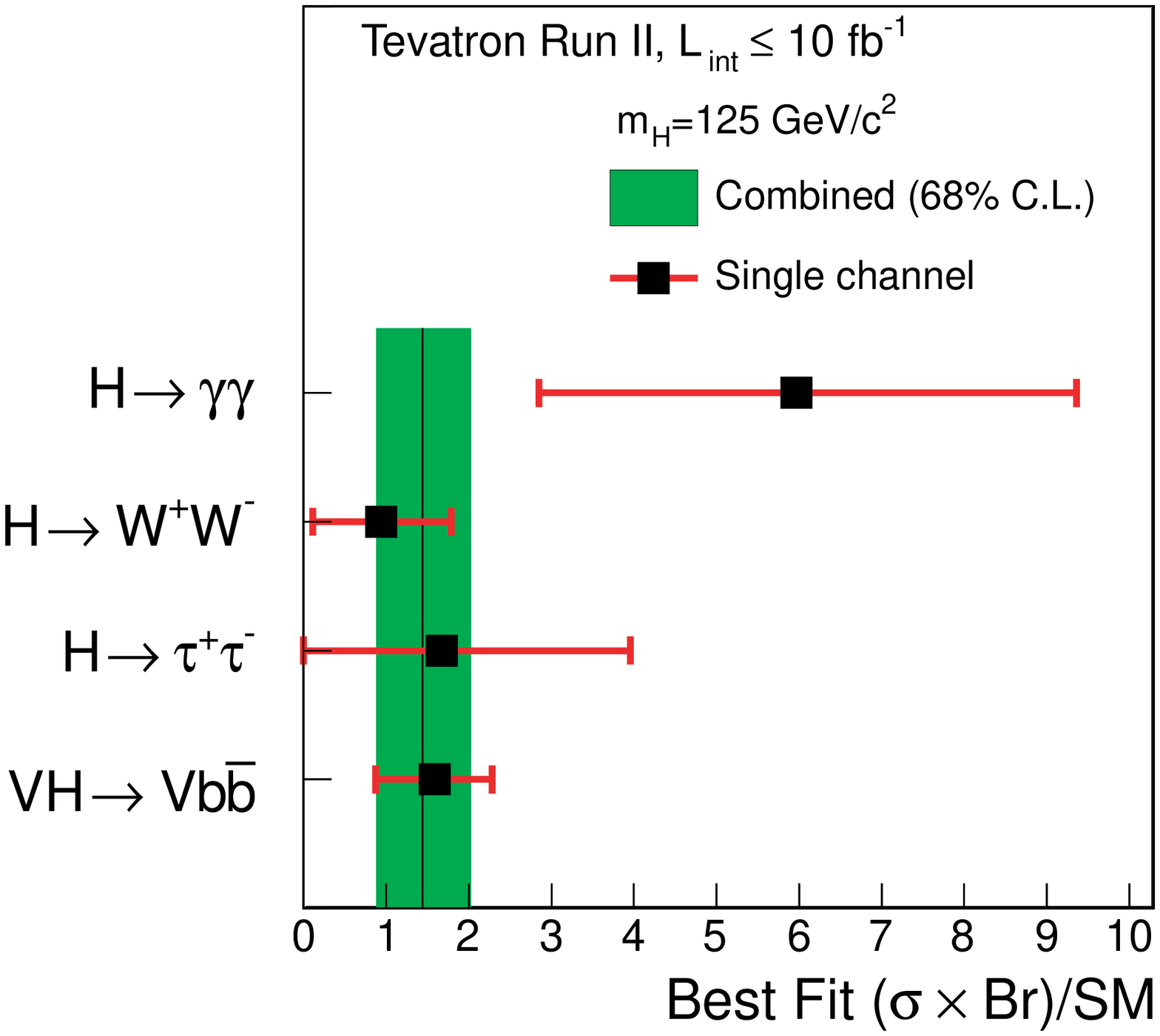}
\caption{
\label{fig:xsectbychannel} (color online)
Best-fit values of $R=(\sigma\times{\cal B})$/SM using the Bayesian method 
for the combinations of CDF and D0's Higgs boson search channels focusing on 
the $H\rightarrow W^+W^-$, $H\rightarrow b{\bar{b}}$, $H\rightarrow\gamma
\gamma$, and $H\rightarrow\tau^+\tau^-$ decay modes for a Higgs boson mass 
of 125~GeV/$c^2$.  The shaded band corresponds to the one s.d.~uncertainty
on the best-fit value of $R$ for all SM Higgs boson decay modes combined.} 
\end{centering}
\end{figure}

\begin{figure}[htb] \begin{centering}
\includegraphics[width=1.0\columnwidth]{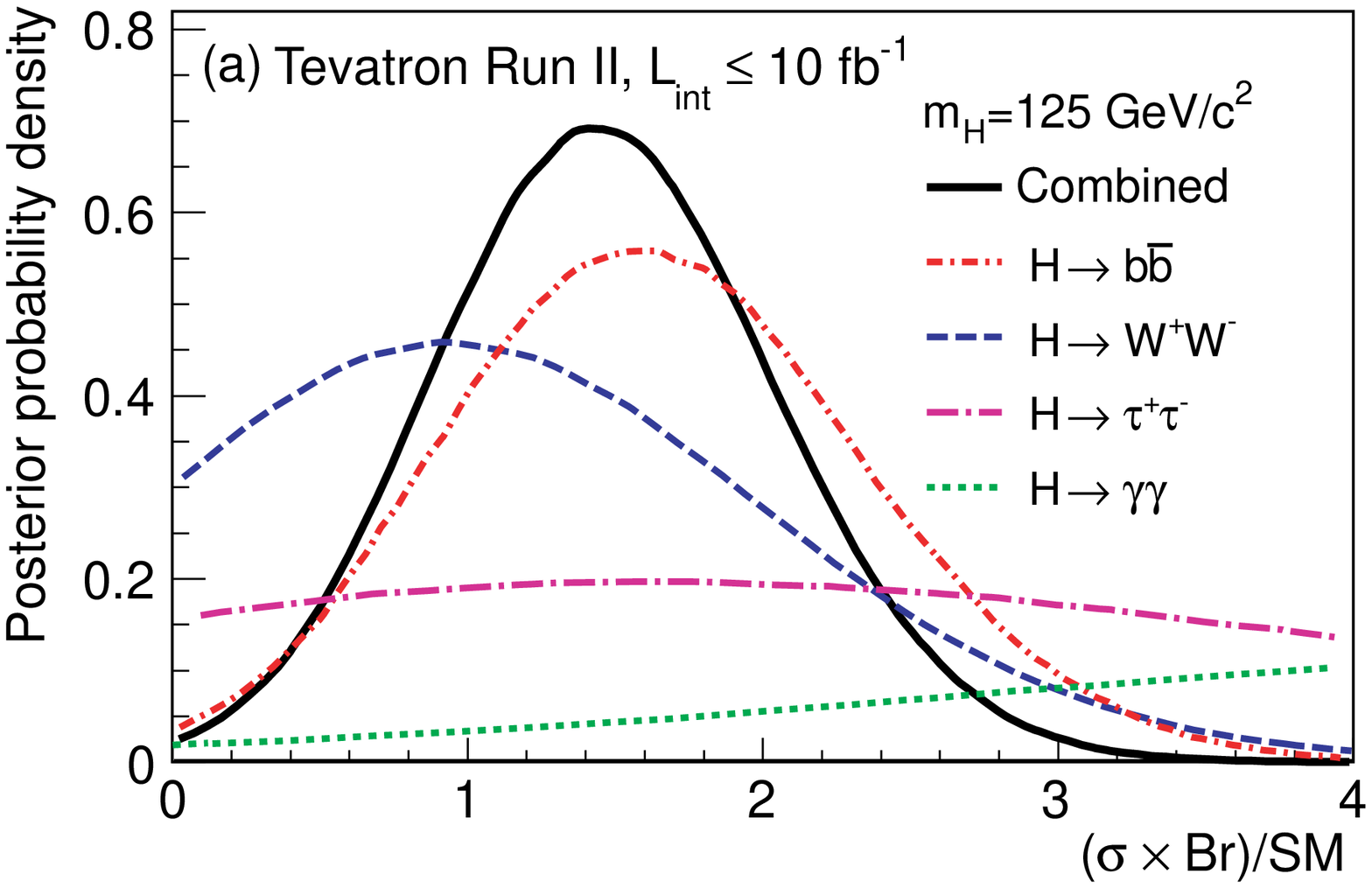}
\includegraphics[width=1.0\columnwidth]{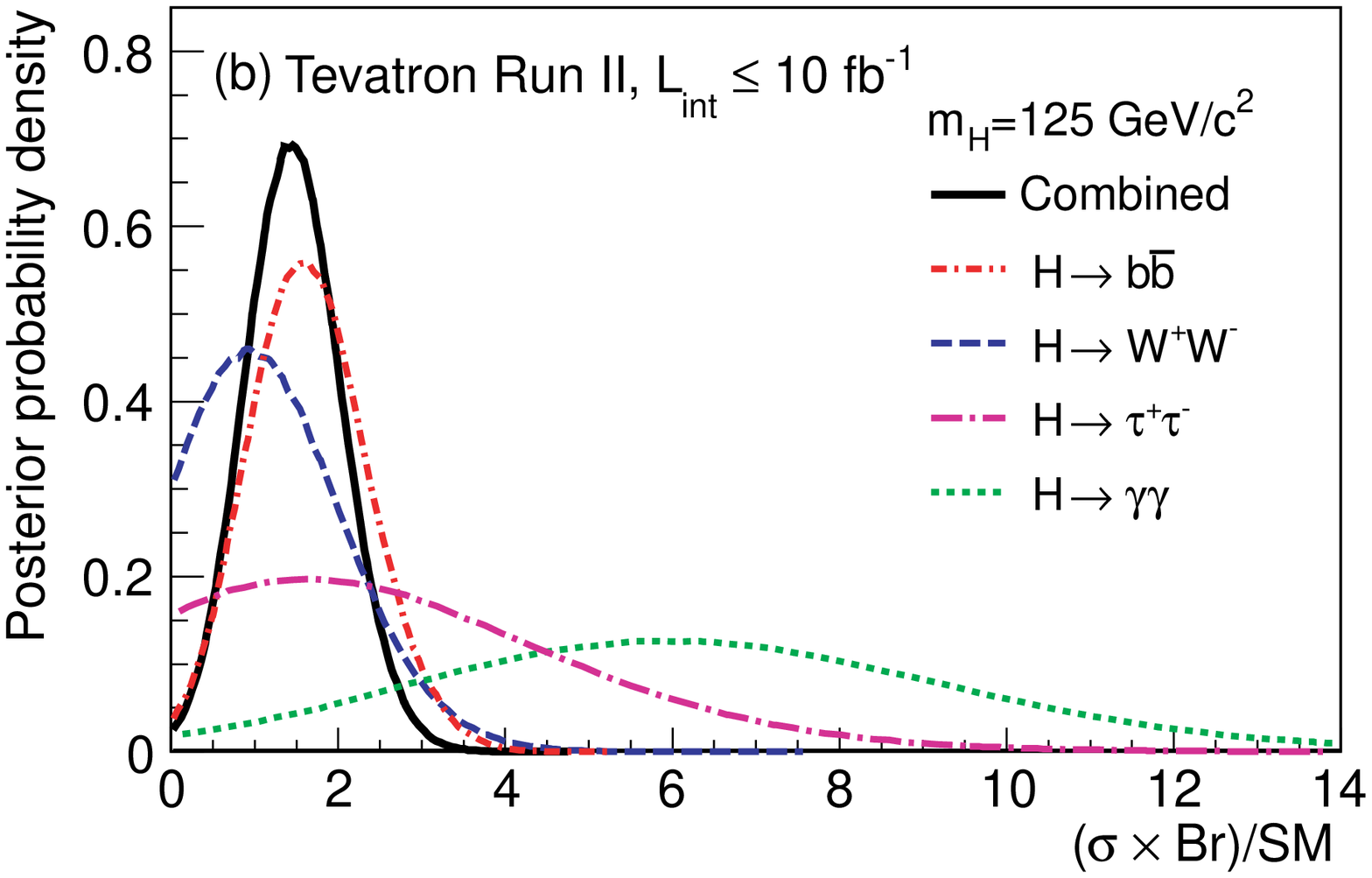}
\caption{
\label{fig:xsposterior125} (color online). (a) Posterior probability densities 
for $R=(\sigma\times{\cal B})$/SM using the Bayesian method from the 
combinations of CDF and D0's Higgs boson search channels focusing on the 
$H\rightarrow W^+W^-$, $H\rightarrow b{\bar{b}}$, $H\rightarrow\gamma
\gamma$, and $H\rightarrow\tau^+\tau^-$ decay modes and for all SM Higgs
boson decay modes combined.   The same curves are shown on an expanded scale in (b).}
\end{centering}
\end{figure}

\clearpage\newpage

\begin{figure} \begin{centering}
\includegraphics[width=1.0\columnwidth]{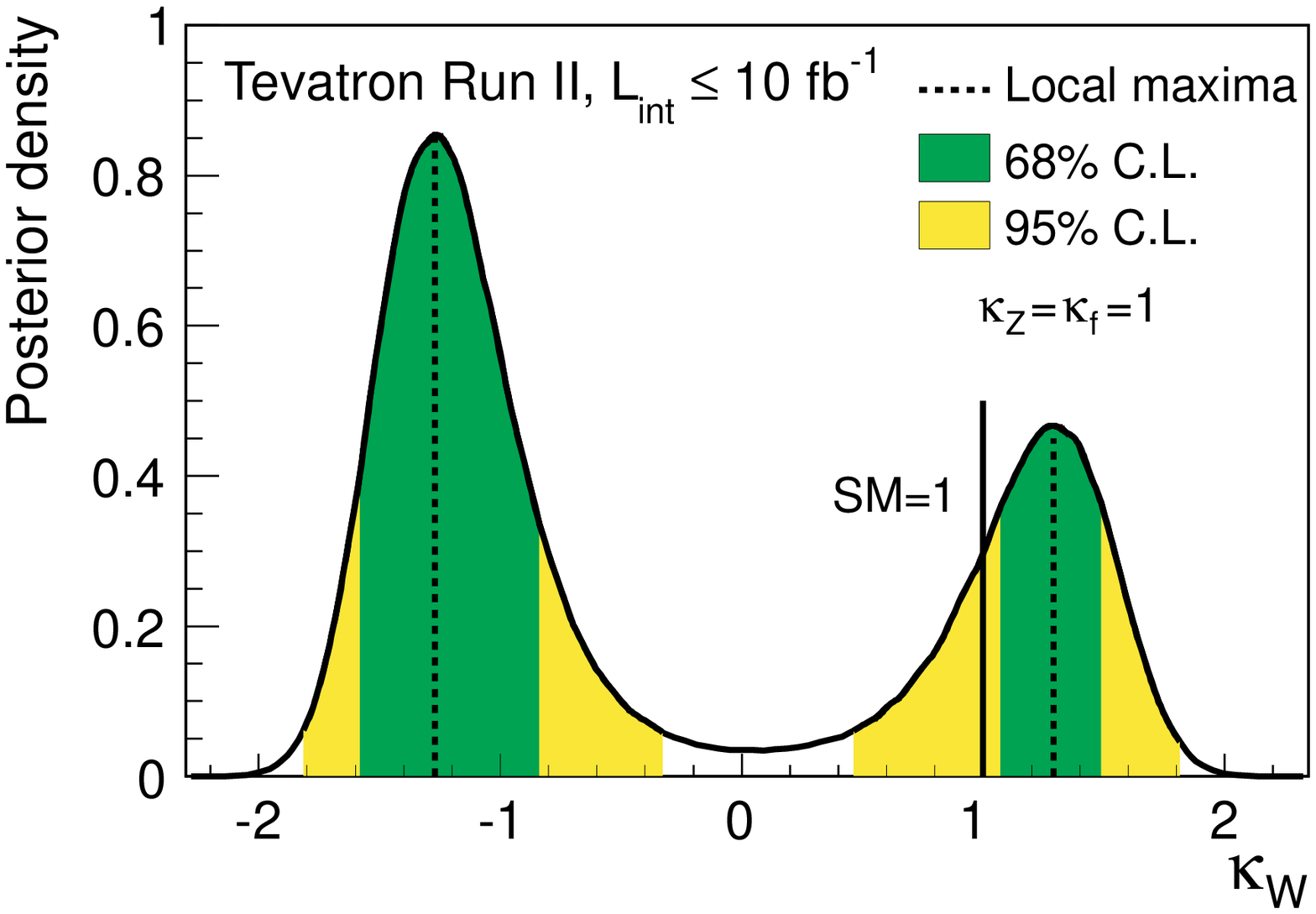}
\caption{
\label{fig:kappaw} (color online). Posterior probability density for $\kappa_W$ from the combination 
of Tevatron searches for a SM-like Higgs boson with $m_H=125$~GeV/$c^2$.  The couplings of the Higgs 
boson to fermions and to the $Z$ boson are assumed to be as predicted by the SM.  The values that 
maximize the local posterior probability densities are shown with dashed lines, and the 68\% and 95\% 
C.L. intervals are indicated with the dark- and light-shaded regions, respectively.  The predicted SM
value of $\kappa_W$ is indicated by the solid vertical line.}
\end{centering}
\end{figure}
\begin{figure} \begin{centering}
\includegraphics[width=1.0\columnwidth]{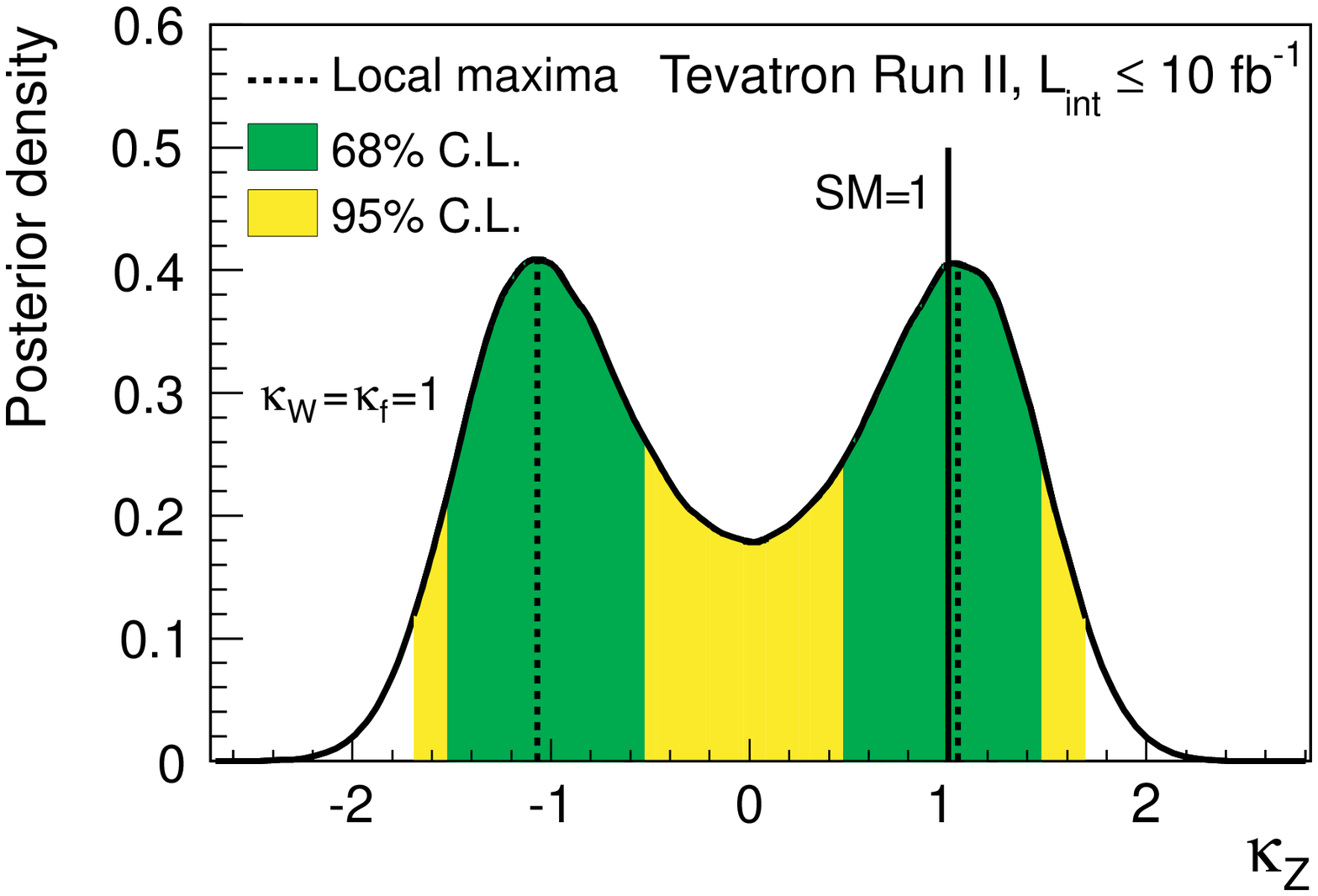}
\caption{
\label{fig:kappaz} (color online). Posterior probability density for $\kappa_Z$ from the combination 
of Tevatron searches for a SM-like Higgs boson with $m_H=125$~GeV/$c^2$.  The couplings of the Higgs 
boson to fermions and to the $W^\pm$ boson are assumed to be as predicted by the SM.  The values that 
maximize the local posterior probability densities are shown with dashed lines, and the 68\% and 95\% 
C.L. intervals are indicated with the dark- and light-shaded regions, respectively.  The predicted SM
value of $\kappa_Z$ is indicated by the solid vertical line.}
\end{centering}
\end{figure}
\begin{figure} \begin{centering}
\includegraphics[width=1.0\columnwidth]{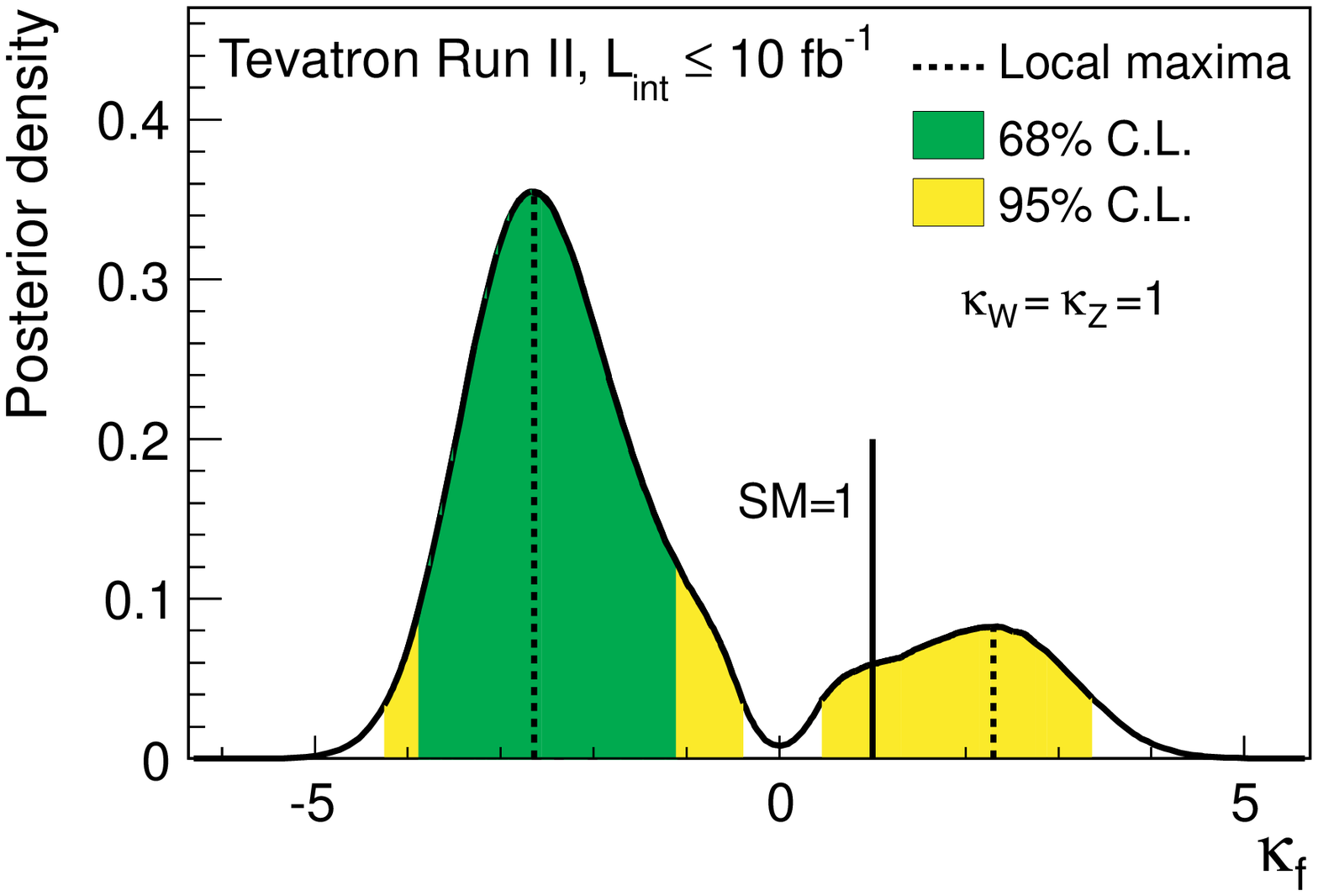}
\caption{
\label{fig:kappaf} (color online). Posterior probability density for $\kappa_f$ from the combination 
of Tevatron searches for a SM-like Higgs boson with $m_H=125$~GeV/$c^2$.  The couplings of the Higgs 
boson to the $W^\pm$ and $Z$ bosons are assumed to be as predicted by the SM.   The values that 
maximize the local posterior probability densities are shown with dashed lines, and the 68\% and 95\% 
C.L. intervals are indicated with the dark- and light-shaded regions, respectively.  The predicted SM
value of $\kappa_f$ is indicated by the solid vertical line.}  
\end{centering}
\end{figure}
\begin{figure} \begin{centering}
\includegraphics[width=0.85\columnwidth]{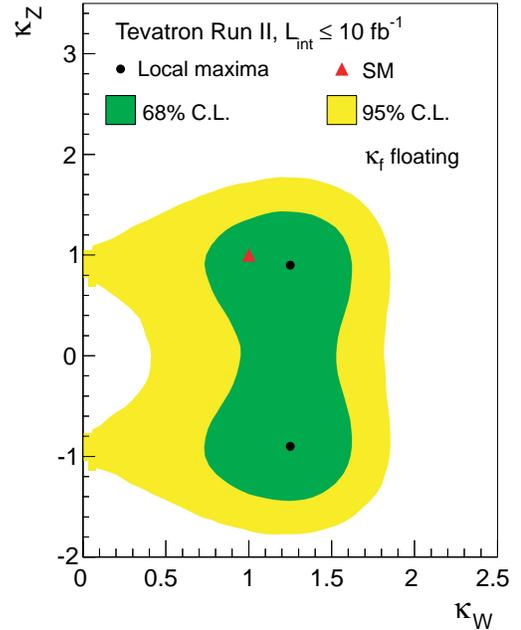}
\caption{
\label{fig:kappawkappaz} (color online). Two-dimensional constraints in the $(\kappa_W,\kappa_Z)$ 
plane, for the combined Tevatron searches for a SM-like Higgs boson with mass 125~GeV/$c^2$ 
allowing $\kappa_f$ to float.  The points that maximize the local posterior probability densities 
are marked with dots, and the 68\% and 95\% C.L. intervals are indicated with the dark- and 
light-shaded regions, respectively.  The SM prediction for $(\kappa_W,\kappa_Z)$ is marked 
with a triangle.}
\end{centering}
\end{figure}
\begin{figure} \begin{centering}
\includegraphics[width=1.0\columnwidth]{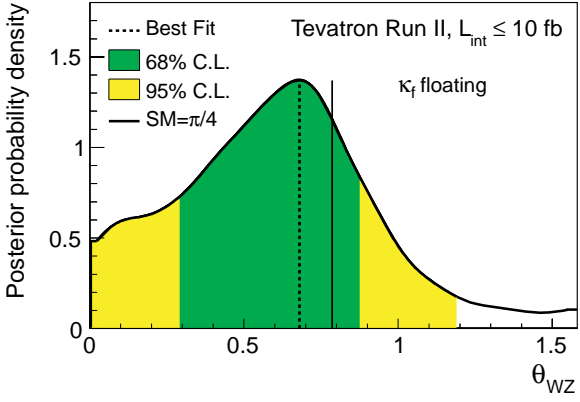}
\caption{
\label{fig:lambdawz}
(color online). Posterior probability density for $\theta_{\it WZ}={\rm{tan}}^{-1}(\kappa_Z/\kappa_W)$, 
from the combination of Tevatron searches for a SM-like Higgs boson with $m_H=125$~GeV/$c^2$ allowing 
$\kappa_f$ to float.  The value that maximizes the posterior probability density is shown with a dashed vertical
line, and the 68\% and 95\% C.L. intervals are indicated with the dark- and light-shaded regions,
respectively.  The predicted SM value of $\theta_{\it WZ}$ is indicated by the solid vertical line.}
\end{centering}
\end{figure}

\begin{figure} \begin{centering}
\includegraphics[width=0.85\columnwidth]{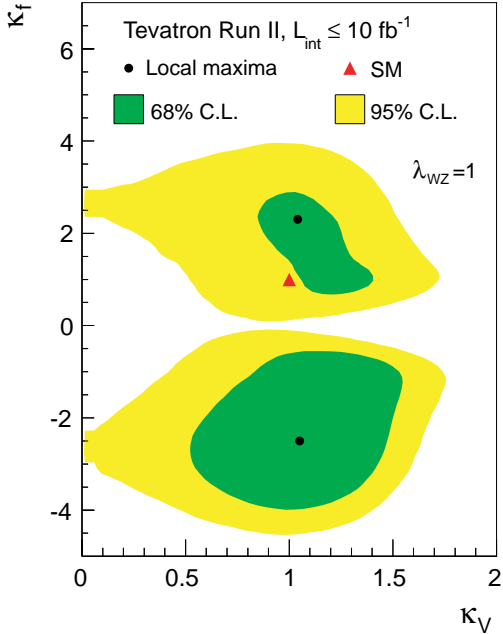}
\caption{
\label{fig:kappavkappaf} (color online). Two-dimensional constraints in the $(\kappa_V,\kappa_f)$ 
plane, for the combined Tevatron searches for a SM-like Higgs boson with mass 125~GeV/$c^2$ 
assuming Custodial symmetry ($\lambda_{\it WZ}=1$).  The points that maximize the local posterior 
probability densities are marked with dots, and the 68\% and 95\% C.L. intervals are indicated 
with the dark- and light-shaded regions, respectively.  The SM prediction for $(\kappa_V,\kappa_f)$ 
is marked with a triangle.}
\end{centering}
\end{figure}

The Higgs boson is expected to couple more strongly to more massive particles 
than to less massive ones, and thus may provide sensitivity to non-SM particles 
whose interactions become more relevant at higher energies.  It is important 
therefore to study in detail the properties of the new particle.  The channel-by-channel
 values of $R=(\sigma\times{\cal B})$/SM 
 provide useful constraints on the possible 
couplings of the particle~\cite{grojean}, but their interpretation is ambiguous because 
signal contributions from multiple sources are simultaneously accepted by each sub-channel.  
For example, the $ZH\rightarrow\nu{\bar{\nu}}b{\bar{b}}$ channels have sensitivity to both 
the $WH$ and $ZH$ production modes, and the $H\rightarrow W^+W^-$ searches are sensitive to 
gluon-gluon fusion, $WH$, $ZH$, and VBF in different mixtures within independent sub-channels 
characterized by the number of reconstructed jets.  

Most of the searches conducted at the Tevatron are sensitive to the product of fermion and 
boson coupling strengths.  In the $VH\rightarrow Vb{\bar{b}}$ searches, the production depends 
on the coupling of the Higgs boson to the weak vector bosons, while the decay is to fermions.  
In the $gg\rightarrow H\rightarrow W^+W^-$ searches, the production is dominated by the Higgs 
boson couplings to fermions via the quark loop processes, but the decay is to bosons.  A large 
enhancement of the Higgs boson's couplings to fermions can thus be masked by a small coupling 
to bosons, and vice versa, as shown in Fig.~2 of Ref.~\cite{grojean}.  However, other 
less-sensitive channels included in this combination provide additional constraints.  The 
same-sign di-lepton searches, the 
tri-lepton searches, and some of the searches with tau leptons as decay products of $W$ bosons
 are primarily sensitive to $VH\rightarrow VW^+W^-$, an entirely bosonic 
process, although their results are customarily reported in combination with the other 
$H\rightarrow W^+W^-$ searches.  The searches for $t{\bar{t}}H\rightarrow t{\bar{t}}b{\bar{b}}$ 
provide constraints on the fermion couplings with minimal masking from the bosonic couplings.

We follow the notation of Ref.~\cite{lhccouplingrec} and introduce multiplicative 
scaling factors for the coupling of the Higgs boson to fermions ($\kappa_f$) and 
either to $W$ bosons ($\kappa_W$) and $Z$ bosons ($\kappa_Z$) or more generically to 
vector bosons ($\kappa_V$). We then search for deviations from the expected SM 
values of $\kappa_i = 1$.

The first test assumes $m_H=125$~GeV/$c^2$, based on the ATLAS and CMS observations,
and fits for the $H\rightarrow W^+W^-$ coupling, holding all other couplings fixed to 
their SM values.
This test corresponds to holding the values of $\kappa_Z=\kappa_f=1$, while
varying $\kappa_W$.  At each value of $\kappa_W$, we recompute the predicted cross sections and decay branching ratios,
as described in Ref.~\cite{lhccouplingrec}.  We assume a uniform prior density in $\kappa_W$, and show the posterior
probability distribution in Fig.~\ref{fig:kappaw}.  A negative sign of $\kappa_W$ is preferred by the Tevatron data
due to the excess seen in the $H\rightarrow\gamma\gamma$ searches.  In the SM, this process proceeds at lowest order
via a $W$-boson loop or a quark loop (dominated by the top quark), with destructive interference between the two 
contributions~\cite{spira95}, as given by 
$ \Gamma(H\rightarrow\gamma\gamma) = \Gamma(H\rightarrow\gamma\gamma)_{SM} \times | 1.28~\kappa_V - 0.28~\kappa_f |^2$.
If the sign of the $H\rightarrow W^+W^-$ coupling is negative, then this interference becomes constructive, 
allowing for a larger prediction of the yield.  We obtain a best-fit value of $\kappa_W=-1.27$.  Our 
procedure for finding the smallest set of intervals that contain 68\% of the integral of the posterior 
results in two intervals, $-1.56<\kappa_W < -0.81$ and $1.04<\kappa_W<1.51$.  We perform a similar test 
for $\kappa_Z$, assuming $\kappa_W=\kappa_f=1$.  The resulting posterior probability density is shown in 
Fig.~\ref{fig:kappaz}.  The Higgs boson searches at the Tevatron are sensitive almost exclusively to the 
square of $\kappa_Z$, and thus the posterior density is nearly symmetric in positive and negative couplings.  
The best-fit values are $\kappa_Z=1.05_{-0.55}^{+0.45}$ and $\kappa_Z=-1.05_{-0.45}^{+0.55}$.  Finally, we 
perform a similar test for $\kappa_f$, the common scale factor on the Higgs boson couplings to fermions, 
holding $\kappa_W = \kappa_Z=1$.  The resulting posterior probability density is shown in Fig.~\ref{fig:kappaf}.  
An asymmetry is seen in this distribution, due again to the outcome in the $H\rightarrow\gamma\gamma$ channels.  
We obtain a best-fit value of $\kappa_f=-2.64_{-1.30}^{+1.59}$.
The large magnitude of the fitted value is due to the excesses seen in the $H\rightarrow b{\bar{b}}$ and 
$H\rightarrow\gamma\gamma$ searches.

We then allow both $\kappa_W$ and $\kappa_Z$ to vary independently, also allowing $\kappa_f$ to vary 
by integrating the likelihood function times a uniform prior in $\kappa_f$ over negative and positive 
values.  The resulting areas in the $(\kappa_W,\kappa_Z)$ plane preferred by the Tevatron data are 
shown in Fig.~\ref{fig:kappawkappaz}.  While we allow either coupling scale factor to be negative, 
only two quadrants are shown in  Fig.~\ref{fig:kappawkappaz} due to an overall sign ambiguity.   
The point $(\kappa_W,\kappa_Z)=(0,0)$ corresponds to no Higgs boson production or decay in the most 
sensitive search modes at the Tevatron and is
 excluded at more than the 95\% C.L. due to the Higgs-boson-like signal in the \hbb\ and \hww\ channels. 
  Our best-fit points are $(\kappa_W,\kappa_Z)=(1.25,\pm 0.90)$.

We study the ratio $\lambda_{\it WZ} = \kappa_W/\kappa_Z$ using the same posterior probability density that is
used in Fig.~\ref{fig:kappawkappaz}.  We choose a projection onto a one dimensional variable that preserves 
the uniformity of the prior probability density in the two-dimensional plane.  This variable is the angle 
with respect to the $\kappa_W$ axis, $\theta_{\it WZ}={\rm{tan}}^{-1}(\kappa_Z/\kappa_W) = {\rm{tan}}^{-1}
(1/\lambda_{\it WZ})$.  Figure~\ref{fig:lambdawz} shows the one-dimensional posterior probability density 
in this variable.  This function is symmetric for positive and negative $\theta_{\it WZ}$.  
We measure $|\theta_{\it WZ}|=0.68_{-0.41}^{+0.21}$, which corresponds to $\lambda_{\it WZ}=1.24^{+2.34}_{-0.42}$.

Assuming that custodial symmetry~\cite{custodial} holds ($\lambda_{\it WZ}=1$),
we allow both $\kappa_V$ and $\kappa_f$ to vary, and show in Fig.~\ref{fig:kappavkappaf}
the regions preferred at the 68\% C.L. and the 95\% C.L. in the two-dimensional plane 
$(\kappa_V,\kappa_f)$.  The asymmetry induced by the excesses in the $H\rightarrow\gamma
\gamma$ searches is visible in this projection as well.  The best-fit point is 
$(\kappa_V,\kappa_f)=(1.05,-2.40)$, but a secondary maximum in the posterior density 
is seen at $(\kappa_V,\kappa_f)=(1.05,2.30)$, consistent with the SM expectation, given the large uncertainties.
  The integral of the posterior density 
in the (+,+) quadrant is 26\% of the total, while the remaining 74\% of the integral 
of the posterior density is contained within the (+,--) quadrant.

\section{Results - Non-standard model interpretations}\label{sec:bsm}

The mechanism of electroweak symmetry breaking
may entail a richer phenomenology than expected in the SM.  
Natural extensions include the addition of a
fourth generation of fermions with masses much larger than those of
the three known generations or models with several Higgs bosons or models in which the
Higgs boson(s) may have modified couplings. 
We interpret our Higgs boson search results in models with a sequential fourth generation
of fermions (SM4) and in the fermiophobic Higgs model (FHM) described below.

\subsection{Fourth Generation Interpretation}\label{sec:4g}

With the inclusion of two additional heavy fourth-generation quarks in the SM4~\cite{fourthgen}, 
the $gg\rightarrow H$ coupling is enhanced by a factor of roughly three relative to the SM 
coupling~\cite{arik,g4_hdecay,abf}.  The partial decay width for $H\rightarrow gg$ is enhanced 
by the same factor as the production cross section.  However, because the $H\rightarrow gg$ 
decay is mediated by a loop amplitude, the $H\rightarrow W^+W^-$ decay continues to dominate 
for Higgs boson masses above 135~GeV/$c^2$.  Since the expected signal yield is larger in the 
SM4 model than the SM, the sensitivity of CDF and D0's Higgs boson searches extends to higher 
masses.  For this reason, the upper end of the search range for the relevant channels is raised  
to 300~GeV/$c^2$ for interpretations associated with this model.

Two scenarios for the masses of the fourth-generation
fermions are considered.  In the first, the {\it low-mass} scenario,
we set the mass of the fourth-generation neutrino $m_{\nu 4}=80$~GeV/$c^2$ 
and the mass of the fourth-generation charged lepton $m_{\ell 4}=100$~GeV/$c^2$,
in order to have the maximum impact on the Higgs boson decay branching ratios and
to be compatible with the experimental constraint on the mass of an unstable $\nu_4$~\cite{L3_lepton}.
In the case that the $\nu_4$ is stable or has a lifetime long enough to escape the search presented in 
Ref.~\cite{L3_lepton}, $m_{\nu_4}$ could be lighter, modifying the decay branching 
ratios~\cite{belotsky}, resulting in weaker mass limits.
In our second scenario, the {\it high-mass} scenario, we set $m_{\nu 4}=m_{\ell 4}=1$~TeV/$c^2$, 
so that the fourth-generation leptons do not modify the
decay branching ratios of the Higgs boson relative to the SM. In both scenarios, 
we choose the masses of the quarks to be those of the second scenario in
Ref.~\cite{abf} ($m_{d4}$ = 400~GeV/$c^2$ and $m_{u4}$ = 450~GeV/$c^2$). The 
next-to-next-to-leading order (NNLO) production cross 
section calculation of Ref.~\cite{abf} is used, which is a modified version of 
the NNLO SM calculation.
Previous interpretations of SM Higgs boson searches within the context of a 
fourth generation of fermions at the Tevatron excluded $131<m_H<207$~GeV/$c^2$~\cite{PRDRC}. 
Similar searches have been performed by the ATLAS~\cite{atlas4g} and CMS~\cite{cms4g} 
Collaborations, excluding $140<m_H<185$~GeV/$c^2$ and $144<m_H<207$~GeV/$c^2$, 
respectively.  A more recent search by the CMS Collaboration excluded the mass 
range $110<m_H<600$~GeV/$c^2$~\cite{cms4g-new}.

We combine our searches for a Higgs boson in the processes $gg\rightarrow H\rightarrow W^+W^-$ 
and $gg\rightarrow H\rightarrow ZZ$.  Limits on the SM4 models and on $\sigma(gg\rightarrow H)
\times\mathcal{B}(H\rightarrow W^+W^-)$ are derived. This result is an update of Ref.~\cite{PRDRC}. 
The analyses are performed equivalently to the SM searches except that $gg\rightarrow H$ production 
only is considered for the signal. The MVA classifiers are retrained accordingly and, for the 
specific case of the D0 $H\rightarrow W^+ W^- \rightarrow \ell^\pm\nu \ell^\mp\nu$ channel, the 
two-jet bin, which is less sensitive to $gg\rightarrow H$ production, is not included.

The branching ratios for $H\rightarrow W^+W^-$ are calculated using {\sc hdecay}~\cite{hdecay} modified 
to include fourth-generation fermions~\cite{g4_hdecay}.  To include the $gg\rightarrow H\rightarrow ZZ$ 
searches, we assume the SM value for $\mathcal{B}(H\rightarrow W^+W^-)/\mathcal{B}(H\rightarrow ZZ)$.  
In setting limits on $\sigma(gg\rightarrow H)\times\mathcal{B}(H\rightarrow W^+W^-)$, the 
$gg\rightarrow H\rightarrow ZZ$ process is included assuming that its signal yield scales 
equivalently to that from the $gg\rightarrow H\rightarrow W^+W^-$ channel.

When setting limits on
$\sigma(gg\rightarrow H)\times \mathcal{B}(H\rightarrow W^+W^-)$,  
the theoretical uncertainty on the prediction of
$\sigma(gg\rightarrow H)\times \mathcal{B}(H\rightarrow W^+W^-)$ 
is not included since these limits are independent of the
predictions.  However, when setting limits on $m_H$ in the context 
of fourth-generation models, uncertainties on the theoretical 
predictions are included as described for the SM searches.

The combined limits on $\sigma(gg\rightarrow H)\times\mathcal{B}
(H\rightarrow W^+W^-)$ obtained using the Bayesian method are shown
in Fig.~\ref{fig:xslimits} along with the theory
predictions for fourth-generation models in the low- and high-mass scenarios.  Limits obtained 
using both the Bayesian and ${\rm CL}_{\rm s}$ methods are listed 
as a function of Higgs boson mass in Table~\ref{tab:sm4limits}.
A broad, moderate excess above the background expectation is seen 
for masses above 200~GeV/$c^2$.

Production limits obtained for the two SM4 scenarios using the Bayesian
method are shown in Fig.~\ref{fig:4glimits}.  The limits are presented  
as ratios relative to SM4 low-mass scenario predictions as a function of 
the Higgs boson mass. 
In the low-mass scenario, which gives the smaller excluded mass range, a SM-like 
Higgs boson with a mass in the range \SMFLobslow --\SMFLobshigh~GeV/$c^2$ is 
excluded at the 95\% C.L.  The expected excluded mass range is \SMFLexplow 
--\SMFLexphigh~GeV/$c^2$.  In the high-mass scenario, the mass range \SMFHobslow 
--\SMFHobshigh~GeV/$c^2$ is excluded, with an expected excluded mass range of
\SMFHexplow --\SMFHexphigh~GeV/$c^2$.

\begin{figure} \begin{center}
\includegraphics[width=1.0\columnwidth]{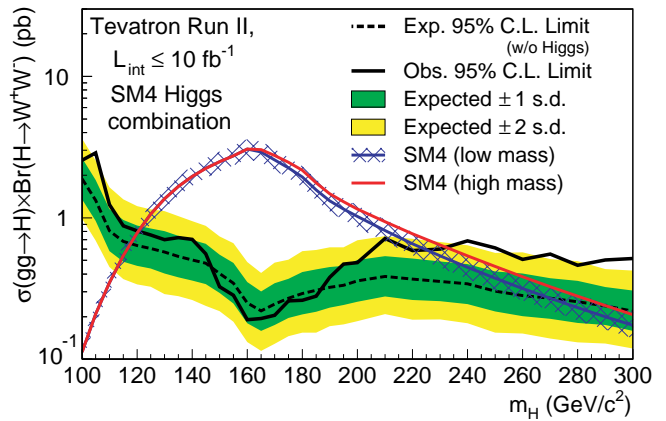}
\end{center}
\caption{\label{fig:xslimits} (color online). 
 Observed and median expected (for the background-only 
hypothesis) 95\% C.L. Bayesian upper limits on the cross section times 
branching ratio $\sigma(gg\rightarrow H)\times \mathcal{B}(H\rightarrow 
W^+W^-)$ from the combination of CDF and D0's Higgs boson search channels 
focusing on this production and decay mode.
The dark and light-shaded bands indicate, respectively, the one and 
two s.d.~probability regions in which the limits are expected to fluctuate in the 
absence of signal.  Theoretical predictions for SM4 in the low- and 
high-mass scenarios are shown with blue and red lines, respectively.
The hatched band indicates the theoretical uncertainty associated 
with the SM4 low-mass scenario.}
\end{figure}

\begin{table}[!]
\caption{\label{tab:sm4limits} Observed and median expected (for 
the background-only hypothesis) 95\% C.L. upper limits on the cross 
section times branching ratio $\sigma(gg\rightarrow H)\times \mathcal{B}
(H\rightarrow W^+W^-)$ from the combination of CDF and D0's Higgs boson 
search channels focusing on this production and decay mode, obtained 
using the Bayesian and ${\rm CL}_{\rm s}$ methods.}
\begin{ruledtabular}
\begin{tabular}{lcccc}
 & \multicolumn{2}{c}{Bayesian} & \multicolumn{2}{c}{${\rm CL}_{\rm s}$} \\ 
$m_H$             & Observed & Expected & Observed & Expected \\
  (GeV/$c^2$)     & limit (pb) & limit (pb) & limit (pb) & limit (pb)  \\ \hline
100 &     2.56 &     1.89 &     2.58 &     1.87 \\    
105 &     2.87 &     1.33 &     2.62 &     1.32 \\
110 &     1.24 &     0.81 &     1.27 &     0.82 \\
115 &     0.88 &     0.68 &     0.90 &     0.70 \\
120 &     0.81 &     0.63 &     0.81 &     0.66 \\
125 &     0.75 &     0.61 &     0.77 &     0.61 \\
130 &     0.70 &     0.57 &     0.70 &     0.58 \\
135 &     0.72 &     0.53 &     0.75 &     0.54 \\
140 &     0.70 &     0.50 &     0.72 &     0.52 \\
145 &     0.56 &     0.48 &     0.55 &     0.48 \\
150 &     0.33 &     0.40 &     0.32 &     0.42 \\
155 &     0.28 &     0.34 &     0.30 &     0.35 \\
160 &     0.19 &     0.25 &     0.20 &     0.26 \\
165 &     0.19 &     0.22 &     0.20 &     0.22 \\
170 &     0.20 &     0.24 &     0.21 &     0.25 \\
175 &     0.26 &     0.27 &     0.26 &     0.27 \\
180 &     0.26 &     0.29 &     0.26 &     0.29 \\
185 &     0.28 &     0.31 &     0.29 &     0.31 \\
190 &     0.38 &     0.32 &     0.40 &     0.33 \\
195 &     0.47 &     0.33 &     0.47 &     0.34 \\
200 &     0.48 &     0.36 &     0.49 &     0.37 \\
210 &     0.71 &     0.38 &     0.73 &     0.39 \\
220 &     0.59 &     0.37 &     0.60 &     0.37 \\
230 &     0.60 &     0.36 &     0.61 &     0.36 \\
240 &     0.69 &     0.34 &     0.69 &     0.34 \\
250 &     0.61 &     0.30 &     0.60 &     0.30 \\
260 &     0.51 &     0.28 &     0.49 &     0.29 \\
270 &     0.55 &     0.27 &     0.56 &     0.27 \\
280 &     0.46 &     0.25 &     0.47 &     0.25 \\
290 &     0.50 &     0.24 &     0.48 &     0.24 \\
300 &     0.52 &     0.22 &     0.50 &     0.22 \\
\end{tabular}
\end{ruledtabular}
\end{table}

\begin{figure}
\begin{center}
\includegraphics[width=1.0\columnwidth]{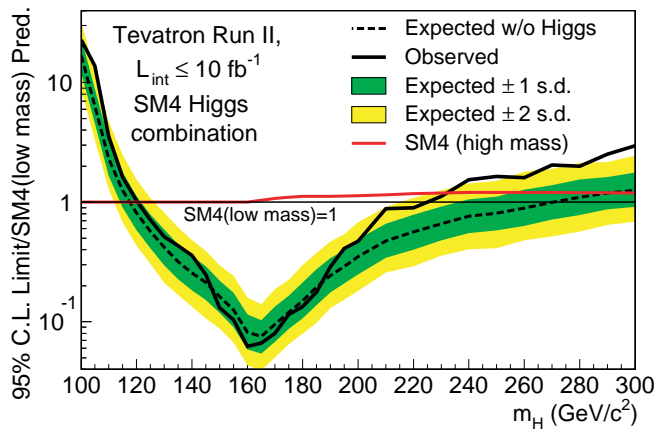}
\end{center}
\caption{
\label{fig:4glimits}  (color online). 
Observed and median expected (for the background-only hypothesis) 95\% 
C.L. Bayesian upper production limits expressed as multiples of the 
SM4 (low-mass scenario) cross section as a function of Higgs boson mass 
from the combination of CDF and D0's Higgs boson searches focusing on the 
$gg\rightarrow H$ production and $H\rightarrow W^+W^-$ decay modes.
Uncertainties associated with theoretical cross section and branching 
ratio predictions are incorporated in the limit.
The dark and light-shaded bands indicate, respectively, the one and 
two s.d.~probability regions in which the limits are expected to fluctuate in the 
absence of signal.  The red line shows the prediction for the signal  
rate in the high-mass scenario, divided by that of the low-mass 
scenario.}
\end{figure}

\subsection{Fermiophobic Interpretation}\label{sec:fhm}

In the FHM, the lightest Higgs boson does not couple to fermions at tree 
level, but aside from this one difference, its behavior is indistinguishable 
from that of the SM Higgs boson. In the FHM, the production of Higgs bosons, 
$H_f$, at hadron colliders via the process $gg\rightarrow H_f$ is suppressed 
to a negligible rate and is ignored in the context of this interpretation. The 
associated production mechanisms $p{\bar{p}}\rightarrow WH_f+X$ and $p{\bar{p}}
\rightarrow ZH_f+X$, as well as the vector-boson-fusion (VBF) processes 
$q\bar{q}\rightarrow q^{\prime}\bar{q}^{\prime}H_f$,
remain nearly unchanged relative to the corresponding processes in the SM. 
Thus, the corresponding SM cross sections and associated uncertainties described 
previously are also used here.
In the FHM, direct decays to fermions are forbidden; the decays to 
$W^+W^-$, $\gamma\gamma$, $ZZ$, and $Z\gamma$ account for nearly the entire decay 
width.  For the mass range under investigation the $W^+W^-$ decay mode has the 
largest branching fraction.
The branching fraction ${\cal{B}}(H_f\rightarrow\gamma\gamma$) is greatly enhanced 
over ${\cal{B}}(H_{\rm{SM}}\rightarrow\gamma\gamma$) for all $m_{H}$, and the clean 
signature and excellent mass resolution of this channel provide most of the search 
sensitivity for $m_{H_f}<120$~GeV/$c^2$.
The analyses combined here seek Higgs boson decays to $W^+W^-$, $\gamma\gamma$, 
and $ZZ$.  Previous searches for a fermiophobic Higgs boson at the Tevatron excluded 
signals with masses smaller than 119~GeV/$c^2$~\cite{prevtevfhm}; the expected 
exclusion was also $m_{H_f}<119$~GeV/$c^2$.  The ATLAS and CMS Collaborations 
excluded $m_{H_f}$ in the ranges 110.0--118.0~GeV/$c^2$ and 119.5--121.0~GeV/$c^2$ 
using diphoton final states~\cite{atlasfhm} and in the range 110--194~GeV/$c^2$ by 
combining multiple final states~\cite{cmsfhm}.

The SM $H \rightarrow\gamma\gamma$ analyses are reoptimized as the 
kinematic distributions of the Higgs bosons, their decay products, and the 
particles produced in association with the Higgs bosons differ between the 
FHM and the SM. Events contain either an associated $W$ or $Z$ boson, or 
recoiling quark jets in the case of VBF and thus the transverse momentum 
($p_T$) of the Higgs boson is on average greater than it is in the SM. The 
analyses combined here update previous searches for the Higgs boson in the 
FHM~\cite{prevcdffhm,prevd0fhm}.  Similarly, SM searches in $H \rightarrow 
W^+W^-$ channels cannot be interpreted directly in the FHM due to the 
different mixture of production modes.  
Signal contributions from $gg\rightarrow H_f$ production to the MVA 
discriminant distributions are ignored, and the remaining contributions 
from other production mechanisms are scaled by the ratio of branching ratio 
predictions ${\cal{B}}(H_f\rightarrow VV)/{\cal{B}}(H_{\rm{SM}}\rightarrow VV)$.
The existing subdivision of channels based on the number of reconstructed jets 
accompanying the leptons and missing transverse energy in the event naturally 
optimizes the search within the FHM interpretation.  Hence, the development 
of a separate set of analysis channels as in the case of $H_f\rightarrow\gamma
\gamma$ is not required, though the MVAs are retrained.

The combined limits on Higgs boson production normalized to FHM predictions
obtained from both the Bayesian and CL$_{\rm s}$ methods are listed in 
Table~\ref{tab:FPlimits} as a function of Higgs boson mass.  The expected 
limits assume no Higgs boson production.  The limits obtained using the 
Bayesian method are shown in Fig.~\ref{fig:fplimits}.
Fermiophobic Higgs bosons in the mass range \FPobslow --\FPobshigh~GeV/$c^2$ 
are excluded at the 95\% C.L.; the expected excluded mass range is \FPexplow --\FPexphigh~GeV/$c^2$.

\begin{table}
\caption{\label{tab:FPlimits} Ratios of observed and median expected 
(for the background-only hypothesis) 95\% C.L. upper 
limits on the production rate of a Fermiophobic Higgs boson
relative to the FHM prediction as a function of the Higgs boson mass 
for the combination of CDF and D0's searches, obtained using the Bayesian and 
${\rm CL}_{\rm s}$ methods.}
\begin{ruledtabular}
\begin{tabular}{lcccc}
 & \multicolumn{2}{c}{Bayesian} & \multicolumn{2}{c}{${\rm CL}_{\rm s}$} \\ 
$m_{H_f}$ (GeV/$c^2$) & $R_{95}^{\mathrm{obs}}$ & $R_{95}^{\mathrm{exp}}$ & $R_{95}^{\mathrm{obs}}$ & $R_{95}^{\mathrm{exp}}$ \\ \hline
  100 & 0.21 & 0.13 & 0.21 & 0.13 \\
  105 & 0.36 & 0.22 & 0.37 & 0.23 \\
  110 & 0.40 & 0.37 & 0.36 & 0.37 \\
  115 & 0.95 & 0.54 & 0.88 & 0.53 \\
  120 & 1.13 & 0.69 & 1.06 & 0.68 \\
  125 & 1.41 & 0.83 & 1.44 & 0.81 \\
  130 & 1.21 & 0.91 & 1.06 & 0.90 \\
  135 & 1.26 & 1.00 & 1.16 & 0.97 \\
  140 & 1.65 & 1.11 & 1.48 & 1.06 \\
  145 & 1.47 & 1.15 & 1.30 & 1.13 \\
  150 & 1.33 & 1.21 & 1.19 & 1.17 \\  
  155 & 1.30 & 1.19 & 1.17 & 1.18 \\
  160 & 1.20 & 1.17 & 1.11 & 1.14 \\
  165 & 0.98 & 1.17 & 0.94 & 1.11 \\
  170 & 1.49 & 1.31 & 1.35 & 1.26 \\
  175 & 1.96 & 1.48 & 1.76 & 1.43 \\
  180 & 2.34 & 1.72 & 2.04 & 1.60 \\
  185 & 3.13 & 1.96 & 2.58 & 1.93 \\
  190 & 3.75 & 2.36 & 3.24 & 2.32 \\
  195 & 4.58 & 2.62 & 3.92 & 2.54 \\
  200 & 5.43 & 2.85 & 4.64 & 2.77 \\
\end{tabular}
\end{ruledtabular}
\end{table}

\begin{figure}
\begin{center}
\includegraphics[width=1.0\columnwidth]{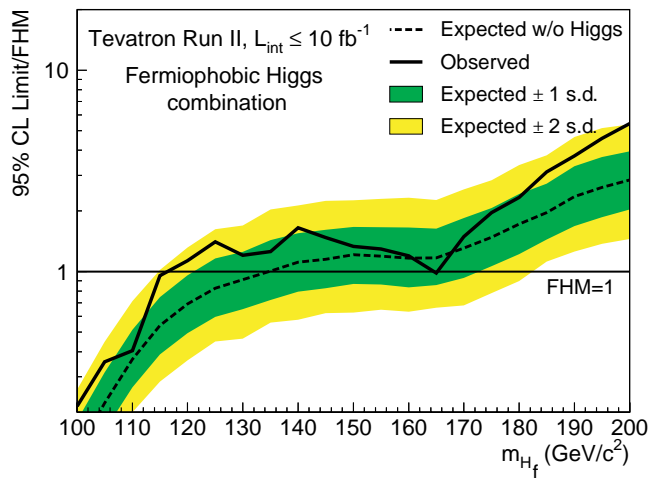}
\end{center}
\caption{\label{fig:fplimits}  (color online).  Observed and median 
expected (for the background-only hypothesis) 95\% C.L. Bayesian 
upper production limits expressed as multiples of the FHM cross 
section as a function of Higgs boson mass for the combined CDF 
and D0 searches.
The dark and light-shaded bands indicate, respectively, the one 
and two s.d.~probability regions in which the limits are expected to fluctuate 
in the absence of signal.}
\end{figure}

\section{\label{sec:conclusion}Conclusions}

The search for the standard model Higgs boson at the Tevatron is challenging due 
to the small expected signal and the need to accurately model large background 
contributions. We have developed advanced tools to search for the Higgs boson in 
the leading production and decay modes predicted by the SM and control the impact 
of systematic uncertainties using constraints from the observed data.  We have combined 
searches by the CDF and D0 Collaborations for the standard model Higgs boson in 
the mass range 90--200~GeV$/c^2$ using Tevatron $p{\bar{p}}$ collision data 
corresponding to up to 10~fb$^{-1}$ of integrated luminosity collected at 
$\sqrt{s}=1.96$~TeV.  The results of searches focusing on the $H\rightarrow 
b{\bar{b}}$, $H\rightarrow W^+W^-$, $H\rightarrow ZZ$, $H\rightarrow\tau^+\tau^-$, 
and $H\rightarrow \gamma\gamma$ decay modes are included in the combination.  
The results are also interpreted in fermiophobic and fourth generation models. 
Fermiophobic Higgs bosons in the mass range \FPobslow --\FPobshigh~GeV/$c^2$ 
are excluded at the 95\% C.L., and a SM-like Higgs boson in the presence 
of a fourth sequential generation of fermions is excluded in the mass range 
\SMFLobslow --\SMFLobshigh~GeV/$c^2$ at the 95\% C.L. The SM Higgs boson is 
excluded, at the 95\% C.L., from \SMLobslow\ to \SMLobshigh~GeV/$c^{2}$, and 
from \SMHobslow\ to \SMHobshigh~GeV/$c^{2}$.  The expected exclusion regions 
in the absence of signal are \SMLexplow --\SMLexphigh~GeV/$c^{2}$ and 
\SMHexplow --\SMHexphigh~GeV/$c^{2}$. The results of the $H \to b\bar{b}$ 
searches were validated through a measurement of the diboson ($WZ+ZZ$) production 
cross section using the same data samples and analysis techniques, treating those 
diboson processes as signal.  The resulting diboson cross-section measurement is 
in agreement with the SM prediction. We observe a significant excess of events 
in the mass range between 115 and 140~GeV/$c^2$. The local significance 
at $m_H=125$~GeV/$c^2$ corresponds to 3.0 standard deviations, with a median expected significance,
assuming the SM Higgs boson is present at $m_H=125$~GeV/$c^2$, of 1.9 standard deviations.
with a best-fit signal strength of $1.44^{+0.59}_{-0.56}$ times the SM expectation. 
We also separately combined searches focusing on the $H \to b\bar{b}$, $H \to W^+W^-$, 
$H\rightarrow\tau^+\tau^-$, and $H\rightarrow \gamma\gamma$ decay modes.  The 
observed best-fit signal strengths obtained from each of these combinations are 
consistent with the expectations for a SM Higgs boson at $m_H=125$~GeV/$c^2$.  We 
performed tests of the compatibility of the observed excess with the expectations 
for the couplings of a SM Higgs boson and saw no significant deviations.  

\section{\label{sec:ackn}ACKNOWLEDGMENTS}
%
We thank the Fermilab staff and technical staffs of the participating institutions for their vital contributions.  
We acknowledge support from the
DOE and NSF (USA),
ARC (Australia),
CNPq, FAPERJ, FAPESP and FUNDUNESP (Brazil),
NSERC (Canada),
NSC, CAS and CNSF (China),
Colciencias (Colombia),
MSMT and GACR (Czech Republic),
the Academy of Finland,
CEA and CNRS/IN2P3 (France),
BMBF and DFG (Germany),
DAE and DST (India),
SFI (Ireland),
INFN (Italy),
MEXT (Japan),
the Korean World Class University Program and NRF (Korea),
CONACyT (Mexico),
FOM (Netherlands),
MON, NRC KI and RFBR (Russia),
the Slovak R\&D Agency, 
the Ministerio de Ciencia e Innovaci\'{o}n, and Programa Consolider-Ingenio 2010 (Spain),
The Swedish Research Council (Sweden),
SNSF (Switzerland),
STFC and the Royal Society (United Kingdom),
the A.P. Sloan Foundation (USA), 
and the EU community Marie Curie Fellowship contract 302103.


\begin{thebibliography}{999}
\bibitem[{hfj()}]{hfjet}
\bibinfo{note}{A heavy-flavor jet is a reconstructed cluster of calorimeter
  energies associated with particles produced in the hadronization and decay of
  a bottom or charm quark.}
\bibitem[{btag()}]{btag}
\bibinfo{note}{A $b$-tagged jet is one identified to have originated from the decay of a heavy flavor quark.}
\bibitem[{mix()}]{mixture}
\bibinfo{note}{As discussed later, a particular decay mode defined by an experimental signature as done here may be an admixture of several decay modes, though
dominated by the one denoted.}
\end{thebibliography}
\end{document}